\documentclass[preprint,11pt,twocolumn,usenatbib,twocolappendix]{aastex631}
\usepackage[utf8]{inputenc}
\usepackage{amsmath}
\usepackage{mathtools}
\usepackage{amsfonts}
\usepackage{amssymb}
\usepackage{makeidx}
\usepackage{indentfirst}
\usepackage{xcolor}
\usepackage{graphicx}
\usepackage{txfonts}
\usepackage[]{hyperref}
\usepackage{enumitem}
\usepackage{color}
\usepackage{lmodern}
\usepackage{url}
\usepackage{soul}
\usepackage{mathrsfs}
\usepackage{longtable}
\usepackage{multirow}
\usepackage [english]{babel}
\usepackage [autostyle, english = american]{csquotes}
\MakeOuterQuote{"}
\usepackage{chemformula}
\usepackage{textcomp}
\usepackage{aas_macros}
\usepackage{natbib}
\usepackage{tabularx}
\usepackage{comment}
\usepackage{soul}
\usepackage{xcolor}
\definecolor{voilet}{RGB}{127,0,255}
\definecolor{KA}{rgb}{0.36, 0.73, 0.58}

\shorttitle{Metal Oxide Clusters in Gas Giant Exoplanet Atmospheres}
\shortauthors{Bisht et al. (2025)}
\begin{document}
\title{Metal Oxide Clusters in Gas Giant Exoplanet Atmospheres}

\author[0009-0006-7546-5402]{Deepak Bisht}
\affiliation{Space Research Institute, Austrian Academy of Sciences, Schmiedlstrasse 6, A-8042 Graz, Austria}
\affiliation{Institute of Theoretical and Computational Physics, TU Graz, NAWI Graz, Petersgasse 16, 8010 Graz, Austria.}
\author[0000-0002-8275-1371]{Christiane Helling}
\affiliation{Space Research Institute, Austrian Academy of Sciences, Schmiedlstrasse 6, A-8042 Graz, Austria}
\affiliation{Institute of Theoretical and Computational Physics, TU Graz, NAWI Graz, Petersgasse 16, 8010 Graz, Austria.}
\author[0000-0002-3443-3416]{David Gobrecht}
\affiliation{Swiss Federal Institute of Intellectual Property (IGE), Stauffacherstrasse 65/59g, 3003 Bern, Switzerland}
\author[0000-0001-7934-0259]{Ludmila Carone}
\affiliation{Space Research Institute, Austrian Academy of Sciences, Schmiedlstrasse 6, A-8042 Graz, Austria}
\author[0000-0002-0469-5191]{Helena Lecoq-Molinos}
\affiliation{Space Research and Planetary Sciences, Physics Institute, University of Bern, Gesellschaftsstrasse 6, 3012 Bern, Switzerland.}
\author[0000-0002-8900-3667]{Peter Woitke}
\affiliation{Space Research Institute, Austrian Academy of Sciences, Schmiedlstrasse 6, A-8042 Graz, Austria}
\author[0000-0003-1034-5187]{Markus Aichhorn}
\affiliation{Institute of Theoretical and Computational Physics, TU Graz, NAWI Graz, Petersgasse 16, 8010 Graz, Austria.}
\author[0000-0001-8619-5096]{Jan Philip Sindel}
\affiliation{Space Research Institute, Austrian Academy of Sciences, Schmiedlstrasse 6, A-8042 Graz, Austria}
\author[0000-0001-7934-0259]{Amit Reza}
\affiliation{Space Research Institute, Austrian Academy of Sciences, Schmiedlstrasse 6, A-8042 Graz, Austria}

\begin{abstract}
This study investigates the thermal stability and absorption of metal oxide clusters in exoplanetary atmospheres. Utilizing our thermochemical data, we analyze eight distinct cluster families: magnesium oxide (MgO), silicon monoxide (SiO), titanium monoxide (TiO), vanadium monoxide (VO), titanium dioxide (TiO$_2$), vanadium dioxide (VO$_2$), aluminum oxide (Al$_2$O$_3$), and vanadium pentoxide (V$_2$O$_5$). Equilibrium cluster populations as a function of gas temperature and pressure reveal distinct stability regimes. Under solar elemental abundances, (TiO$_2$)$\rm\rm_N$ and (Al$_2$O$_3$)$\rm_N$ are favored at higher temperatures, while (MgO)$\rm_N$ and (SiO)$\rm_N$ dominate at lower temperatures. Computed absorption spectra exhibit strong size- and composition-dependent absorption features in the mid-infrared (8--50~$\mu$m), many of which fall within the wavelength range accessible to \texttt{JWST/MIRI}. We further coupled cluster thermodynamics with 3D general circulation model (GCM) outputs to investigate the cluster stability across the ultra-hot Jupiters (UHJs) WASP-121 b and WASP-18 b, the hot Jupiter (HJ) WASP-39 b, and the warm Jupiter (WJ) WASP-69 b. In WASP-121 b and WASP-18 b, extreme dayside temperatures suppress large-cluster stability, yielding atmospheres dominated by metal ions at low pressures and neutral metals at depth, with limited cluster survival on the nightside and morning terminator. In WASP-39 b, larger clusters are not thermochemically favoured despite the enhanced metallicity; instead, equilibrium chemistry stabilises smaller species, with only TiO showing a tendency toward stable larger cluster forms, likely due to its open d-orbitals. In contrast, WASP-69 b favors the formation of larger metal oxide clusters across an extended pressure range, highlighting WJs as a favorable environment for metal oxide cluster stability.
\end{abstract}
\section{Introduction} 
\label{sec:intro}
Cloud formation in exoplanetary and brown dwarf atmospheres differs significantly from that on Earth due to variations in chemical composition and physical conditions \citep{helling2019exoplanet}. On Earth, clouds form when supersaturated gases condense onto small rocky particles known as cloud condensation nuclei (CCN) \citep{hudson1993cloud}. In the case of rocky exoplanets, these CCN are typically supplied by the planetary sources, such as sulfite particles from volcanic activity \citep{andres1998time} or sandstorms. However, in gaseous exoplanets, such sources are absent. As a result, cloud formation might begin with the nucleation process, where gaseous atoms or molecules undergo chemical reactions, progressively increasing in size until they can undergo a phase transition from gas to solid, thereby forming macroscopic solid particles \citep{gail1984dust, 1999IAUS..191..233J, 2001A&A...376..194H, gail2014physics, lee2018dust, helling2019exoplanet, boulangier2019developing}. Numerous studies have employed a quantum-mechanical bottom-up approach to investigate nucleation \citep{chang2005inorganic, chang2013small, plane2013nucleation, patzer2014density, lam2015atomistic, bromley2016under, gobrecht2022bottom, sindel2022revisiting, lecoqhelena2024vanadium}. A central question in the study of cloud (or dust) formation is which gas-phase species nucleates first to form the primary condensate (\citealt{1986A&A...166..225G,1998FaDi..109..303G,2003A&A...407..191J,2023ASSP...59...89G}). For homogeneous nucleation, a viable candidate must meet three criteria: the least abundant element should be sufficiently concentrated, the condensed phase should have a high vaporization temperature, and the species must associate during molecular collisions \citep{gail2014physics}.

SiO has a high bond energy and can achieve substantial gas-phase abundances, making it a candidate for seed particle nucleation \citep{nuth1981vibrational, nuth1982experimental, gail1986primary}. Early doubts arose due to low apparent vaporization temperatures, but \citet{nuth2006silicates} revised the SiO vapor pressure, reinstating it as a potential silicate dust precursor \citep{reber2006silicon, reber2008sio, 2013A&A...555A.119G}. Nonetheless, gas-phase SiO presence does not necessarily imply active cloud formation and may indicate cloud-free conditions. Although Mg and Fe binding molecules are abundant (in a solar element composition) in the gas phase and can form condensed phases with high vaporization temperatures, their dimers exhibit unusually low bond energies, making them inefficient nucleation initiators \citep{gail2014physics}. From the perspective of elemental abundance and chemical stability, aluminum- and titanium-bearing oxides therefore emerge as the most promising nucleation candidates. Quantitative predictions of cluster number densities, however, require accurate thermochemical functions for clusters over a wide range of sizes. Quantum chemical calculations (QCC) have proven essential in providing such data, enabling recent studies to integrate cluster thermochemistry into the equilibrium code \texttt{GGchem} and to explore the abundance of larger clusters \citep{lee2015dust,gobrecht2022bottom, sindel2022revisiting, sindel2023infrared, lecoqhelena2024vanadium}. 
Considering multiple candidate species identifies those thermodynamically favoured across temperature regimes and likely to initiate condensation. With the \textit{James Webb Space Telescope} (\texttt{JWST}), particularly MIRI ($\sim$5--28~$\mu$m), vibrational spectra of molecular clusters are increasingly relevant for exoplanet observations. As absorption depends on cluster abundances, multi-species models are essential to interpret mixed spectral signatures.

3D global circulation models (GCMs) have become important tools for investigating the atmospheric structure in exoplanetary atmospheres \citep{2010ApJ...710.1395D,2014A&A...561A...1M, kataria2016atmospheric, carone2018stratosphere, parmentier2018thermal,2018ApJ...869..107M, 2019ApJ...883....4S,2023arXiv230108492C,2021MNRAS.502.2198T,PlaschzugW121b}. By self-consistently resolving the coupled dynamics, radiative transfer, and thermodynamics, GCMs capture the strong day–night temperature contrasts, equatorial jets, and vertical mixing that characterize irradiated giant exoplanets. These spatial and temporal variations play a critical role in determining where clouds can form, grow, and survive, particularly on tidally locked planets where longitudinal asymmetries dominate the atmospheric circulation \citep{2021A&A...649A..44H,lee2023modelling}. GCMs have been widely used to study key cloud-related properties, such as nucleation rates, particle sizes, and cloud vertical extent, and to explore how cloud formation and the resulting global cloud distributions depend on fundamental system parameters, including stellar irradiation, planetary gravity, rotation rate, and atmospheric composition \citep{helling2023exoplanet}. By providing 3D temperature–pressure structures, GCMs identify distinct cloud-forming regions—such as the nightside, terminators, and deep atmosphere—that are not captured by 1D models. They are therefore essential for linking cloud microphysics to phase-dependent transmission and emission spectra and for interpreting exoplanet observations.

The paper presents an inventory of thermodynamic data for metal oxide clusters. It provides insight into which of these cluster species may be relevant triggers for cloud formation and into which of these cluster species may need to be considered by a retrieval approach to interpret \texttt{JWST} observations \citep{baeyens2024detecting}. Therefore, this study addresses two central objectives. First, without resolving the long-standing question of the primary condensate \citep{1986A&A...166..225G}, we provide a systematic assessment of the relative abundances of metal oxide clusters across increasing sizes. For each cluster family and size, wavelength-dependent vibrational absorption spectra are computed and analyzed. Such thermodynamic and spectroscopic data are currently absent from standard databases (e.g., JANAF) and atmospheric retrieval frameworks, e.g., \citep{2020A&A...642A..28M,2022PSJ.....3...82B,2020A&A...639A...3C}, yet are essential for interpreting observations from current and upcoming facilities, including \texttt{JWST} \citep{2024jwst.prop.6045B}. This analysis further enables an investigation of how different cluster families behave in a mixed, solar-metallicity atmosphere, including the preferential stability of thermodynamically stable “magic-number” clusters under varying temperature and pressure conditions. Second, we apply our data to well-studied extrasolar planets and examine the metal cluster distribution in the atmospheres of two UHJs (WASP-121 b and WASP-18 b), a HJ (WASP-39 b), and a WJ (WASP-69 b). By combining 3D GCM outputs with chemical equilibrium calculations, we explore how locally changing thermodynamic conditions, planetary regime, and metallicity shape the spatial distribution of clusters and control the emergence of metal oxide clusters across UHJ, HJ, and WJ atmospheres.

\section{Approach}
\label{sec:approach}
This paper explores our thermodynamic data for Al-, Mg-, Si-, Ti-, and V-bearing oxide clusters to investigate their relative abundances and size distributions under local thermodynamic equilibrium (LTE). Following a summary of the thermochemical data, the clusters are analyzed at representative gas pressures of $p_{\rm gas} = 1, 10^{-4},$ and $10^{-8}$~bar over a temperature range of $T_{\rm gas} = 100\,\ldots\,6000$~K. These pressure levels correspond to the deep, collision-dominated atmosphere (1 bar), the observable atmosphere accessible to \texttt{JWST} ($10^{-4}$ bar), and a high-altitude, low-density regime ($10^{-8}$~bar). Cluster formation is implicitly treated as a collisional process under LTE conditions, and the results provide a first-order assessment of where metal oxide clusters may be stable\footnote{The calculations are performed assuming gas-phase chemical equilibrium only, without inclusion of non-LTE effects, time-dependent kinetics, transport processes (e.g. mixing or diffusion), photochemistry, or condensation.}. Additionally, the absorption for the considered metal oxide clusters is calculated in the wavelength range of 5-100 $\mu$m. Building upon these first results, four representative exoplanets are explored (WASP-18 b, WASP-39 b, WASP-69 b, and WASP-121 b) based on their 3D GCM \texttt{ExoRad} \citep{Carone2020,Schneider2022a} ($T_{\rm gas}$, $p_{\rm gas}$) distribution to study which clusters may be available to trigger the formation of CCN throughout their atmospheres. Four specific profiles (equatorial day side,  night side, morning, and evening terminator) are selected to study if and how the outermost atmospheric regions ($p_{\rm gas}< 10^{-4}$) that are not accessible through remote sensing techniques play a role in the formation of CCN. This systematic approach allows us to identify if and which metal clusters can contribute to the \texttt{JWST} observations \citep{baeyens2024detecting}. The initial calculations were performed assuming oxygen-rich solar elemental abundances from \citet{asplund2009chemical}, whereas for the planetary cases, the element abundances reported in the literature for each individual planet are adopted.

\smallskip
\textbf{Data Collection:}
This study analyzes QCC data for eight cluster species: MgO, SiO, TiO, VO, TiO$_2$, VO$_2$, Al$_2$O$_3$, and V$_2$O$_5$. Global minimum (GM) candidates for (TiO$_2$)$\rm\rm_N$ clusters (N$ = 1$--$15$) are provided in \citet{sindel2022revisiting}, where DFT calculations were performed at the B3LYP/cc-pVTZ level, and the absorption spectra data are provided in \citet{sindel2023infrared}. For vanadium oxides, GM structures of (VO)$\rm\rm_N$ (N$ = 1$--$10$), (VO$_2$)$\rm\rm_N$ (N$ = 1$--$10$), and (V$_2$O$_5$)$\rm\rm_N$ (N$ = 1$--$4$) are provided in \citet{lecoqhelena2024vanadium}, absorption data are provided in \citet{lecoq2025microphysics}. GM data for (SiO)$\rm\rm_N$ (N$ = 1$--$20$) are provided in \citet{bromley2016under} (initial structures) and \citet{lecoq2025microphysics} (DFT calculations), computed at the B3LYP/cc-pVTZ level; both thermochemical and absorption properties are provided in \citet{lecoq2025microphysics}. For (TiO)$\rm\rm_N$ (N$ = 1$--$10$), GM structures, thermochemical properties, and absorption spectra are provided in \citet{lecoq2025microphysics}, based on B3LYP/cc-pVTZ calculations. For (Al$_2$O$_3$)$\rm_N$ (N$ = 1$--$10$), GM structures and thermochemical properties are provided in \citet{gobrecht2022bottom}, computed using CBS-QB3, B3LYP, and PBE0 methods with cc-pVTZ and 6-311+G(d) basis sets, depending on cluster size. Absorption spectra are calculated in this work following \citet{sindel2023infrared}. For (MgO)$\rm_N$ (N $ = 1$--$10$), GM structures and thermochemical properties are adopted from \citet{chen2014structures} and \citet{boulangier2019developing}, based on B3LYP/6-311+G$\ast$ calculations, with absorption spectra computed following \citet{sindel2023infrared}.

\smallskip
\textbf{Cluster number densities:} 
The gas-phase equilibrium chemistry code \texttt{GGchem} (\citealt{woitkeggchem2018}) is used to compute the number densities, n$_{\rm i}$ [cm$^{-3}$], of a cluster $i$  at a given gas temperature, $T_{\rm gas}$ [K]. The calculations assume solar element abundances \citep{asplund2009chemical}, except for WASP-39b, where 10$\times$ of the individual solar abundances are adopted. The details of the equilibrium calculation are outlined in Sect. 2.1. in \citet{woitkeggchem2018} and the role of the element abundances for mass conservation follow from their Eq. 5. Molecular equilibrium constants ($k_p$) for all clusters are fitted following \citet{woitkeggchem2018}, based on standard Gibbs free energies of formation, $\Delta G_{\rm f}^\circ(T)$, from QCC data in the literature. The fitted values for all 89 GM clusters are incorporated. In addition, the thermal stability temperature for each metal oxide solid (TiO$_2$[s] (Rutile), SiO[s] (metastable SI-monoxide), SiO$_2$[s] (Quartz), MgO[s] (Periclase), Al$_2$O$_3$[s] (Corundum ($\alpha$)), V$_2$O$_5$[s] (Shcherbinaite), TiO[s] (TI-Monoxide ($\beta$))) is determined. Thermal stability refers to the temperature at which the rate of evaporation, defined as the number of atoms, molecules, or ions leaving the surface per unit time, is equal to the rate of growth, i.e., the number of atoms, molecules, or ions contributing to the increase in volume \citep{goeres1996chemistry, vehkamaki2006classical, helling2019exoplanet}. At thermal stability of the material (e.g., TiO[s], MgO[s]),  the supersaturation ratio $S(T_{\rm gas}$) = 1. The approach outlined in \cite{woitkeggchem2018}  is used to derive the supersaturation ratio (Eqs. 8 and 11). It is derived from the Gibbs free energy of formation of the condensed phase of interest and uses the atomic partial pressure of the elements involved. The supersaturation ratios are derived for each material individually, without accounting for element depletion.

\smallskip
\textbf{Absorption Cross Sections:} 
To compute the absorption cross sections, we follow \citet{sindel2023infrared}. In addition to the thermodynamic properties, QCC can be used to perform a vibration analysis to retrieve vibrational frequencies and their corresponding infrared (IR) intensities—required for calculating molecular absorption features—are adopted from the literature. Frequencies ($\nu$) and IR intensities (I$_{\rm IR}$) are expressed in units of [cm$^{-1}$] and [km mol$^{-1}$]. All vibrational modes are treated as purely harmonic, following \citet{sindel2023infrared}. Electronically excited states are not included in the calculations, and the focus is placed solely on internal vibrations and rotations. Furthermore, no coupling between rotational and vibrational modes, nor any vibration–vibration coupling, is considered. The rotational constants for (TiO$_2)_{\rm N}$, (TiO)$_{\rm N}$, (SiO)$_{\rm N}$, and vanadium-containing species clusters are provided in \citet{lecoq2025microphysics}, while those for  (Al$_2$O$_3$)$_{\rm N}$ and (MgO)$_{\rm N}$ are listed in Tables~\ref{al2o3rotationalconstants} and \ref{mgorotationalconstants}. The IR intensities are converted into molecular absorption coefficients as described by \citet{spanget2015ir}. A key parameter in this conversion is the line width associated with each IR-active mode. To quantify this, four broadening mechanisms are considered: thermal, collisional, rotational, and natural broadening. Each mechanism is evaluated for all vibrational transitions of every cluster, and the dominant contribution is adopted to determine the full width at half maximum (FWHM). The resulting FWHM is then used to compute the absorption cross section, $\sigma_{\text{cluster}}$ [cm$^2$], as follows:

\begin{equation} \label{eq:crosssection}
(\sigma_{\text{cluster}})_{\text{max}} = 27648 \frac{\rm I_{\rm IR}}{w} \, ,
\end{equation}
where $(\sigma_{\text{cluster}})_{\text{max}}$ is the maximum cross-section in [cm$^2$ mole$^{-1}$], I$_{\rm IR}$ is the IR intensity in [km mol$^{-1}$], and \textit{w} is FWHM in [cm$^{-1}$]. The wavelength-dependent absorption, $\kappa (\lambda)$ in [cm$^{-1}$],  for the metal oxide clusters is then derived using the result from the chemical equilibrium calculations $n_{\rm i}=n_{\rm cluster}$ [cm$^{-3}$]
\begin{equation} \label{eq:totalabs}
\kappa (\lambda) = n_{\text{cluster}}\sigma_{\text{cluster}}.
\end{equation}

\smallskip
\textbf{3D \texttt{ExoRad} GCM and extrapolation of selected 1D profiles:} 
3D \texttt{ExoRad} GCM results for the local ($T_{\rm gas}$,$p_{\rm gas}$) atmosphere distribution are the input for calculating the number densities, $n_{\rm i}$, of the metal oxide clusters with \texttt{GGchem}. The following four planets were selected that have been observed with JWST: WASP-18\,b (global temperature 2392~K; \citet{DelineW18b}), WASP-39\,b (global temperature 1117~K; \citet{SteinrueckW39b}), WASP-69\,b (global temperature 964~K; Bangera et al. in prep), and WASP-121\,b (global temperature 2360~K; \citet{PlaschzugW121b}). WASP-18 b and WASP-121 b are UHJs with similar global temperatures but very different surface gravities (190.5~m/s$^2$ and 8.44 m/s$^2$) (Planetary parameters listed in Table \ref{Exoradparameters} in the appendix). The gas pressure of the 3D \texttt{ExoRad} GCM covers the interval $p_{\rm gas}=750\,\ldots\,10^{-4}\,$ bar within its computational volume where the respective conservation equations are considered to be unaffected by boundary conditions (i.e., excluding the ghost cells). 

Four representative 1D equatorial profiles (day, night, morning, and evening) are extracted from the 3D \texttt{ExoRad} GCM and extended to  $p_{\rm gas} < 10^{-3}$~bar for WASP-121 b, WASP-18 b, WASP-39 b, and WASP-69 b. This extension enables an exploration of cluster stability in upper atmospheric regions that are not resolved by the GCM. The extrapolation is performed on a logarithmically spaced pressure grid using a parameterized temperature structure \citep{madhusudhan2009temperature}, with boundary conditions guided by previous studies of irradiated exoplanets. For substellar regions, the temperature is assumed to increase toward an upper-atmosphere asymptote of $T_{\rm gas} \sim 10^{4}$~K, consistent with thermospheric heating driven by stellar irradiation and hydrodynamic escape \citep{yelle2004aeronomy, munoz2007physical}. Nightside profiles adopt an outer $T_{\rm gas} \sim 100$~K, representing efficient radiative cooling in the absence of direct irradiation \citep{helling2023exoplanet}. For the terminator regions, an isothermal structure is assumed for $p_{\rm gas} < 10^{-3}$~bar, consistent with the flattening of temperature gradients expected in low-density upper atmospheres. These assumptions are broadly consistent with trends reported in observational studies of highly irradiated exoplanets, including strong day--night contrasts and heated upper atmospheres \citep[e.g.][]{DelineW18b, garai2025kelt}. The extrapolation is carried down to $p_{\rm gas} = 10^{-12}$~bar, where the gas can still be approximated as collisionally coupled and treated within a hydrodynamic framework, as described in Section~5 of \citep{helling2023exoplanet}.

\section{Metal oxide cluster abundances and their absorption spectra}
\label{clusterabun}
Metal oxide clusters are among the first solids to form in hot, oxygen-rich environments due to favorable energetics and growth kinetics \citep{goeres1996chemistry, helling2013modelling, lee2018dust}. While various clusters such as (TiO$_2$)$_N$, (TiO)$_N$, (Al$_2$O$_3$)$_N$, and vanadium and silicon oxides have been studied \citep{sindel2022revisiting, lecoq2025microphysics, gobrecht2022bottom, bromley2016under}, their relative importance in mixed atmospheres depends on elemental abundances, C/O ratio, and local thermodynamic conditions \citep{mahapatra2017cloud}. Here, we examine a chemically mixed, solar-metallicity atmosphere to identify thermodynamically favored clusters and their spectral signatures.

\subsection{Concentrations of Metal oxide clusters}
\label{numberdensity}
This section studies the concentrations of (Al$_2$O$_3$)$\rm\rm_N$ and (MgO)$\rm\rm_N$ at $p_{\rm gas}$=1~bar for varying gas temperatures, and their impact on other ionic and gaseous species containing the same metal is evaluated in this section. This study may be supplemented by previous publications:(TiO$_2$)$\rm\rm_N$ by \citet{sindel2023infrared, sindel2022revisiting}, for Vanadium species by \citet{lecoqhelena2024vanadium}, and for (SiO)$\rm\rm_N$ and (TiO)$\rm\rm_N$ by \citet{lecoq2025microphysics}. The results are presented in terms of concentrations, $n_{\rm i}$/$n_{\rm tot}$, and discussed in comparison to the cluster of highest thermal stability, which is derived from $\Delta G^\circ_f(\rm N)/\rm N$. A comparison of different monomer data shows that the uncertainties in those thermodynamic data may be negligible for the study of larger metal-oxide cluster concentration and, hence, for their absorption coefficients (Sects.~\ref{wavelengthdepopa},~\ref{results-total}) as input for retrieving information from observations of present (HST, \texttt{JWST}; e.g., \citealt{2020A&A...639A...3C}) and future (e.g., Ariel) space missions (e.g., \citealt{2024RASTI...3..636C}). 

\subsubsection{(Al$_2$O$_3$)$\rm\rm_N$}
\label{al2o3abun}
\begin{figure}[ht!]
\centering
\begin{minipage}[b]{0.45\textwidth}
\centering
\includegraphics[width=\linewidth]{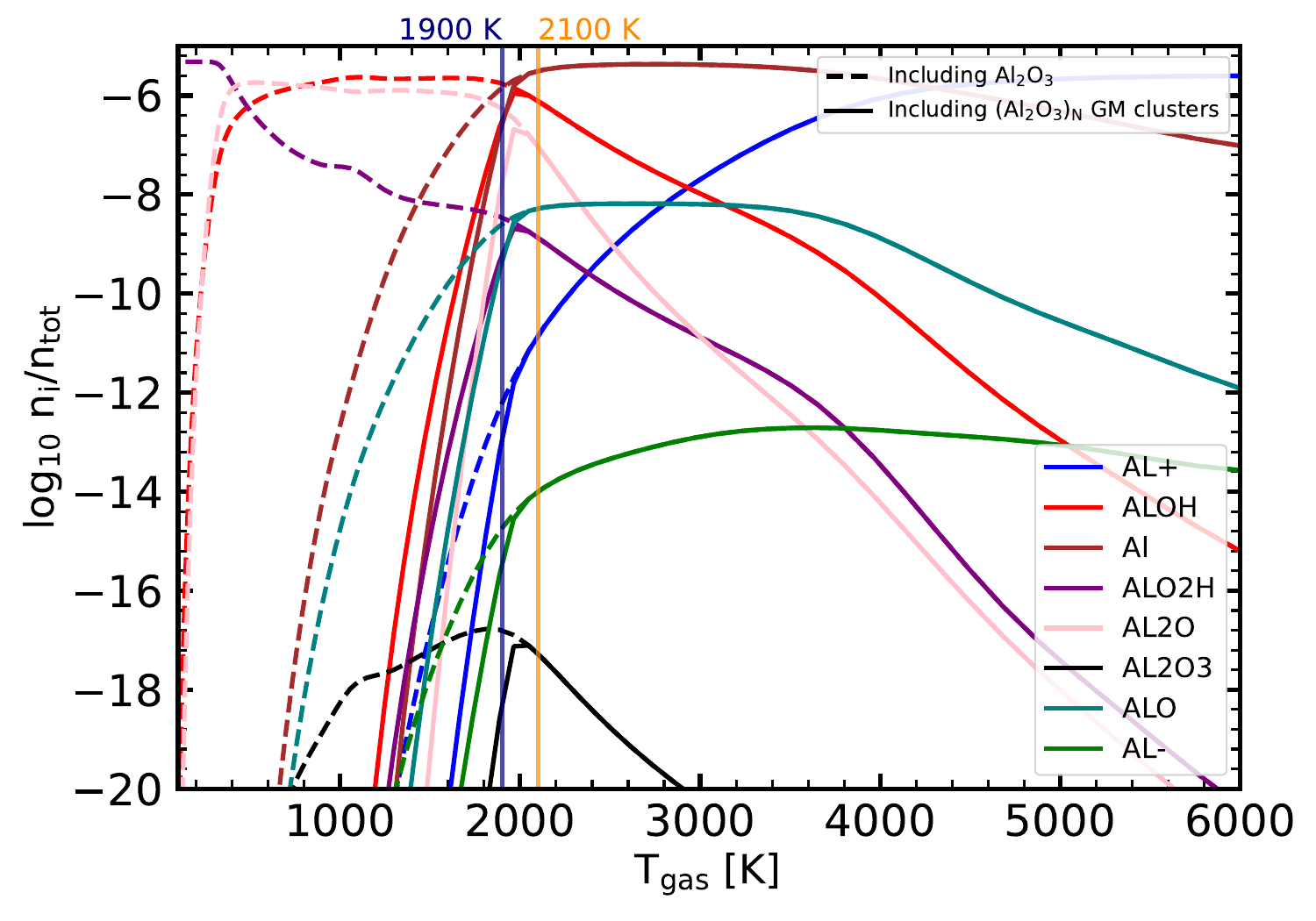}
\end{minipage}
\begin{minipage}[b]{0.45\textwidth}
\centering
\includegraphics[width=\linewidth]{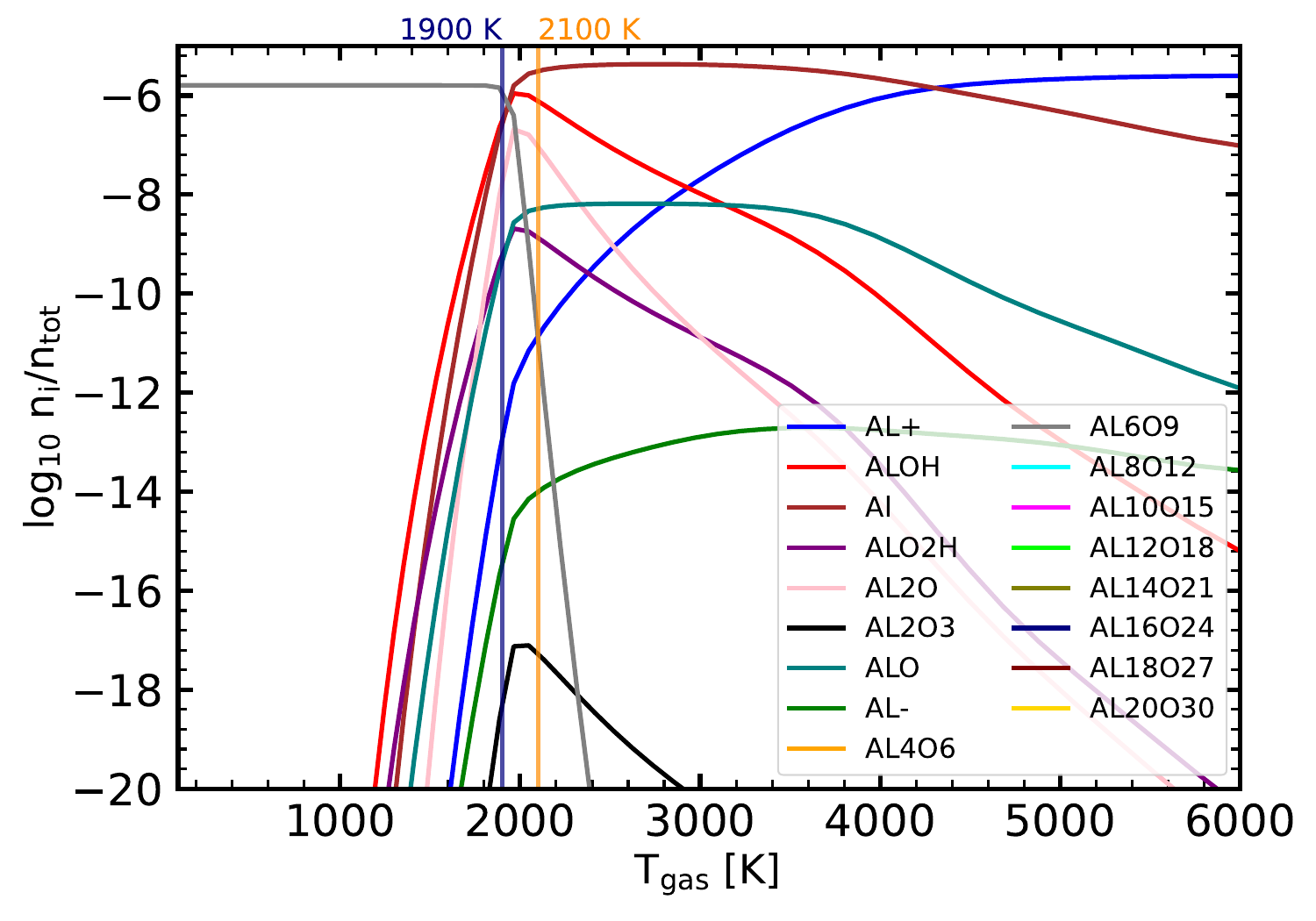}
\end{minipage}
\caption{Concentrations, $n_{\rm i}$/$n_{\rm tot}$ in chemical equilibrium at $p_{\rm gas}=1$ bar \textbf{Top}: Al ions and molecules, \textbf{Bottom:} Al ions, molecules and clusters.}
\label{al2o3numb}
\end{figure}

For our analysis, we incorporate the Al$_2$O$_3$ monomer and GM structures of larger clusters directly obtained from \citet{gobrecht2022bottom}. In the top panel of Fig.~\ref{al2o3numb}, we first consider only the updated thermochemical data for Al$_2$O$_3$. Under these conditions, AlO$_2$H, Al$_2$O, and AlOH emerge as the most abundant Al-bearing species at gas temperatures of $T_{\rm gas} \lesssim 2100\,\mathrm{K}$, while Al$_2$O$_3$ remains the least abundant species. This result indicates that, even under chemical equilibrium, only a small fraction of the available aluminium is partitioned into Al$_2$O$_3$ monomers over much of the temperature range considered. Such behaviour is qualitatively consistent with previous studies of alumina chemistry. For example, \citet{boulangier2019developing} found that Al$_2$O$_3$ molecules do not form efficiently from an initially atomic gas in their kinetic network, preventing the growth of large (Al$_2$O$_3$)$_N$ clusters. Similarly, \citet{2007ASPC..378..181P,2013RSPTA.37110581H,decin2017study} and \citet{gobrecht2022bottom} highlighted the importance of AlO-, AlOH-, and related Al$_x$O$_y$ species in alumina formation pathways. When all the (Al$_2$O$_3$)$\rm \rm_N$ (N = 1 to 10) are considered in chemical equilibrium the number densities of AlO$_2$H, Al$_2$O, and AlOH change significantly, and this change occurs only up to temperatures close to where Al$_2$O$_3$[s] is thermally stable ($T_{\rm gas}$ \(\lesssim 1900\) K), showing that the cluster size increases in the temperature interval where the (Al$_2$O$_3$)$\rm\rm_N$ clusters are supersaturated \citep{helling2019exoplanet}. This decrease in the abundances of the originally prevalent Al-bearing molecules occurs because most of the aluminium is instead consumed by $(\mathrm{Al}_2\mathrm{O}_3)_3$ (grey solid line; Fig.~\ref{al2o3numb}, middle), which becomes the most abundant aluminium-containing species at gas temperatures T$_{\mathrm{gas}} \lesssim 1900\,\mathrm{K}$. For most cluster families considered in this study, the largest cluster of a given species becomes the most abundant at lower $T_{\rm gas}$. In contrast, for aluminum oxide, the trimer $(\mathrm{Al}_2\mathrm{O}_3)_3$  represents the most abundant species under these conditions. The seemingly enhanced stability at N = 3 can be attributed to a methodological artefact arising from the transition between zero-point energy calculations: CBS-QB3 values are employed for smaller clusters, whereas B3LYP zero-point energies are used for N $\geq 4$ due to computational constraints \citep{gobrecht2022bottom}.

\subsubsection{(MgO)$\rm\rm_N$}
\label{ss:MgO}
\begin{figure}[ht!]
\centering
\begin{minipage}[b]{0.45\textwidth}
\centering
\includegraphics[width=\linewidth]{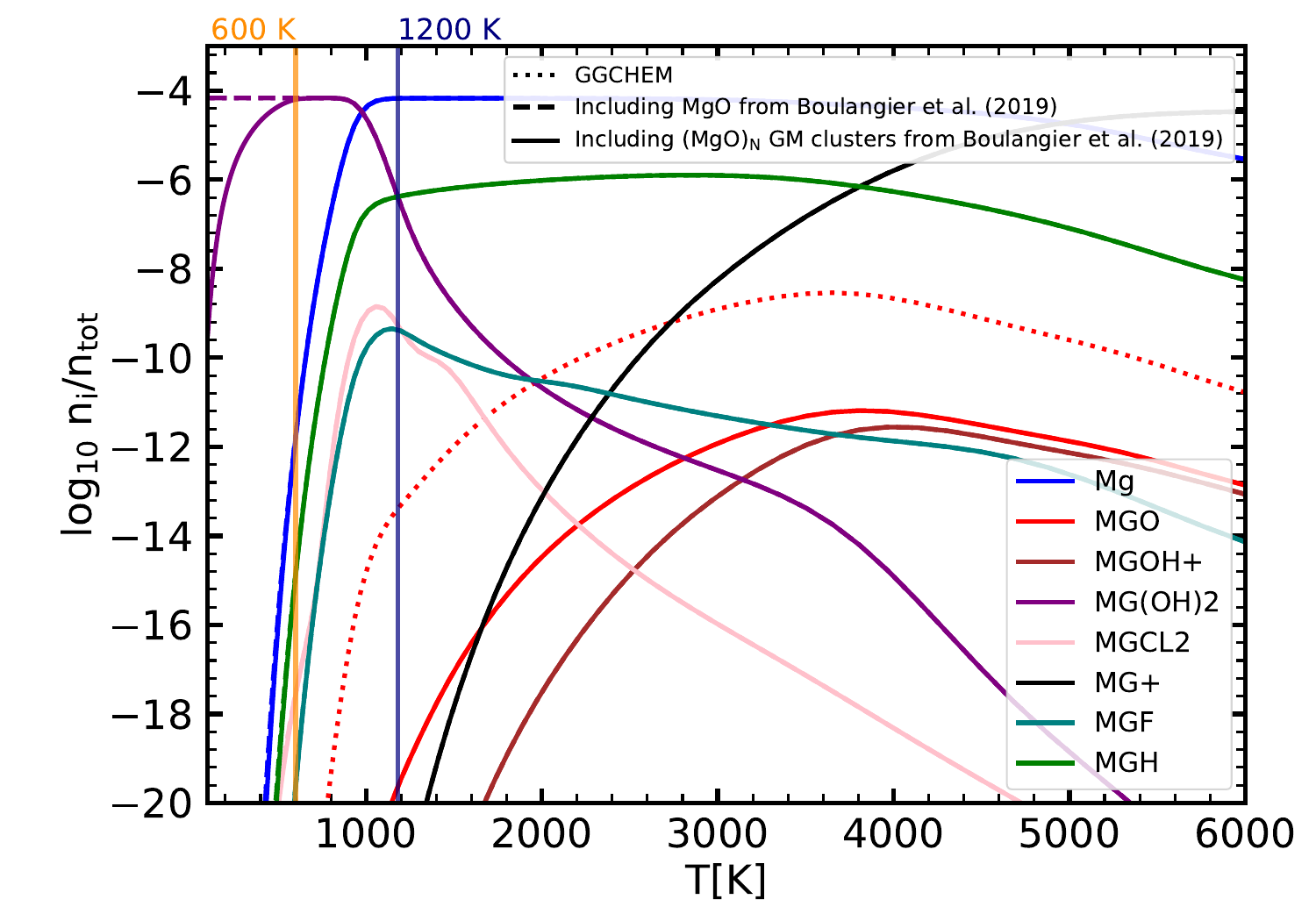}
\end{minipage}
\begin{minipage}[b]{0.45\textwidth}
\centering
\includegraphics[width=\linewidth]{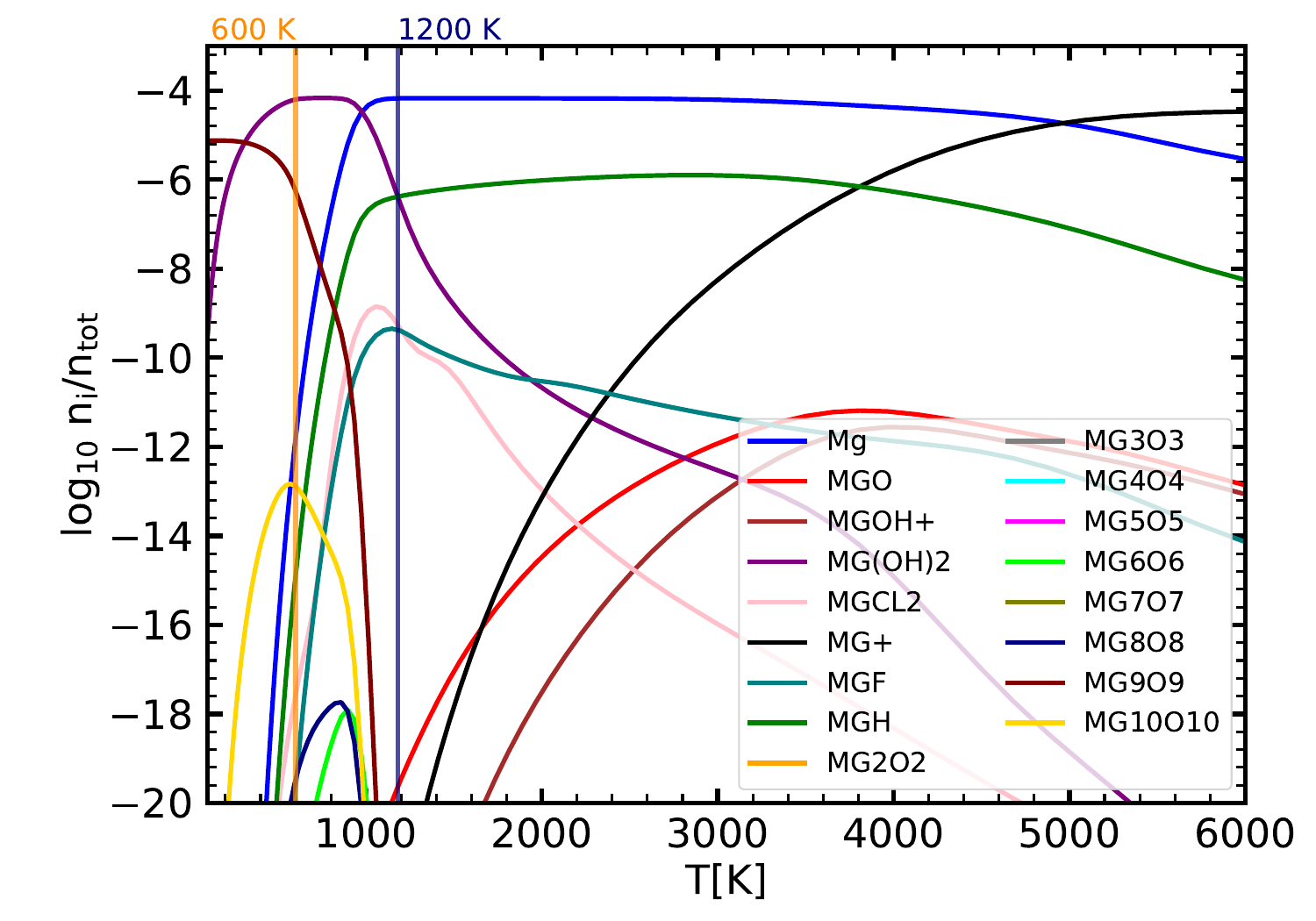}
\end{minipage}
\begin{minipage}[b]{0.44\textwidth}
\centering
\includegraphics[width=\linewidth]{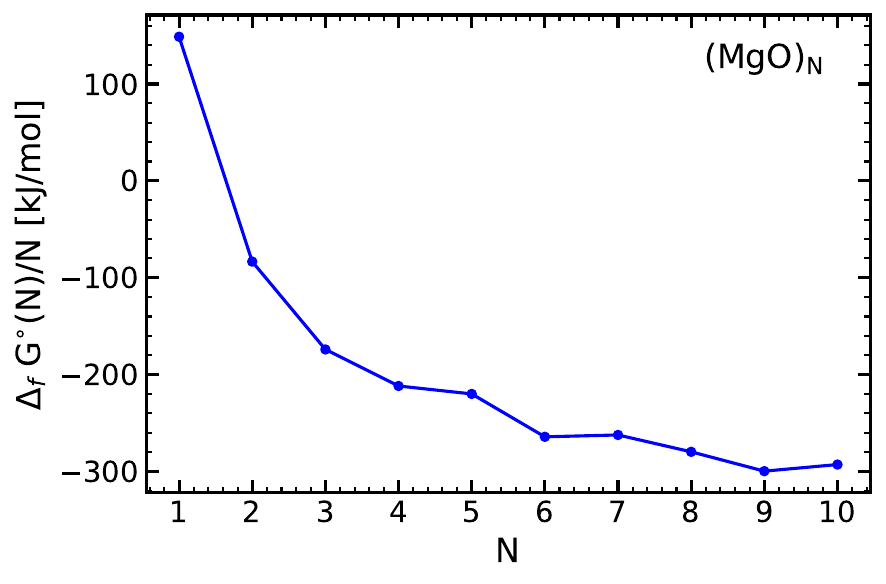}
\end{minipage}
\caption{Concentrations, $n_{\rm i}$/$n_{\rm tot}$ in 
chemical equilibrium at $p_{\rm gas}=1$ bar \textbf{Top}: Mg ions and molecules, \textbf{Middle:} Mg ions, molecules and clusters. \textbf{Bottom:} $\Delta G^\circ_f(N)/\rm N$ of the GM candidate of (MgO)$\rm\rm_N$ clusters at T$_{\rm gas}$ \(\sim 500\) K.}
\label{MgOnumb}
\end{figure}

We used the $\Delta G^\circ_f(N)$ values for (MgO)$_{\rm N}$ (N=$1 - 10$) clusters obtained from \citet{boulangier2019developing} (based on \citealt{1997A&A...320..553K,chen2014structures}) to calculate the number densities of Mg-bearing species at $p_{\rm gas}$=1~bar. The updated $\Delta G^\circ_f(N)$ values for MgO (dotted red line), relative to the JANAF data, primarily affect the number densities of MgO, while having minimal impact on the other Mg-bearing species (Fig.~\ref{MgOnumb}, top). The change in MgO number densities arises from differences in the adopted MgO cluster structures between the JANAF database and those reported by \citet{boulangier2019developing}. Additionally, when (MgO)$\rm\rm_N$ clusters (N$ = 1$--$10$) are included, only Mg(OH)$_2$ shows a change in number density, limited to very low temperatures ($T_{\rm gas} \lesssim 600$~K). In this regime, the (MgO)$\rm\rm_N$ clusters are supersaturated, while solid MgO (MgO[s]) remains thermally stable up to $T_{\rm gas} \lesssim 1200$~K. This behaviour is driven mainly by the larger clusters, particularly (MgO)$_9$ and (MgO)$_{10}$. (MgO)$_9$ becomes the most abundant only at very low temperatures, $T_{\rm gas}$ \(\lesssim 300\) K. Above this temperature, MgOH remains the most abundant Mg-bearing species up to $T_{\rm gas}$ \(\sim 600\) K. Additionally, (MgO)$_9$ is found to be more stable than (MgO)$_{10}$. Unlike (Al$_2$O$_3$)$_N$, for which different levels of theory are employed, the calculations for MgO clusters are performed consistently using the same basis set and functional. Therefore, the observed behavior can be attributed to enhanced stability at specific cluster sizes of (MgO)$_N$ relative to both smaller and larger sizes, a feature commonly referred to as a ``magic cluster'' \citep{harbola1992magic}. In this context, enhanced stability is associated with electronic shell effects that lead to increased stability at particular sizes. Such clusters are thus more stable than neighboring sizes within the considered range. For metal oxide clusters, this stability can additionally be assessed through quantities such as the free energy per atom \citep{wang2018magic}. In this work, we evaluate the Gibbs free energy per cluster size, $N$, as shown in the bottom panel of Fig.~\ref{MgOnumb}. The figure clearly indicates that $(\mathrm{MgO})_9$ possesses the lowest $\Delta G^\circ_f(N)/N$, confirming its classification as a magic cluster. However, the present dataset for MgO clusters extends only one size beyond the identified magic cluster, and the difference in Gibbs free energy per monomer between (MgO)$_{10}$ and (MgO)$_9$ remains relatively small. It is therefore plausible that the next particularly stable cluster appears at sizes $N = 11$ or $12$, consistent with trends observed at smaller sizes, where (MgO)$_6$ is more stable than (MgO)$_7$.

\subsection{Wavelength dependent absorption}
\label{wavelengthdepopa}
To complement existing data, we provide wavelength-dependent absorption for (Al$_2$O$_3$)$_N$ and (MgO)$_N$, while absorption data for other clusters are taken from the literature \citep{sindel2023infrared, lecoq2025microphysics}. These two species are selected due to their contrasting stability regimes: (Al$_2$O$_3$)$_N$ remains stable at higher temperatures ($T_{\rm gas} \sim 2000$~K), whereas (MgO)$_N$ clusters are stabilized at lower temperatures ($T_{\rm gas} \lesssim 600$~K). Absorption is evaluated at $p_{\rm gas} = 1$, $10^{-4}$, and $10^{-8}$~bar to probe variations with atmospheric depth. The results indicate that spectral features from different clusters often overlap, making the identification of individual species challenging, particularly at low spectral resolution.

\subsubsection{(Al$_2$O$_3$)$_{\rm N}$}
\begin{figure}[ht!]
\centering
\begin{minipage}[b]{0.45\textwidth}
\centering
\includegraphics[width=\linewidth]{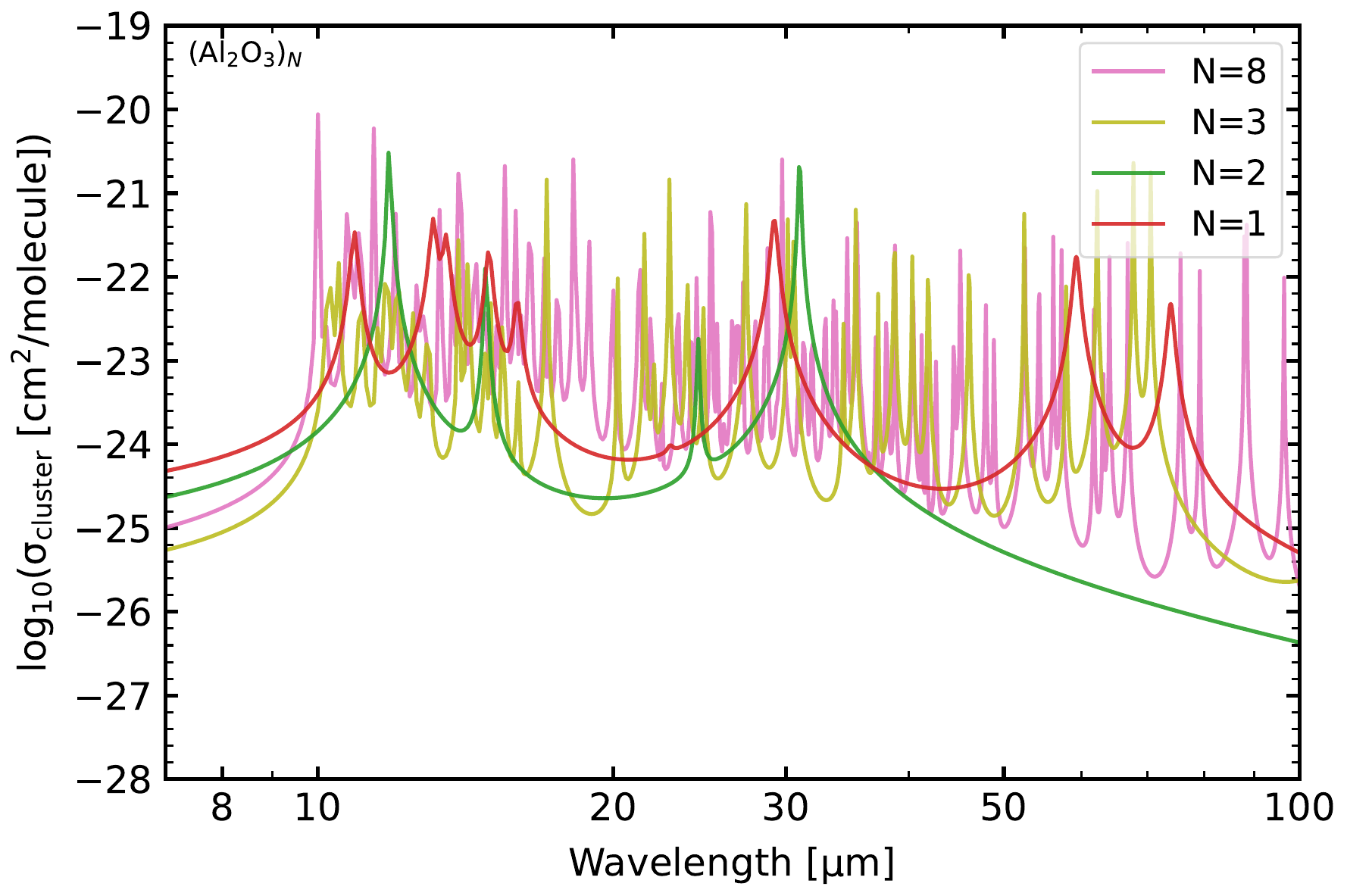}
\end{minipage}
\begin{minipage}[b]{0.45\textwidth}
\centering
\includegraphics[width=\linewidth]{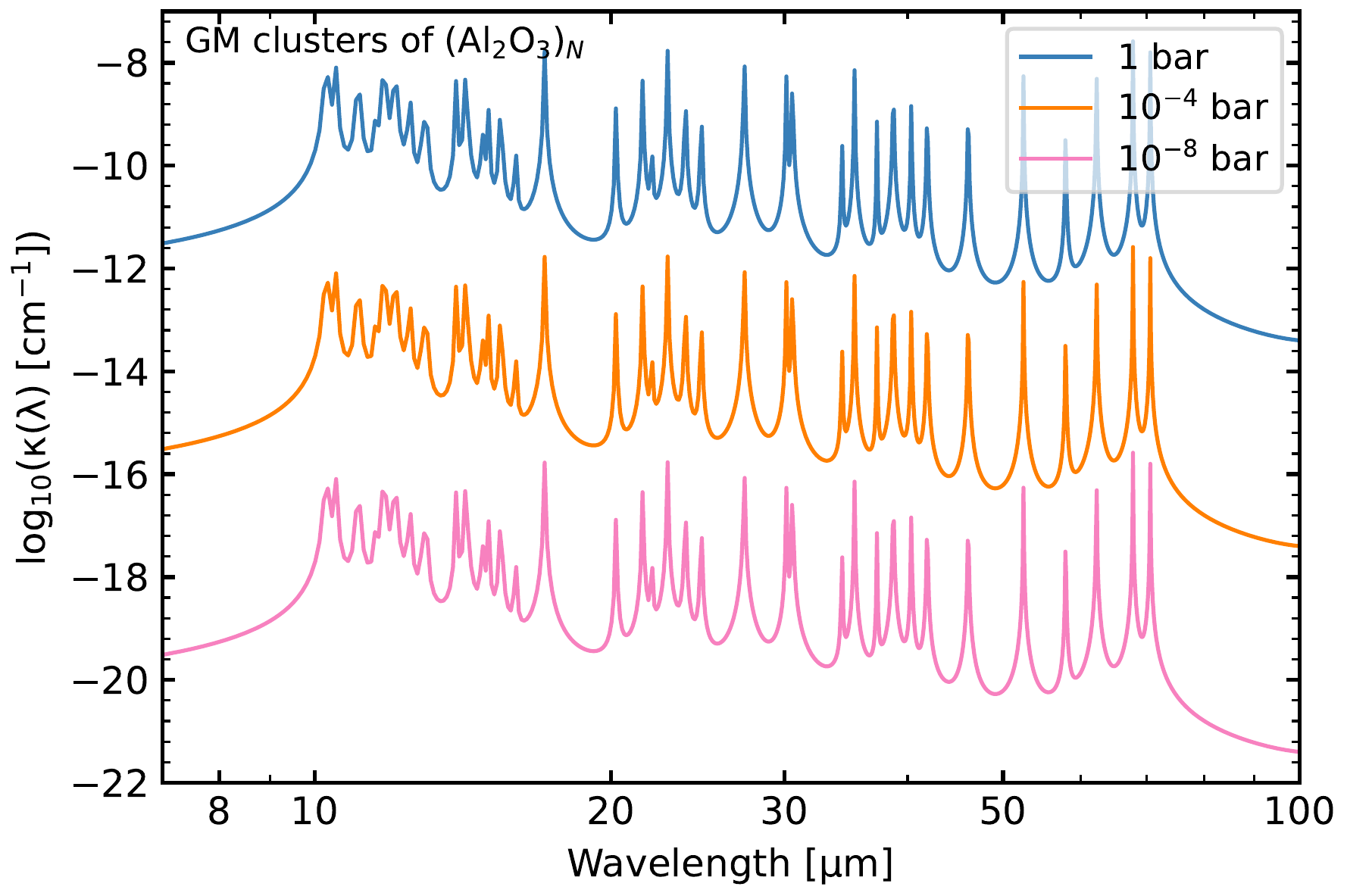}
\end{minipage}
\caption{Top: Wavelength-dependent cross-sections, $\sigma_{\text{cluster}}(\lambda)$, of (Al$_2$O$_3$)$\rm\rm_N$ clusters with N = 1, 2, 3, and 8, evaluated at $T_{\rm gas}$= 1000 K. Bottom: Absorption spectra of (Al$_2$O$_3$)$\rm\rm_N$ clusters for N = 1–8 under varying $p_{\rm gas}$ conditions, each at their corresponding $T_{\rm gas}$ = 1000 K.}
\label{absorspecal2o3}
\end{figure}

The wavelength-dependent cross-sections for (Al$_2$O$_3$)$\rm_N$ clusters with N$ = 1, 2, 3,$ and $8$ at $p_{\rm gas} = 1$~bar and $T_{\rm gas} = 1000$~K are shown in the top panel of Fig.~\ref{absorspecal2o3}. These sizes are selected to illustrate the variation in cross-sections from smaller to larger clusters. We limit our analysis to clusters up to N = 8 because IR intensity data for (Al$_2$O$_3$)$_{\rm N}$ clusters are currently available only for this size range. Additionally, N = 3 is included because (Al$_2$O$_3$)$_3$ is identified as the most abundant species among all (Al$_2$O$_3$)$_{\rm N}$ clusters, as discussed in Section~\ref{al2o3abun}. The top panel of Fig.~\ref{absorspecal2o3} clusters, as discussed in Section ~\ref{al2o3abun} shows that smaller (Al$_2$O$_3$)$_{\rm N}$ clusters exhibit high absorption at shorter wavelengths. Notable spectral features appear for N = 1, 2 around $\lambda$ \(\approx 11 \, \mu\text{m}\), with additional features emerging near $\lambda$ \(\approx 30 \, \mu\text{m}\). The N = 2 cluster also shows distinct features at $\lambda$ \(\approx 60 \) and \( 75 \, \mu\text{m}\). These results indicate that small (Al$_2$O$_3$)$_{\rm N}$ clusters contribute to spectral features across a broad $\lambda$ range. The most abundant cluster, (Al$_2$O$_3$)$_3$, exhibits several strong features between $\lambda$ \(\approx 10-71 \, \mu\text{m}\). In particular, a dense cluster of features appears at $\lambda$ \(\approx 11-45 \, \mu\text{m}\) range, where multiple peaks dominate specific wavelengths. For $\lambda$ \(\gtrsim 45 \, \mu\text{m}\), the features become more isolated, with four prominent absorption peaks appearing at $\lambda$ \(\approx 52 \,, 62 \,, 67 \,\), and  \(71 \, \mu\text{m}\). For N = 8, the spectral features are distributed across $\lambda$ \(\approx 10-100 \, \mu\text{m}\). Within $\lambda$ \(\approx 10-50 \, \mu\text{m}\), the features are densely packed, while at $\lambda$ \(\gtrsim 50 \, \mu\text{m}\), they are more sparsely distributed. 

The combined absorption spectrum for cluster sizes N$ = 1$--$8$ of (Al$_2$O$_3$)$_N$ is shown in the bottom panel of Fig.~\ref{absorspecal2o3}. The strongest absorption features occur at deeper atmospheric levels ($p_{\rm gas} \sim 1$~bar), with overall intensity decreasing toward lower pressures, while the number of spectral features remains largely unchanged. The enhanced absorption at higher pressures reflects the greater number densities of stable larger clusters under elevated gas densities. In an isothermal atmosphere (possibly in the terminator regions, e.g. \citealt{DelineW18b}), this implies that fewer large clusters remain thermochemically favoured at lower pressures, although stable cluster populations persist throughout the column. Notably, the absence of features beyond $\lambda \gtrsim 70 \, \mu\mathrm{m}$ indicates that the N$ = 3$ cluster dominates the spectrum, due to its relatively high number density compared to other sizes. This is consistent with observations of small gas-phase alumina clusters in AGB winds, which suggest that clusters up to N$ = 4$ are present before efficient dust condensation occurs \citep{evans2016detection}.
\subsubsection{(MgO)$_{\rm N}$}
\begin{figure}[ht!]
\centering
\begin{minipage}[b]{0.45\textwidth}
\centering
\includegraphics[width=\linewidth]{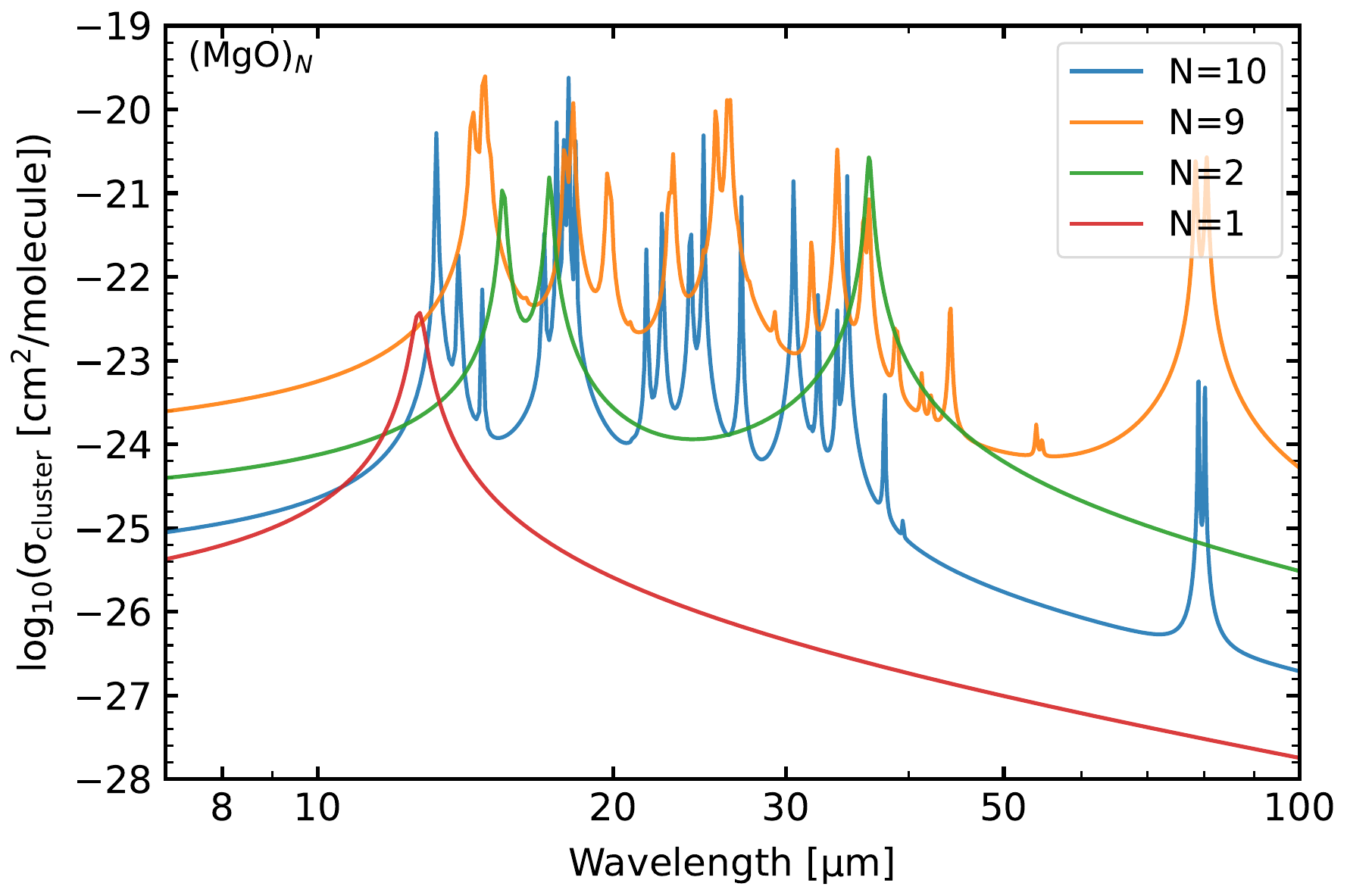}
\end{minipage}
\begin{minipage}[b]{0.45\textwidth}
\centering
\includegraphics[width=\linewidth]{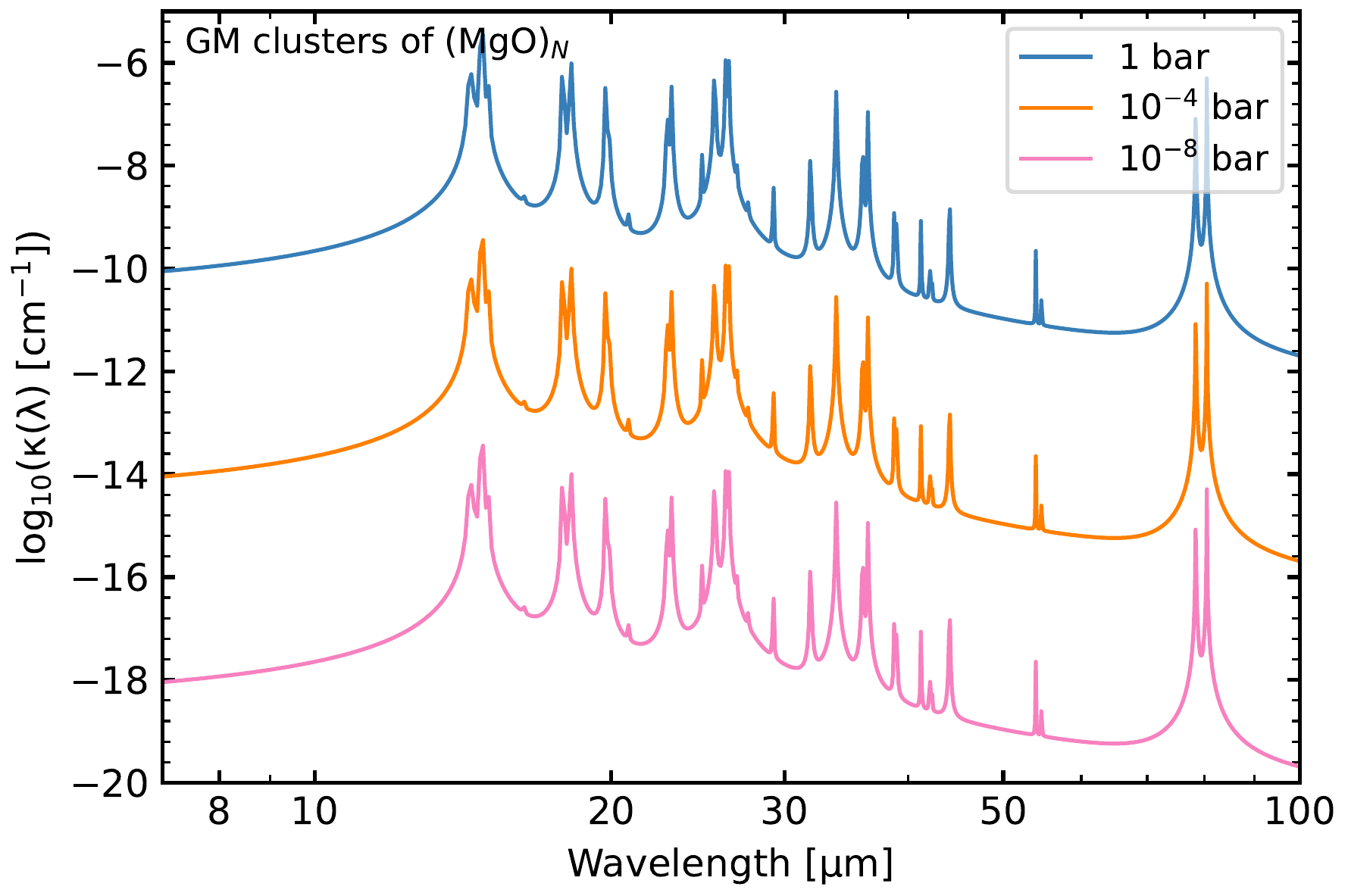}
\end{minipage}
\caption{Top: Wavelength-dependent cross-sections, $\sigma_{\text{cluster}}(\lambda)$, of (MgO)$\rm\rm_N$ clusters with N = 1, 2, 9, and 10, evaluated at T$_{\rm gas's}$= 250 K. Bottom: Absorption spectra of (MgO)$\rm\rm_N$ clusters for N = 1–10 under varying $p_{\rm gas}$ conditions, each at their corresponding $T_{\rm gas}$ = 250 K.}
\label{absorspecamgo}
\end{figure}

The wavelength-dependent cross-sections for (MgO)$\rm_N$ clusters with N$ = 1, 2, 9,$ and $10$ at $p_{\rm gas} = 1$~bar and $T_{\rm gas} = 250$~K are shown in the top panel of Fig.~\ref{absorspecamgo}. The observed trends are similar to those seen for TiO clusters by \citet{lecoq2025microphysics}, with most features appearing in the $\lambda$ \(\approx 15-45 \, \mu\text{m}\) range. The smaller clusters (N = 1, 2) do not exhibit any features for $\lambda$ \(\gtrsim 38 \, \mu\text{m}\). In the range $\lambda$ \(\approx 45-76 \, \mu\text{m}\), only (MgO)$_9$ shows one or two faint features, while at $\lambda$ \(\approx 76-84 \, \mu\text{m}\), some prominent features emerge from the larger clusters, particularly in the spectra of N = 9, 10. Additionally, since (MgO)$_9$ is the most abundant cluster at this $T_{\rm gas}$, its absorption cross-sections are the most prominent. The combined absorption spectrum for cluster sizes N$ = 1$--$10$ of (MgO)$\rm_N$, shown in the bottom panel of Fig.~\ref{absorspecamgo}, exhibits the strongest features at deeper atmospheric layers ($p_{\rm gas} \sim 1$~bar), with the intensity decreasing toward lower pressures. Similar to the behaviour of (Al$_2$O$_3$)$\rm\rm_N$ clusters, an isothermal atmosphere across different altitudes results in fewer thermochemically favoured large clusters at lower pressures, although stable cluster populations remain present. The figure shows that the highest number of features appears within $\lambda$\(\approx 15-45 \, \mu\text{m}\). In the $\lambda$ \(\approx 45-76 \, \mu\text{m}\) range, only a few features are present, primarily due to (MgO)$_9$, and two prominent features observed at $\lambda$ \(\approx 76-84 \, \mu\text{m}\) are attributed to the larger clusters.

\subsection{Cumulative concentration and Absorption of Al/Ti/V/Mg/Si-O Metal Oxide Clusters}
\label{results-total}
\begin{figure}[ht!]
\centering
\begin{minipage}[b]{0.48\textwidth}
\centering
\includegraphics[width=\linewidth]{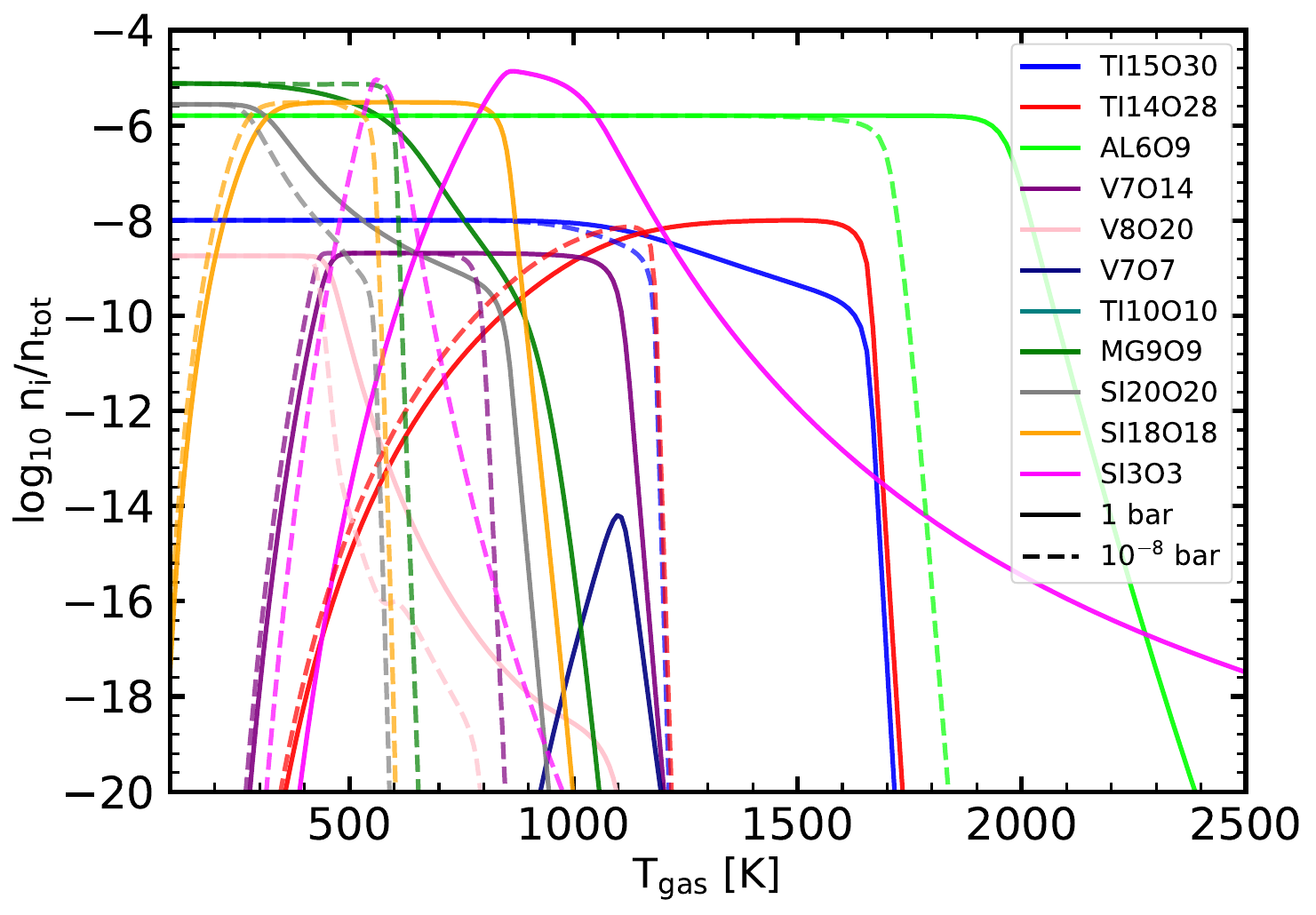}
\end{minipage}
\begin{minipage}[b]{0.48\textwidth}
\centering
\includegraphics[width=\linewidth]{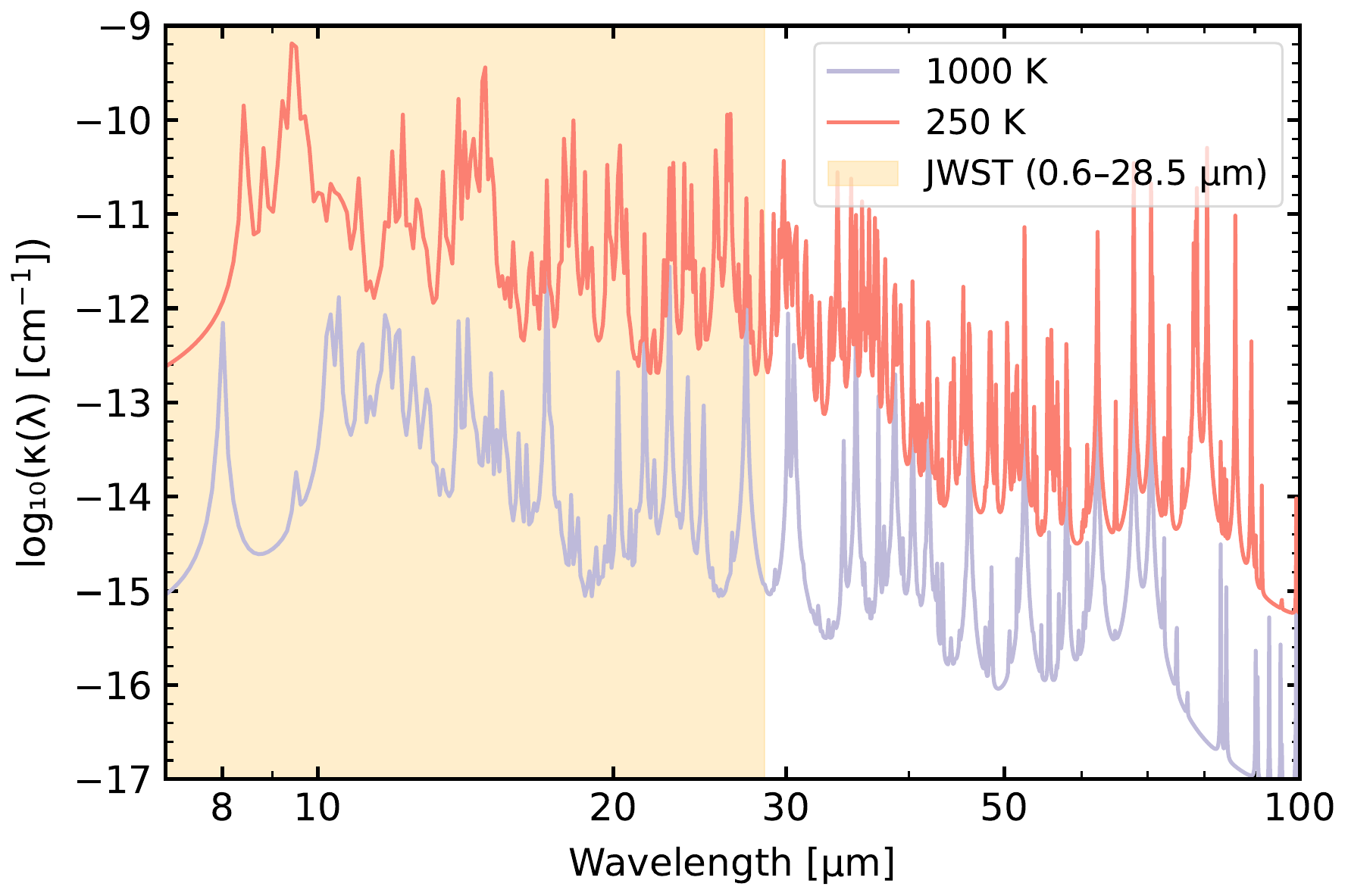}
\end{minipage}
\caption{Top: Concentrations, $n_{\rm i}$/$n_{\rm tot}$ of the most abundant molecular clusters. The calculations include all 89 GM clusters of various species considered in this study and the default species in \texttt{GGchem}. Bottom: Combined absorption plot for all 89 clusters analyzed in this study at $p_{\rm gas}=$ 10$^{-4}$~bar.}
\label{Combinedanalysis}
\end{figure}

The combined impact of all the considered Al/Ti/V/Mg/Si-O metal oxide clusters on the equilibrium gas-phase chemistry composition is studied in order to verify the most abundant clusters within each species at gas temperatures $T_{\rm gas}$ $\lesssim$ 2000 K and to derive cumulative absorption coefficients for test cases relevant for, for example,  \texttt{JWST} low-resolution transmission spectroscopy. When considering all metal oxide clusters (Fig.~\ref{Combinedanalysis}, top), the concentrations of these most abundant clusters remain largely similar to those obtained when only a single species is present at p$_{\mathrm{gas}} = 1$~bar and $10^{-8}$~bar. This can be attributed to the high abundance of oxygen (e.g., solar composition), which is sufficient to stabilise larger clusters composed of elements that are much less abundant. In particular, CO formation effectively sequesters carbon, leaving excess oxygen available. As a result, no single element acts as a limiting reagent, and the concentrations of the individual species are largely unaffected when they coexist in the atmosphere. The only notable deviation occurs with (TiO)$_{10}$, whose concentration appears to be influenced by the presence of (TiO$_2$)$\rm\rm_N$ clusters. Additionally, there is a slight increase in the concentration of (TiO$_2$)$_{15}$ at temperatures where (TiO)$_{10}$ is highly abundant. 

This study confirms that metal oxide clusters composed of simple monoxides, such as MgO and SiO, are thermochemically favoured as larger clusters at low gas temperatures. (MgO)$_9$ remains the most abundant cluster up to $T_{\rm gas}$ $\approx$ 600 K. (SiO)$_{20}$ is the second most abundant cluster for $T_{\rm gas}\lesssim 300$ K, while (SiO)$_{18}$ becomes the second most abundant in the range $T_{\rm gas}$ \(\approx 300-600 \) K, and the most abundant cluster between  $T_{\rm gas}$ \(\approx 600-800 \) K. (SiO)$_3$ is the most abundant species at $T_{\rm gas}$ \(\approx 800-1000 \) K. This finding aligns with previous studies, which suggest bigger clusters of (SiO)$_{\rm N}$ are formed at lower temperatures \citep{gail2013seed, gail2014physics}. \citet{komatsu2018first} reported evidence of silica (SiO)$_2$ condensation in the early solar system, supporting the possibility of (SiO)$_{\rm N}$ stability at lower $T_{\rm gas}$ values. For (MgO)$\rm \rm_N$, \citet{boulangier2019developing} reported (MgO)$_{\rm N}$ to be less abundant than (Al$_2$O$_3$)$_{\rm N}$ and (TiO$_2$)$_{\rm N}$ clusters but they only considered  $T_{\rm gas}$ $\gtrsim$ 500 K where we observe the same trend where (Al$_2$O$_3$)$_{\rm N}$ and (TiO$_2$)$_{\rm N}$ clusters are more abundant than (MgO)$\rm \rm_N$ clusters. Top panel of fig.~\ref{Combinedanalysis} shows further that for $T_{\rm gas}$ $\gtrsim$ 600 K, (Al$_2$O$_3$)$_3$ is the most abundant cluster, while (TiO$_2$)$_{15}$ and (TiO$_2$)$_{14}$ are the second most abundant clusters. Moreover, we confirm the sharp decrease in (TiO$_2$)$_{\rm N}$ cluster concentrations for $T_{\rm gas}$ \(\approx 1000-1200 \) K which was also observed by \citet{sindel2023infrared}. 

Testing the concentrations of these cluster species at lower gas pressure, p$_{\mathrm{gas}} = 10^{-8}$~bar (dashed lines in top panel of Fig.~\ref{Combinedanalysis}), shows an increase of larger cluster sizes at gas pressures typical for the upper atmosphere of exoplanets or AGB star envelopes: The concentration of (MgO)$_9$ is higher at $10^{-8}$~bar compared to 1~bar for T$_{\mathrm{gas}} \lesssim 600$~K. Similarly, an increased concentration of (TiO$_2$)$_{14}$ at T$_{\mathrm{gas}} \lesssim 1200$~K, and  (VO$_2$)$_7$ at T$_{\mathrm{gas}} \lesssim 400$~K occurs. For other species, the concentrations remain similar or decrease at lower p$_{\mathrm{gas}}$. Top panel of fig.~\ref{Combinedanalysis} also demonstrates that larger clusters require a lower temperature to be thermally stable at the lower of p$_{\mathrm{gas}} = 10^{-8}$~bar compared to p$_{\mathrm{gas}} = 1$~bar. Or conversely, larger clusters remain thermally stable until higher gas temperatures for higher pressures (see Fig. 3 in \citealt{1996ASPC...96...69G}). This, hence, points to the need for a supercooled gas below the classical thermal stability consideration to kick-start cloud formation (\citealt{2019AREPS..47..583H}).

{\bf Summary:} Table~\ref{tab:dominantclustersateachtemp} summarizes the most abundant metal oxides and clusters for the elements Al, Ti, V, Si, and Mg in three different temperature regimes: cold, intermediate, and hot. The cold temperature is an example for the night side, the uppermost atmosphere of a tidally locked exoplanet, or quite generally a cold gas giant with a global temperature T$_{\rm global}<800$ K. The intermediate temperature range, $T_{\rm gas} = 250 - 1000$ K, represents an example of terminator regions of such planets. The high gas temperature represents the dayside of a hot extrasolar gas giant with global temperatures T$_{\rm global}> 1800$ K. The element abundances are solar, where Mg, Si, and Al are two orders of magnitude more abundant than Ti and V (e.g., Fig. 5 in \citealt{2020A&A...636A..71H}). Mg and Si form low-temperature condensate precursors (rows 5, 6 in Table~\ref{tab:dominantclustersateachtemp}), Ti, V, and Al (rows 2,3,4) form high-temperature condensate precursors.

Table~\ref{tab:dominantclustersateachtemp} demonstrates that any atmosphere or gas for $T_{\rm gas}\geq 2000$ K will have a diminishing number of complex metal oxide clusters, and instead, the monoxides TiO and SiO, as well as VO, Al, and Mg, will be present as the most abundant gas species of this particular element. Since no (or very low abundant) metal-oxide clusters are available and the gas is undersaturated $(S(T_{\rm gas}) < $ 1) for the relevant solids if $T_{\rm gas}\geq 2000$ K, no cloud particle can form. The retrieval of SiO, TiO, or VO from observation may indicate a cloud-free atmosphere/gas. Conversely, for $T_{\rm gas}<1000$ K, the gas may be highly supersaturated $(S(T_{\rm gas}) \gg$1) and metal-oxide clusters do exist in the gas phase, which may continue to grow into CCNs eventually. Table~\ref{tab:dominantclustersateachtemp} summarizes the most abundant metal oxide clusters. However, for Ti, V, and Si, the largest clusters coincide with the maximum cluster size for which thermodynamic data are available, such that these most likely grow to larger sizes. The extremely low densities encountered in tenuous atmospheric regimes lead to reduced particle growth rates and weak gas–particle coupling, limiting the efficiency of cloud and cluster evolution. This behaviour is consistent with coagulation and growth treatments in meteoric smoke particle models, where particle growth occurs under free‑molecular conditions and is constrained by infrequent collisions at low densities \citep{Plane2018}.
\begin{deluxetable}{lccc} 
\tabletypesize{\scriptsize}
\tablewidth{0pt} 
\tablecaption{The most abundant gas species for the elements Al, Ti, V, Si, and Mg at low, intermediate, and high $\rm{T}_{\rm gas}$ for $\rm{p}_{\rm gas} = 1$ bar. \label{tab:dominantclustersateachtemp}}
\tablehead{
\colhead{Element} & \colhead{$\rm{T}_{\rm gas} \leq 250$\,K} & \colhead{$\rm{T}_{\rm gas}=250$--$1000$\,K} & \colhead{$\rm{T}_{\rm gas}=2000$\,K}
} 
\startdata
Al & (Al$_2$O$_3$)$_3$ & (Al$_2$O$_3$)$_3$ & Al \\
Ti & TiCl$_3$, (TiO$_2$)$_{15}$ & (TiO$_2$)$_{15}$ & TiO \\
V & (V$_2$O$_5$)$_4$ & (V$_2$O$_5$)$_4$, (VO$_2$)$_7$ & VO \\
\hline
Si & (SiO)$_{20}$ & (SiO)$_{18}$, (SiO)$_3$ & SiO \\
Mg & (MgO)$_9$ & (MgO)$_9$ & Mg \\
\hline
Supersat. ratio 
& \multicolumn{2}{c}{$\rm{S}(\rm{T}_{\rm gas}) \gg 1$} & $\rm{S}(\rm{T}_{\rm gas}) < 1$ \\
\enddata
\end{deluxetable}

Finally, the combined absorption spectra based on the joint gas-phase chemistry for all cluster species are derived. For clarity, we have computed the absorption spectra at $p_{\rm gas} = 10^{-4}$~bar, corresponding to an atmospheric region observable by \texttt{JWST}, for two temperatures, $T_{\rm gas} = 1000$~K and 250~K, as shown in the bottom panel of Fig.~\ref{Combinedanalysis}. A clear distinction emerges between the two regimes: the combined absorption spectrum exhibits a higher magnitude at $T_{\rm gas}$ = 250~K than at $T_{\rm gas}$ = 1000~K. At $T_{\rm gas}$ = 1000~K, the spectra are predominantly influenced by larger clusters that remain thermochemically stable at high temperatures, such as (Al$_2$O$_3$)$\rm\rm_N$, (TiO$_2$)$\rm\rm_N$, and (VO$_2$)$\rm\rm_N$. In contrast, at $T_{\rm gas}$ = 250~K, additional contributions arise from (MgO)$\rm \rm_N$ and (SiO)$\rm \rm_N$. A strong absorption feature near $\lambda \approx 8\,\mu\mathrm{m}$ is attributed to the SiO monoxide. Moreover, for $\lambda$ \(\approx 30-38 \, \mu\text{m}\), a dense forest of absorption features appears at $T_{\rm gas}$ = 250~K, whereas at $T_{\rm gas}$ = 1000~K, only a single prominent feature by (Al$_2$O$_3$)$_3$ is present. Finally, at $T_{\rm gas}$ = 250~K, almost no absorption is observed for $\lambda$ \(\approx 90-100 \, \mu\text{m}\). Hence, these specific observations indicate that cooler, low-pressure atmospheric regions are more conducive to detecting the diverse cluster-induced mid-infrared spectral features, particularly within 5--28~$\mu$m wavelength ranges of the \texttt{JWST/MIRI}.
\section{Metal oxide clusters in UHJ, HJ, and WJ atmospheres}
\label{section:hazesinexoplanets}
Gas giants offer the best opportunity to probe atmospheric gas-phase composition in detail, providing key insights into cloud formation in chemically complex environments. Sect.~\ref{clusterabun} shows that metal oxide clusters emerge in distinct temperature regimes, while Sect.~\ref{section:hazesinexoplanets} examines their abundances across UHJ, HJ, and WJ to explore potential chemical asymmetries and spatial variations in cluster distributions.

We consider four representative planets—WASP-121\,b, WASP-18\,b, WASP-39\,b, and WASP-69\,b—spanning a wide range of temperatures and metallicities. WASP-121\,b ($T_{\rm eq} \approx 2400$~K; solar metallicity; \citealt{evans2016detection, parmentier2018thermal}) and WASP-18\,b ($T_{\rm eq} \approx 2429$~K; solar metallicity; \citealt{cortes2020tramos, coulombe2023broadband}) are UHJs, where strong day--night contrasts and high temperatures provide ideal conditions to test cluster stability \citep{helling2021cloud}. WASP-39\,b ($T_{\rm eq} \approx 1000$~K; $10\times$ solar metallicity; \citealt{espinoza2024inhomogeneous}) probes enhanced elemental abundances and intermediate temperature regimes conducive to cluster growth \citep{carone2023wasp}. WASP-69\,b ($T_{\rm eq} \approx 963$~K; solar metallicity; \citealt{guilluy2022gaps, bangera2025kinetic}) represents a cooler atmosphere, allowing investigation of cluster stability under more favorable growth conditions.

Our objective is to assess how thermochemical stability and preferred cluster sizes vary with atmospheric conditions and to identify regions where metal oxide clusters persist. By analyzing substellar, antistellar, and terminator regions, we map the spatial distribution of clusters and examine how temperature–pressure variations regulate their stability, dissociation, and ionization. This approach provides insight into regions where clusters remain stable or break down, thereby constraining potential cloud-seed formation zones.

\subsection{The local thermodynamic conditions of WASP-121 b, WASP-18 b, WASP-39 b, WASP-69 b}
\begin{figure*}
\centering
\includegraphics[width=0.83\linewidth]{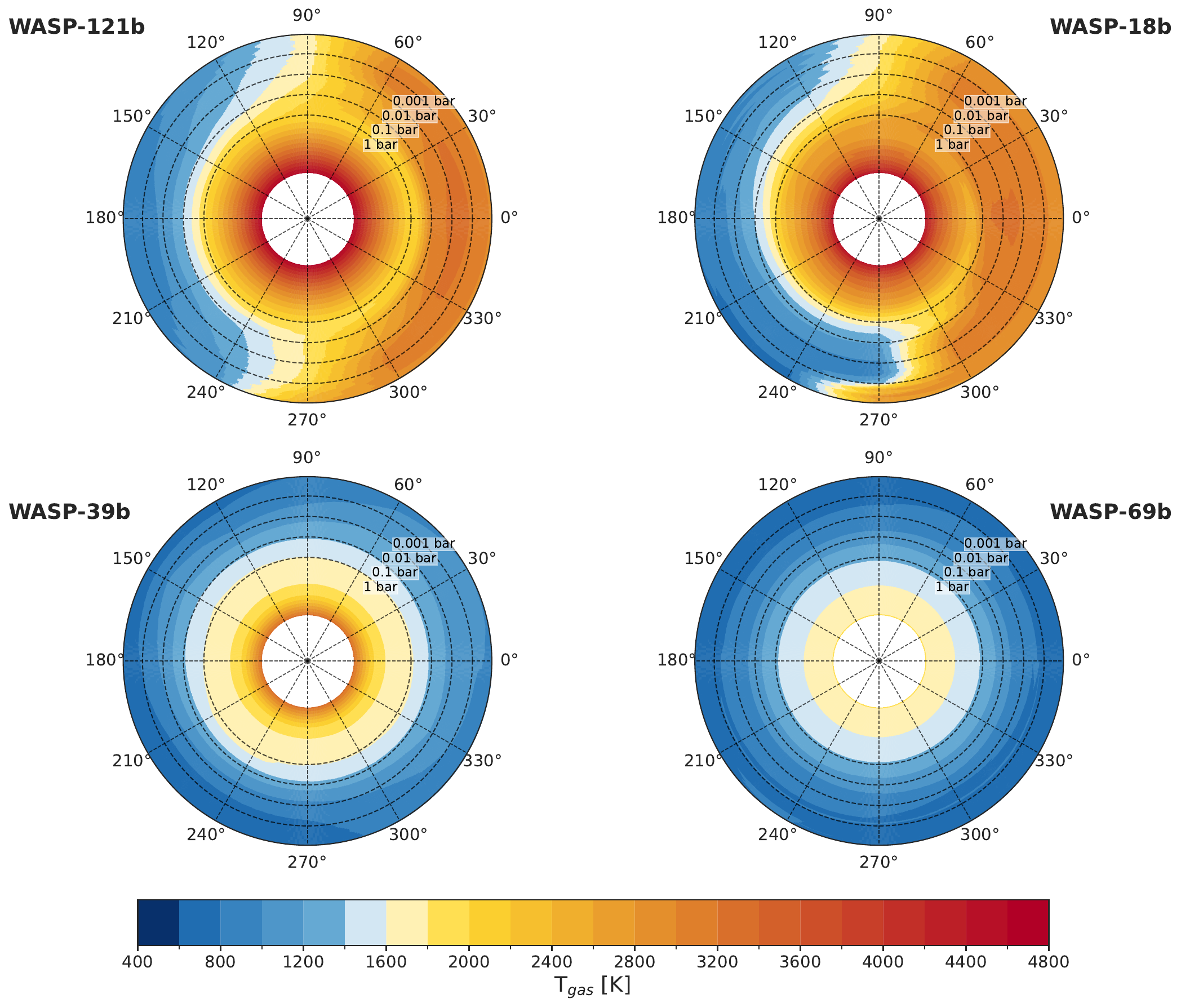}
\caption{Equatorial temperature maps: 2D slices through the equatorial plane ($\theta = 0^{\circ}$) of the 3D GCM \texttt{ExoRad} $(T_{\rm gas},\, p_{\rm gas})$ solutions for WASP-121 b (top left),  WASP-18 b (top right), WASP-39 b (bottom left) and WASP-69 b (bottom right).}
\label{2Dtemp}
\end{figure*}

The differences in the 3D ($T_{\rm gas}$, $p_{\rm gas}$)-structures (Fig. \ref{2Dtemp}) of the four selected planets demonstrates the first trends that will translate into trends in the global cloud structure of these planetary atmospheres. In both UHJs, WASP-121 b (top left) and WASP-18 b (top right), pronounced day–night temperature asymmetries are evident, with extremely hot daysides and much cooler nightsides. This behavior is consistent with the findings of \citet{helling2023exoplanet}, who reported strong thermal asymmetries for planets with global temperatures $T_{\rm eq} \gtrsim 800$~K. For both WASP-121 b  and WASP-18 b, the local gas temperature contrast between the dayside and nightside reaches up to $\sim 2000$~K. In addition, cooler air is advected from the nightside onto the dayside, extending across the evening terminator at pressures around 1~bar in both UHJs, as also suggested by \citet{helling2023exoplanet}. Additionally, we identify a temperature inversion on the dayside of both WASP-121 b and WASP-18 b at pressures of $p_{\rm gas} \sim 1$--$10^{-1}$~bar, consistent with observations reported in the literature \citep{parmentier2018thermal, helling2021cloud, brogi2023roasting}. 

The morning terminator is up to $\sim 500$~K cooler than the evening terminator, in agreement with both observations and GCMs \citep{espinoza2024inhomogeneous, kataria2016atmospheric}. The bottom-left panel of Fig.~\ref{2Dtemp} shows the three-dimensional ($T_{\rm gas}$, $p_{\rm gas}$) structure of the HJ WASP-39 b. In contrast to the UHJs, WASP-39 b exhibits relatively small temperatures on the dayside and nightside at deeper atmospheric levels. However, at lower pressures ($p_{\rm gas} \sim 10^{-4}$~bar), the dayside develops a temperature inversion, with temperatures increasing with altitude, while the nightside temperatures continue to decrease. As in the UHJs studied here, the morning terminator is cooler than the evening terminator, with a more modest temperature contrast of approximately 200~K; this difference is consistent with \textit{JWST} observations reported by \citet{espinoza2024inhomogeneous}. Finally, the bottom-right panel of Fig.~\ref{2Dtemp} presents the three-dimensional ($T_{\rm gas}$, $p_{\rm gas}$) structure of the WJ WASP-69 b. The overall temperature trends are similar to those of WASP-39 b, including comparable day–night and morning–evening temperature differences. However, temperatures across all atmospheric regions are systematically lower than those of WASP-39 b, which may affect the local chemical composition.
\subsection{Equatorial Distribution of Metal oxide clusters in WASP-121 b, WASP-18 b, WASP-39 b, and WASP-69 b}
\label{equatorialslicepalnets}
Temperature asymmetries in exoplanet atmospheres give rise to corresponding asymmetries in the number densities of larger metal oxide clusters; as shown in Sect.~\ref{results-total}, these species can be broadly classified into two categories—high-temperature and low-temperature condensate precursors. The variation in gas temperature across different regions of a planet leads to distinct distributions of larger, thermochemically stable cluster species at different locations. Among the five metal oxide species considered, differences in elemental abundances cause some clusters to be more favoured than others. However, in a chemically mixed atmosphere (e.g., following solar abundances), multiple species can simultaneously stabilise as large clusters. To explore this, we analyse 2D equatorial-plane slices ($\theta = 0^\circ$) to identify which cluster species are thermochemically favoured at different altitudes and longitudinal positions in the atmospheres of the considered planets.

\begin{figure*}[!htbp]
\centering
\begin{minipage}{0.30\textwidth}
\centering
\includegraphics[width=1\linewidth]{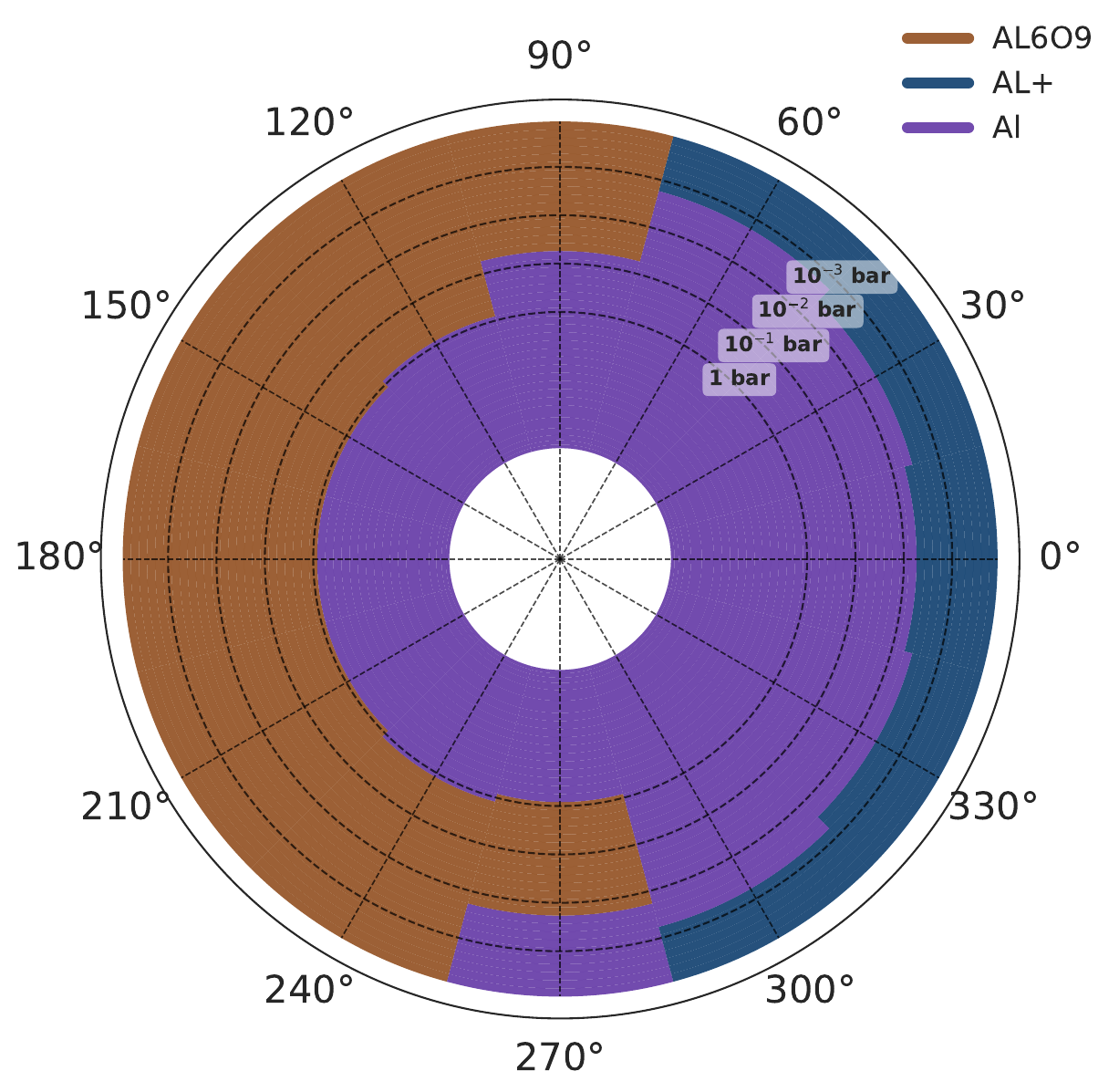}
\end{minipage}
\begin{minipage}{0.30\textwidth}
\centering
\includegraphics[width=1\linewidth]{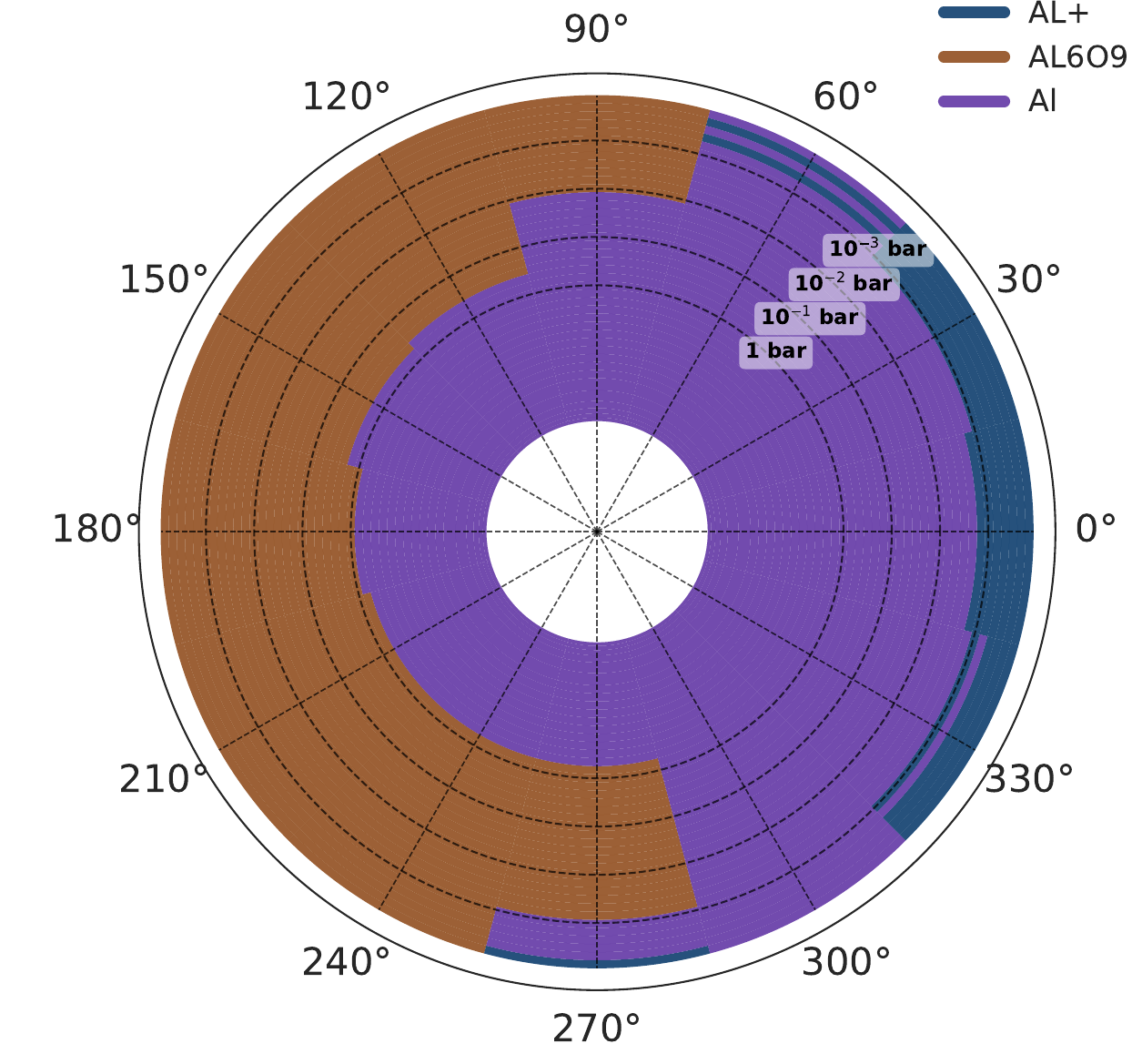}
\end{minipage}
\begin{minipage}{0.30\textwidth}
\centering
\includegraphics[width=1\linewidth]{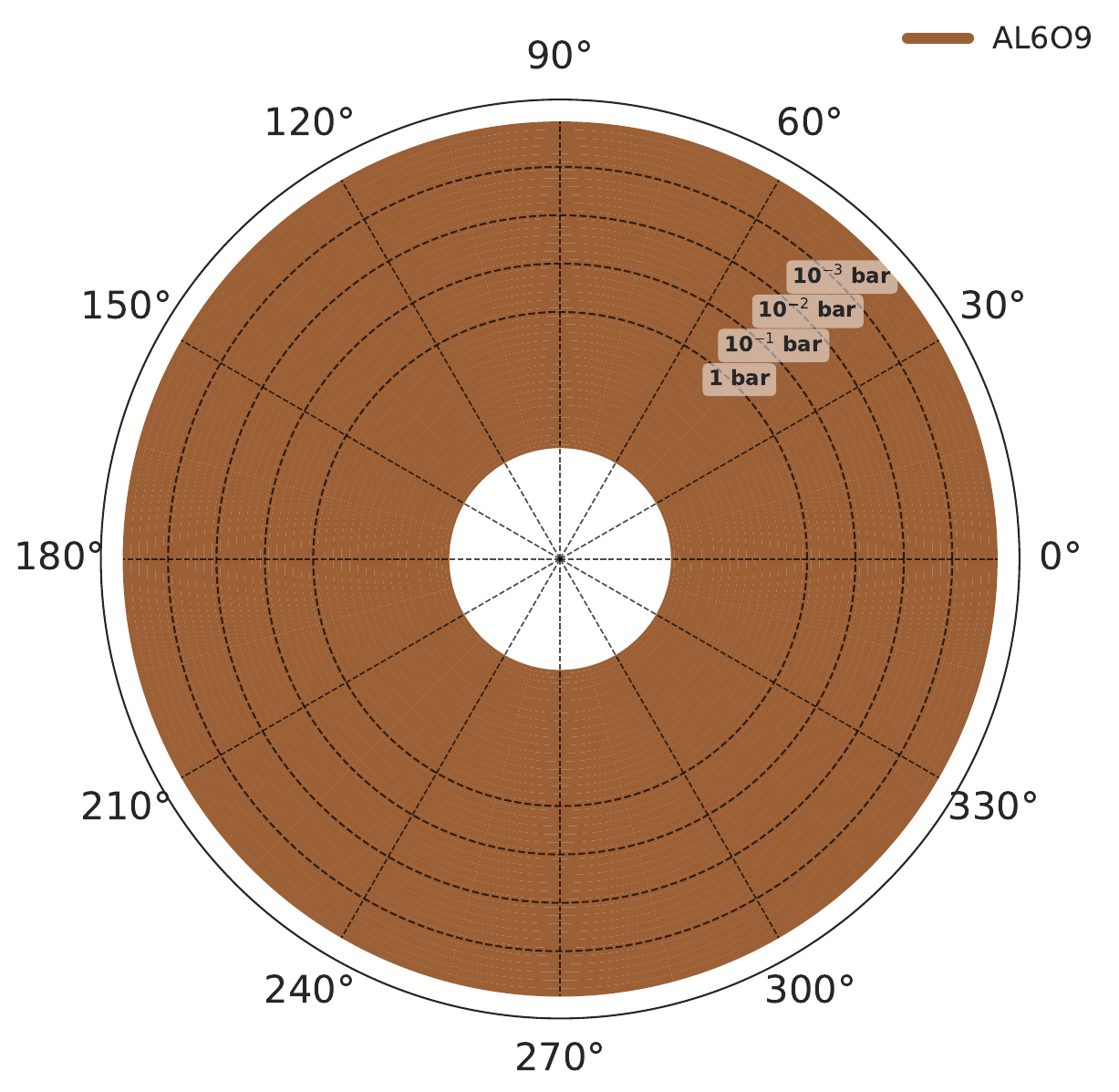}
\end{minipage}
\vspace{0.2cm}
\begin{minipage}{0.30\textwidth}
\centering
\includegraphics[width=1\linewidth]{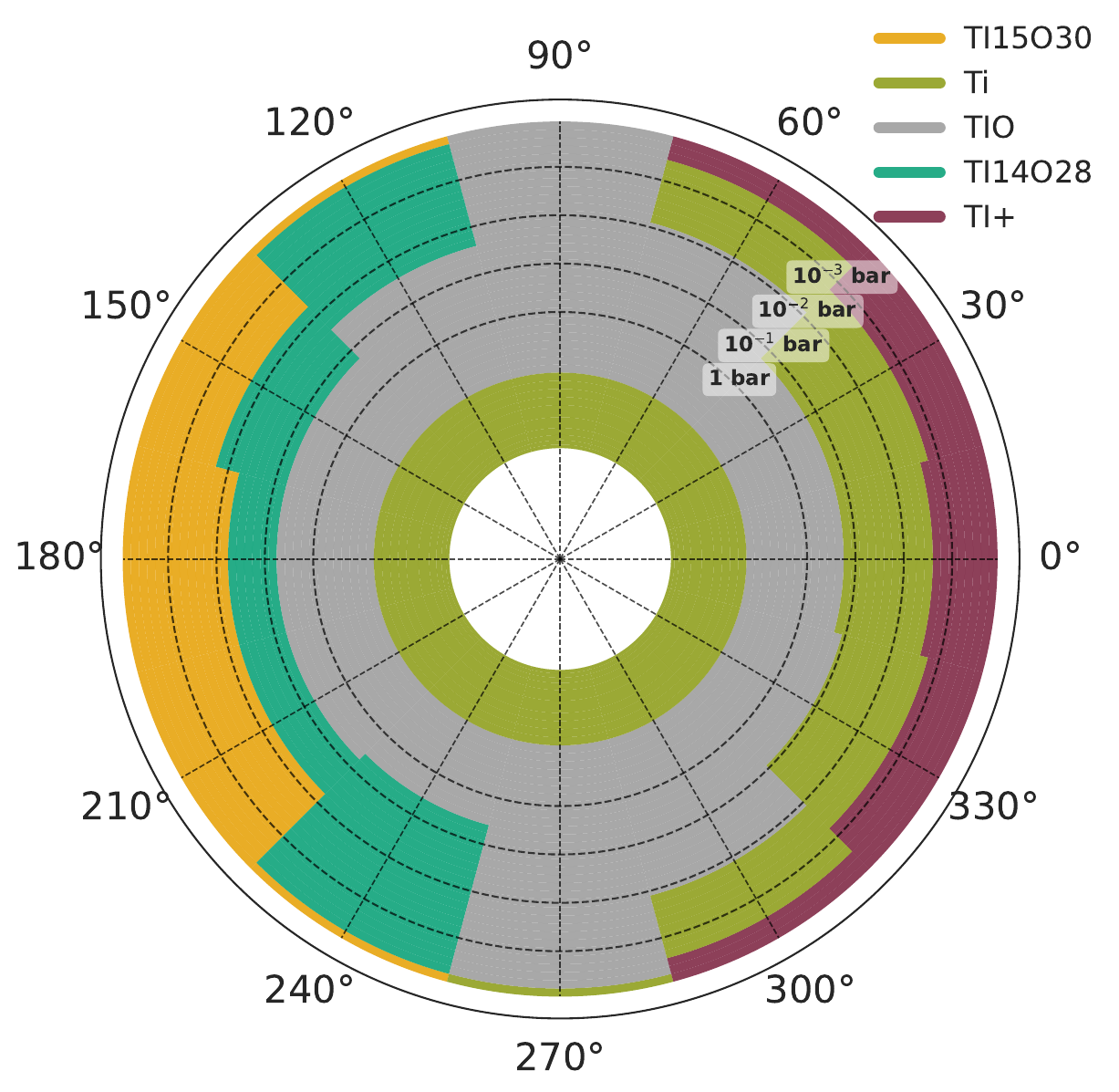}
\end{minipage}
\begin{minipage}{0.30\textwidth}
\centering
\includegraphics[width=1\linewidth]{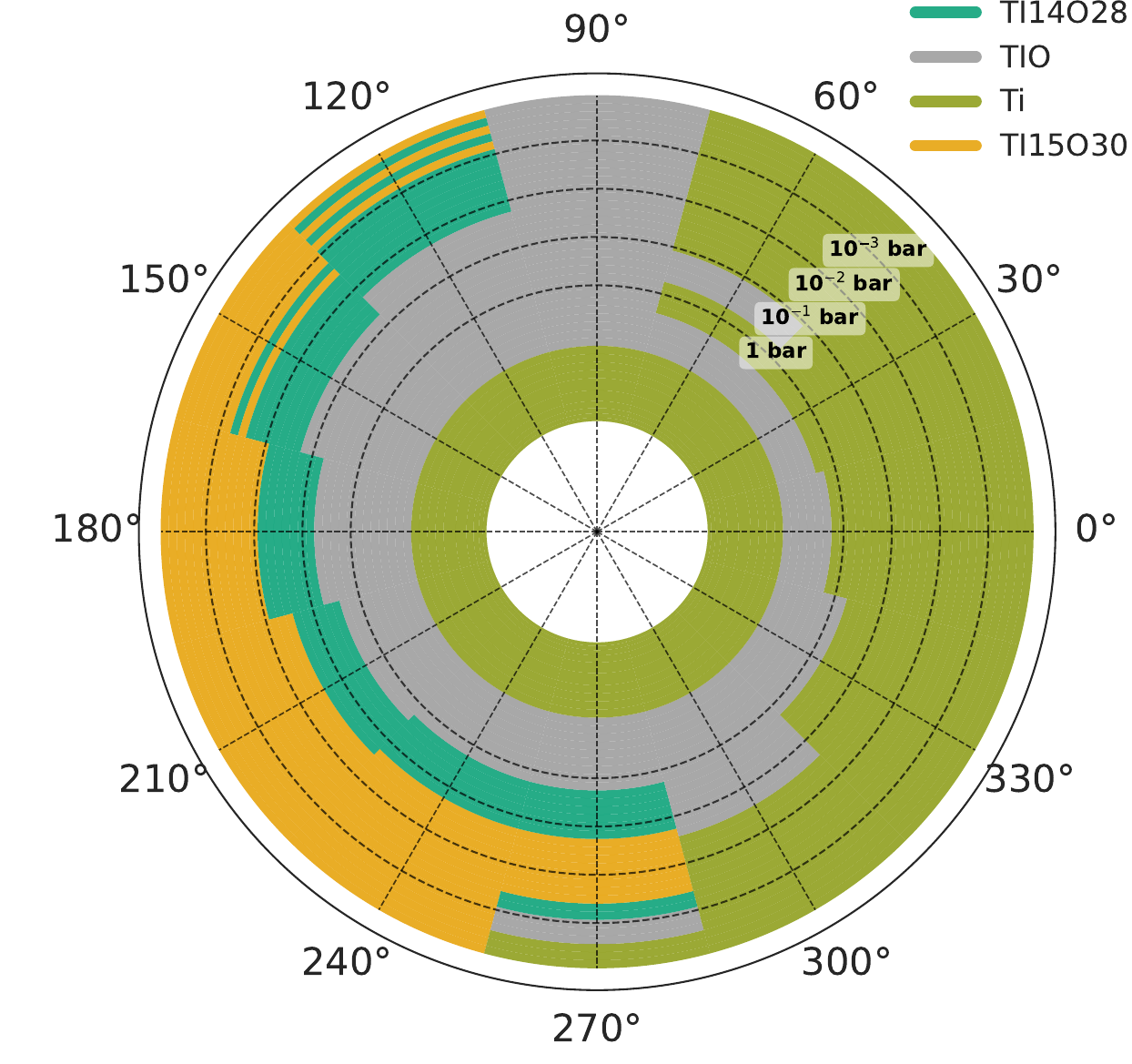}
\end{minipage}
\begin{minipage}{0.30\textwidth}
\centering
\includegraphics[width=1\linewidth]{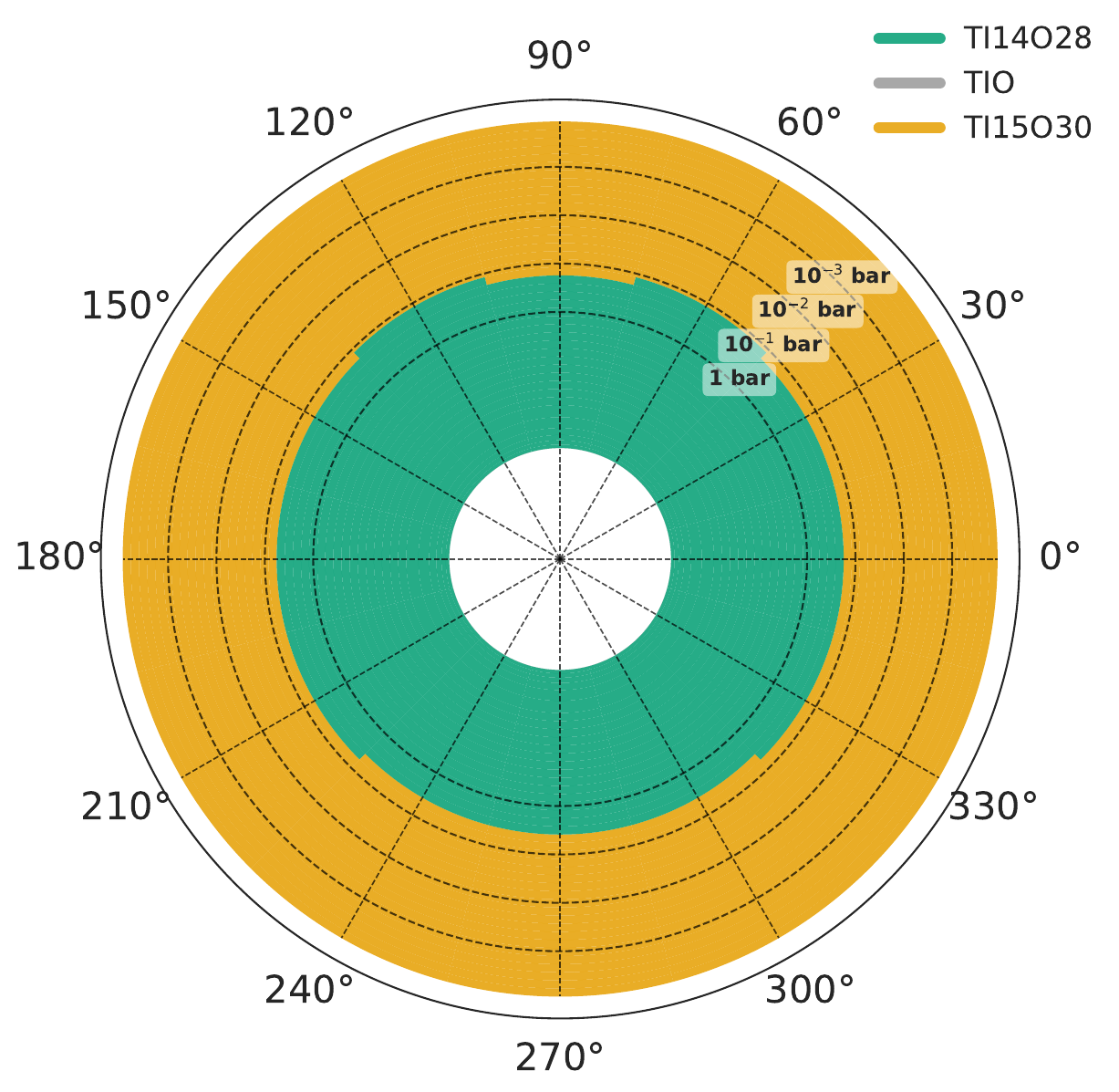}
\end{minipage}

\vspace{0.2cm}
\begin{minipage}{0.30\textwidth}
\centering
\includegraphics[width=1\linewidth]{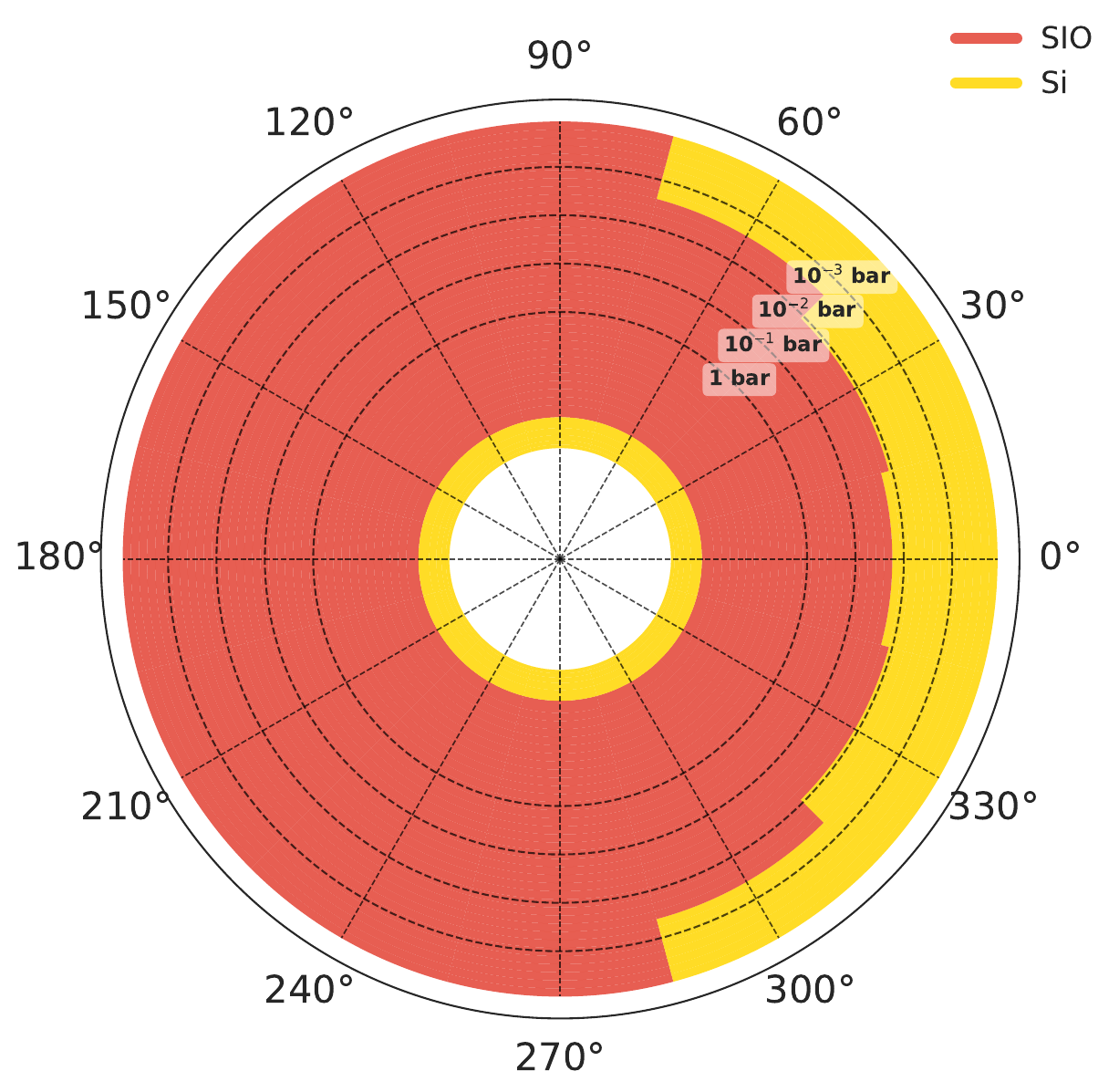}
\end{minipage}
\begin{minipage}{0.30\textwidth}
\centering
\includegraphics[width=1\linewidth]{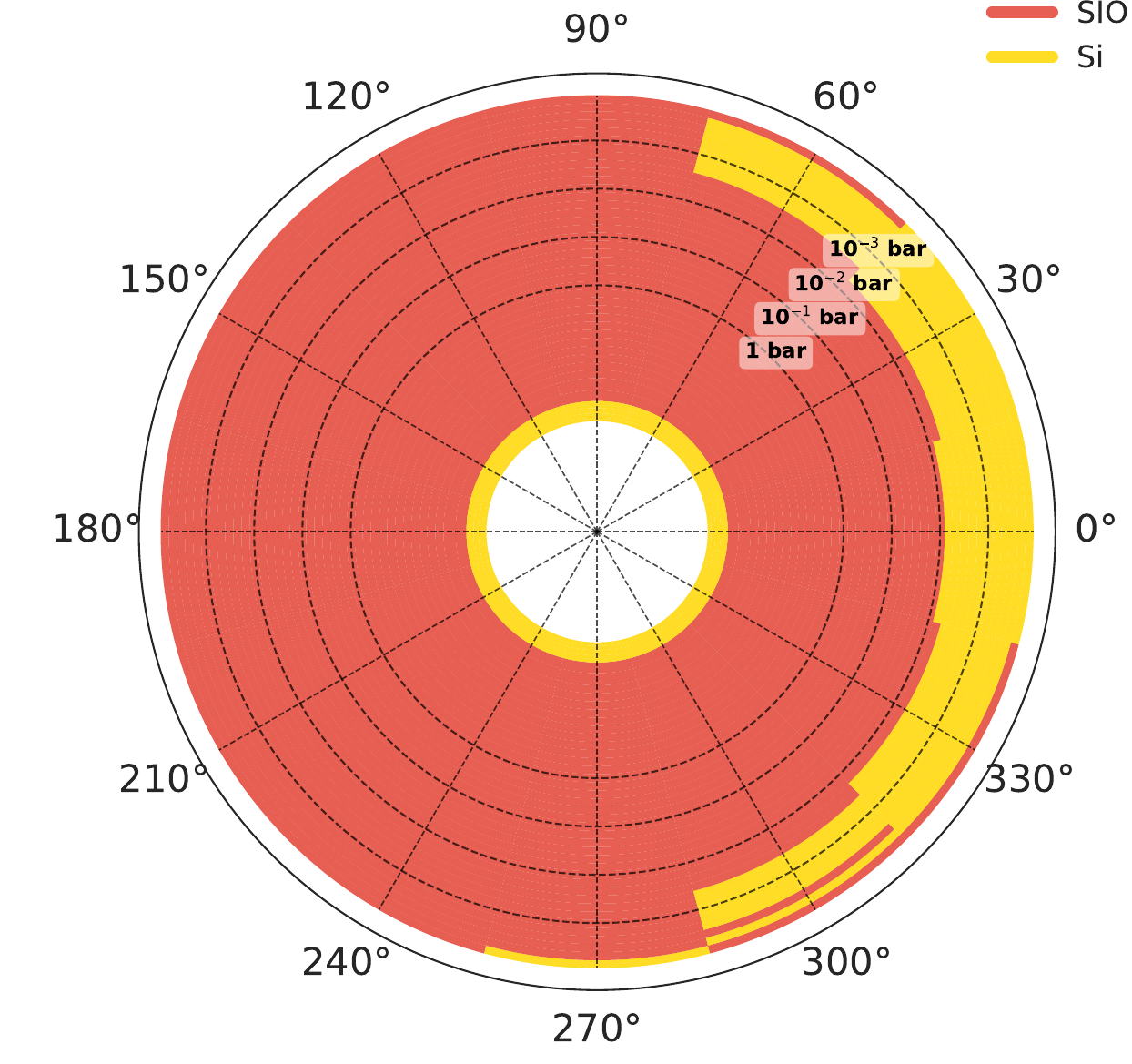}
\end{minipage}
\begin{minipage}{0.30\textwidth}
\centering
\includegraphics[width=1\linewidth]{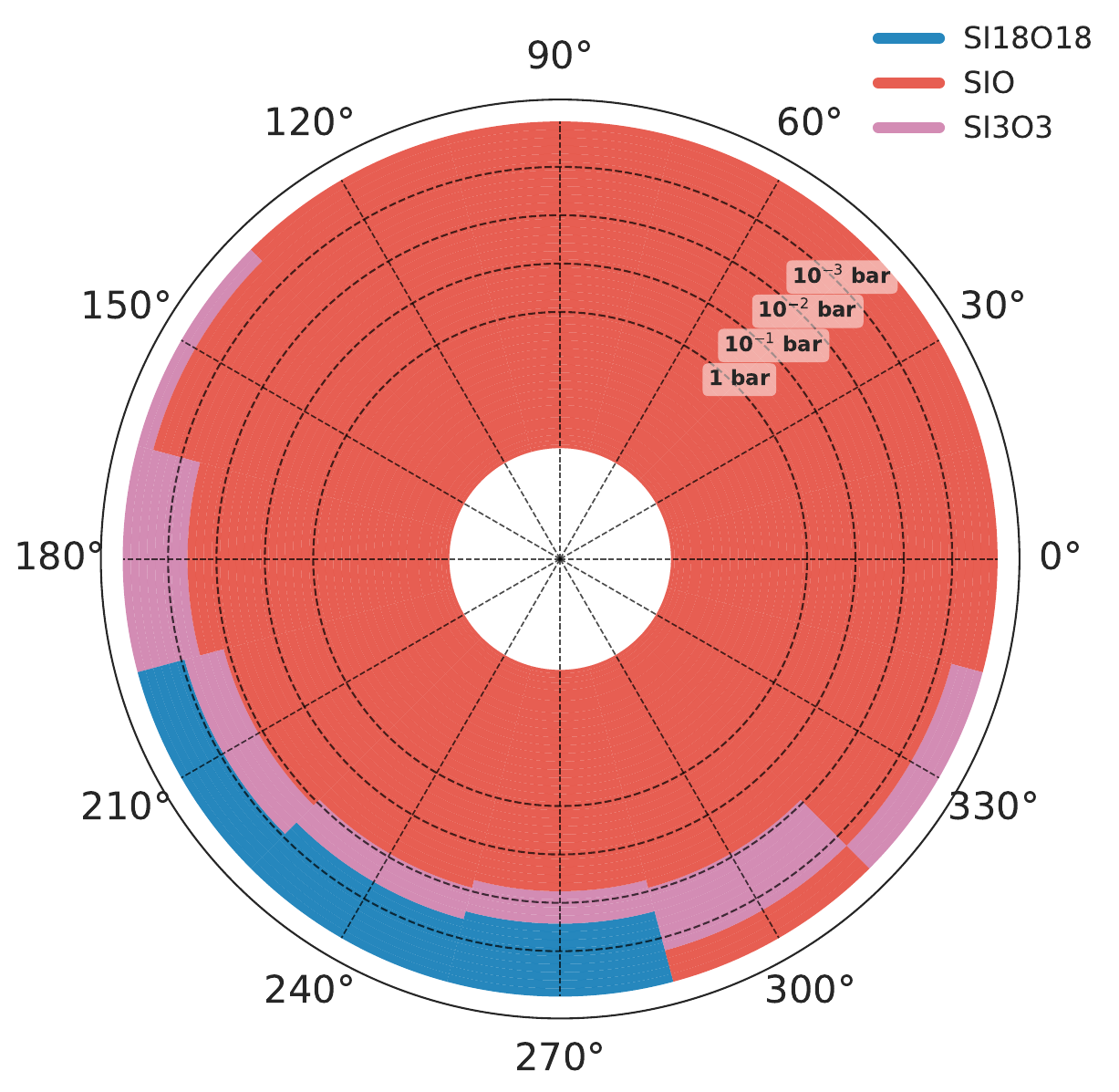}
\end{minipage}

\vspace{0.2cm}
\begin{minipage}{0.30\textwidth}
\centering
\includegraphics[width=1\linewidth]{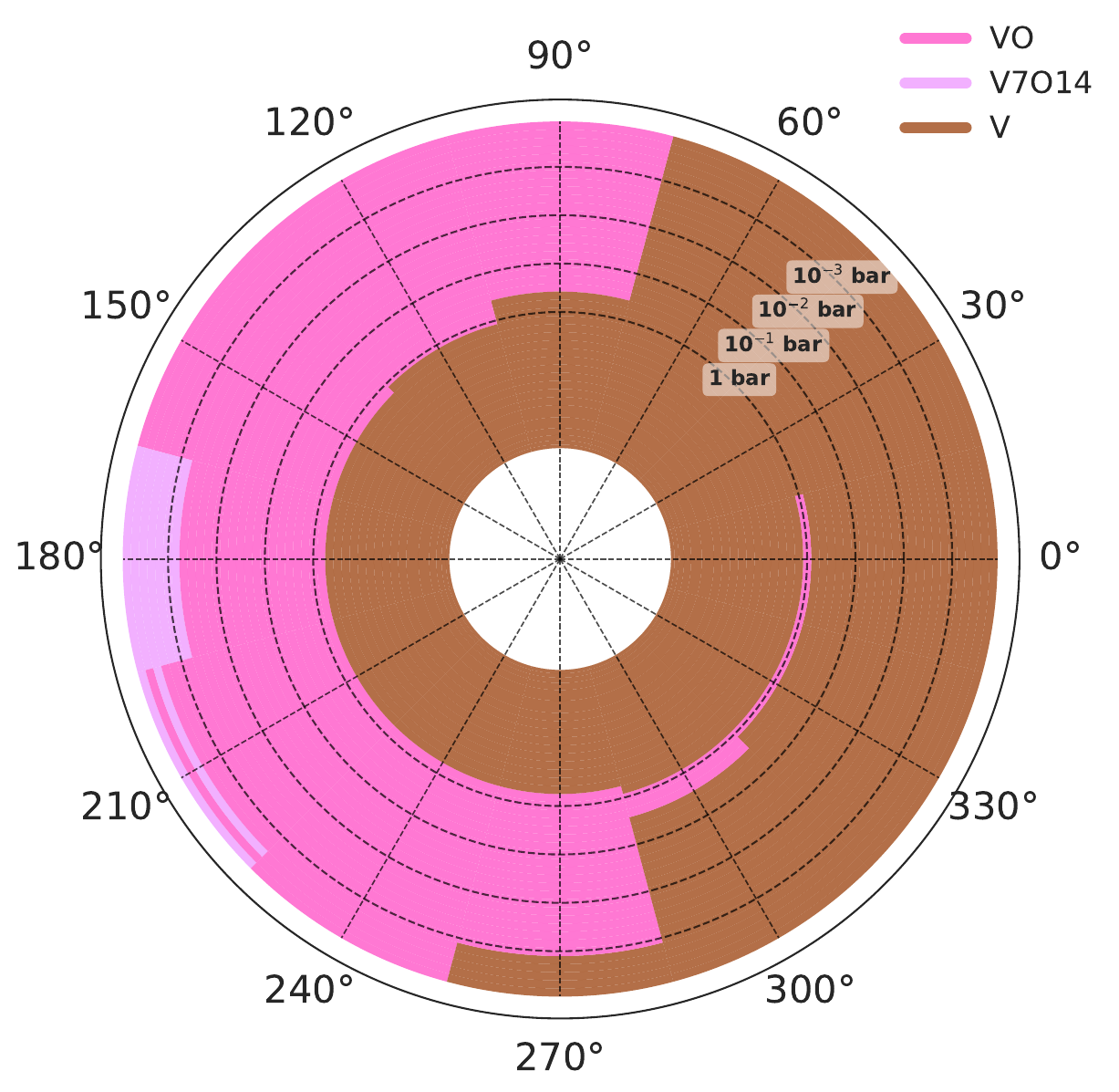}
\end{minipage}
\begin{minipage}{0.30\textwidth}
\centering
\includegraphics[width=1\linewidth]{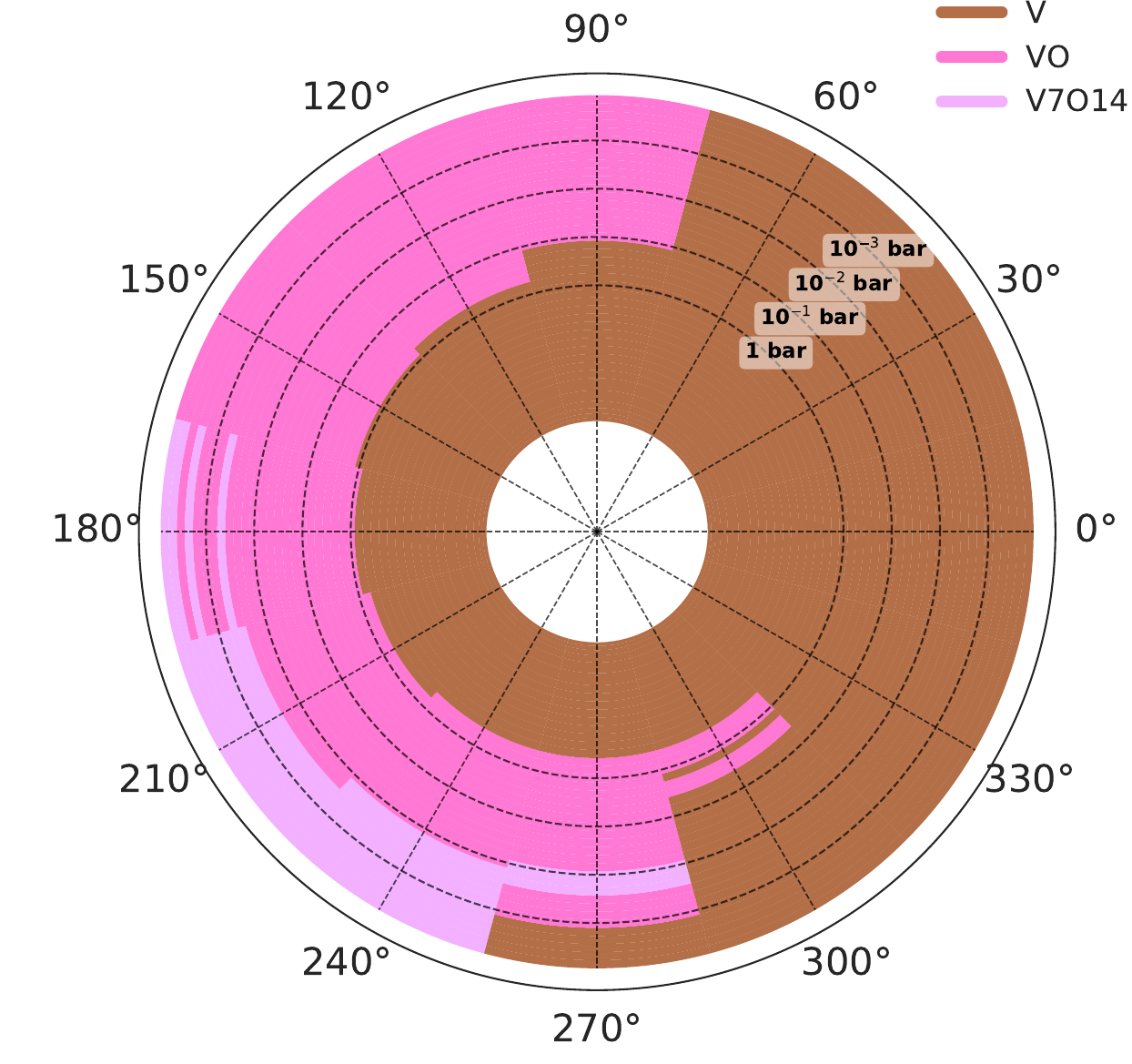}
\end{minipage}
\begin{minipage}{0.30\textwidth}
\centering
\includegraphics[width=1\linewidth]{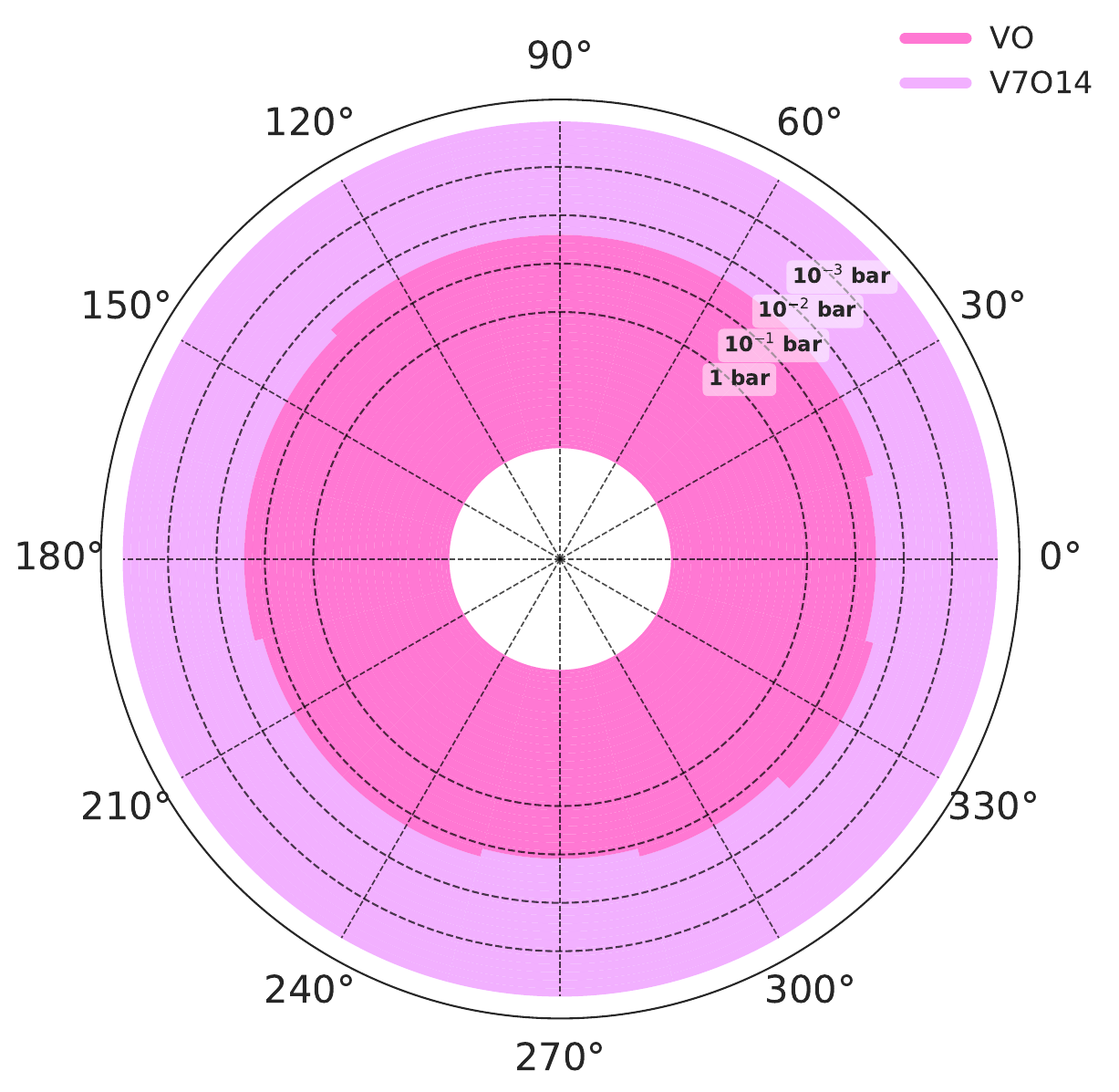}
\end{minipage}
\caption{2D equatorial plane slices ($\theta$ = 0$^\circ$) corresponding to 1D profiles extracted from the 3D GCM models. The panels show the most abundant species among metals, ions, and metal-oxide clusters for Al, Ti, Si, and V across three exoplanets. Left: WASP-121 b; Middle: WASP-18 b; Right: WASP-69 b.}
\label{fig:GCM3dall}
\end{figure*}

Our analysis shows that three of the four planets—WASP-121 b (left panel, Fig.~\ref{fig:GCM3dall}), WASP-18 b (middle panel, Fig.~\ref{fig:GCM3dall}), and WASP-69 b (right panel, Fig.~\ref{fig:GCM3dall})—exhibit significant abundances of larger, thermochemically stable clusters from one or more species. In contrast, the atmosphere of WASP-39 b was dominated by either metals or metal monoxides (Fig.~\ref{fig:GCMwap39b} in the appendix). Interestingly, WASP-39 b is the only planet modeled with a $10\times$ solar metal abundance. At high temperatures, even with enhanced elemental abundances, chemical equilibrium favors small molecules over larger aggregates in order to minimize Gibbs free energy \citep{woitkeggchem2018}. According to the law of mass action, cluster-to-monoxide ratios are primarily dependent on temperature, not absolute abundance; therefore, increasing the number of monoxides does not necessarily increase the fraction of larger clusters. Additionally, competition from other stable small molecules buffers the effective chemical potential of monoxides, reducing the thermodynamic drive for cluster growth \citep{agundez2020chemical}. Entropic penalties further disfavour the formation of larger aggregates unless a strong driving force exists \citep{doye2002entropic}.

Fig.~\ref{fig:GCM3dall} shows the most abundant species in the 2D equatorial-plane slices ($\theta = 0^\circ$) for WASP-121 b, WASP-18 b, and WASP-69 b. We analyze four elements individually—Al, Ti, Si, and V—although the atmosphere includes all considered cluster species simultaneously; the interpretation is performed separately for clarity. Magnesium is not shown, as atomic Mg dominates at all locations. This is consistent with our findings in Section.~\ref{ss:MgO}  (Fig~\ref{MgOnumb}). For WASP-121 b, Al-bearing species are dominated by (Al$_2$O$_3$)$_3$ clusters from the morning to the evening terminator, extending across the entire nightside, while the dayside is predominantly characterised by Al$^+$ ions. A similar spatial distribution is observed for Ti-bearing species. In contrast, for Si-bearing species, neither large clusters nor ions dominate; instead, small molecular species prevail. A similar behaviour is found for most regions of the V-bearing species. However, (VO$_2$)$_7$—identified as a magic-number cluster (see Fig.~\ref{VO2nmagiccluster})—becomes abundant at higher altitudes on the nightside. This trend is consistent with the analysis of \citet{lecoqhelena2024vanadium}. For WASP-18 b, the behavior of Al-bearing species closely resembles that of WASP-121 b. A subtle difference appears for Ti, where the dayside is dominated by neutral Ti rather than Ti+ ions. The Si-bearing species exhibit trends similar to those in WASP-121 b. However, for V-bearing species, (VO$_2$)$_7$ becomes the most abundant species at higher altitudes from the morning terminator through to the nightside, while the remaining regions exhibit distributions comparable to those of WASP-121 b.

Overall, both UHJs are characterized by metal ions dominating the dayside, while larger clusters of high-temperature condensate precursors remain thermochemically favoured at the morning terminator and throughout the nightside. This behavior is consistent with previous studies, such as \citet{helling2021cloud}, which found that nucleation is favored on the nightside and at the morning terminator, with cloud-free daysides. The dayside temperatures are so high that neutral metals and metal oxides become thermally unstable, whereas metal ions remain stable. Cooler conditions on the nightside and morning terminator promote cluster stability. Finally, we consider the WJ WASP-69 b. In this case, high-temperature condensate precursors exist as large, thermochemically stable clusters across all longitudes, with (Al$_2$O$_3$)$_3$ dominating the Al-bearing species and (TiO$_2$)$_{14}$ and (TiO$_2$)$_{15}$ dominating the Ti-bearing species at various locations throughout the atmosphere. At higher altitudes, (SiO)$_{18}$ and (SiO)$_3$ are the most abundant Si-bearing species, extending from the morning terminator to the antistellar point, while (VO$_{2}$)$_7$ are the most abundant V-bearing species at all longitudes at higher altitudes for $p_{\rm gas} \lesssim 10^{-2}$~bar. These results indicate that, for similar metallicities, WJs are expected to host more extensive metal oxide cluster populations than UHJs.

\section{Metal Oxide clusters in the Extrapolated Upper Atmosphere}
This section highlights the extrapolation of the ($T_{\rm gas}$, $p_{\rm gas}$) profiles of WASP-121\,b, WASP-18\,b, WASP-39\,b, and WASP-69\,b following the methodology presented in Sect.~\ref{sec:approach}. We have analyzed both the original GCM profiles (bold black lines) and the extrapolated profiles (dotted black lines) at four representative locations on each planet: the substellar point (Fig.~\ref{Substellar}), antistellar point (Fig.~\ref{Antistellar}), morning terminator (Fig.~\ref{morningt}), and evening terminator (Fig.~\ref{Eveningt}). For each location, we have computed the abundances of the most stable species within each cluster family to identify which clusters persist as larger aggregates at varying atmospheric levels. We further map regions where clusters dissociate into neutral metals and oxygen, which may subsequently ionize, tracing chemical transitions driven by local temperature and pressure variations in the upper atmosphere.

\subsection{Metal ions on the Dayside,\\ Metal oxide clusters on the Nightside}
\label{subsection:daynightextra}
\begin{figure*}[!htbp]
\centering
\begin{minipage}{0.45\textwidth}
\centering
\includegraphics[width=1\linewidth]{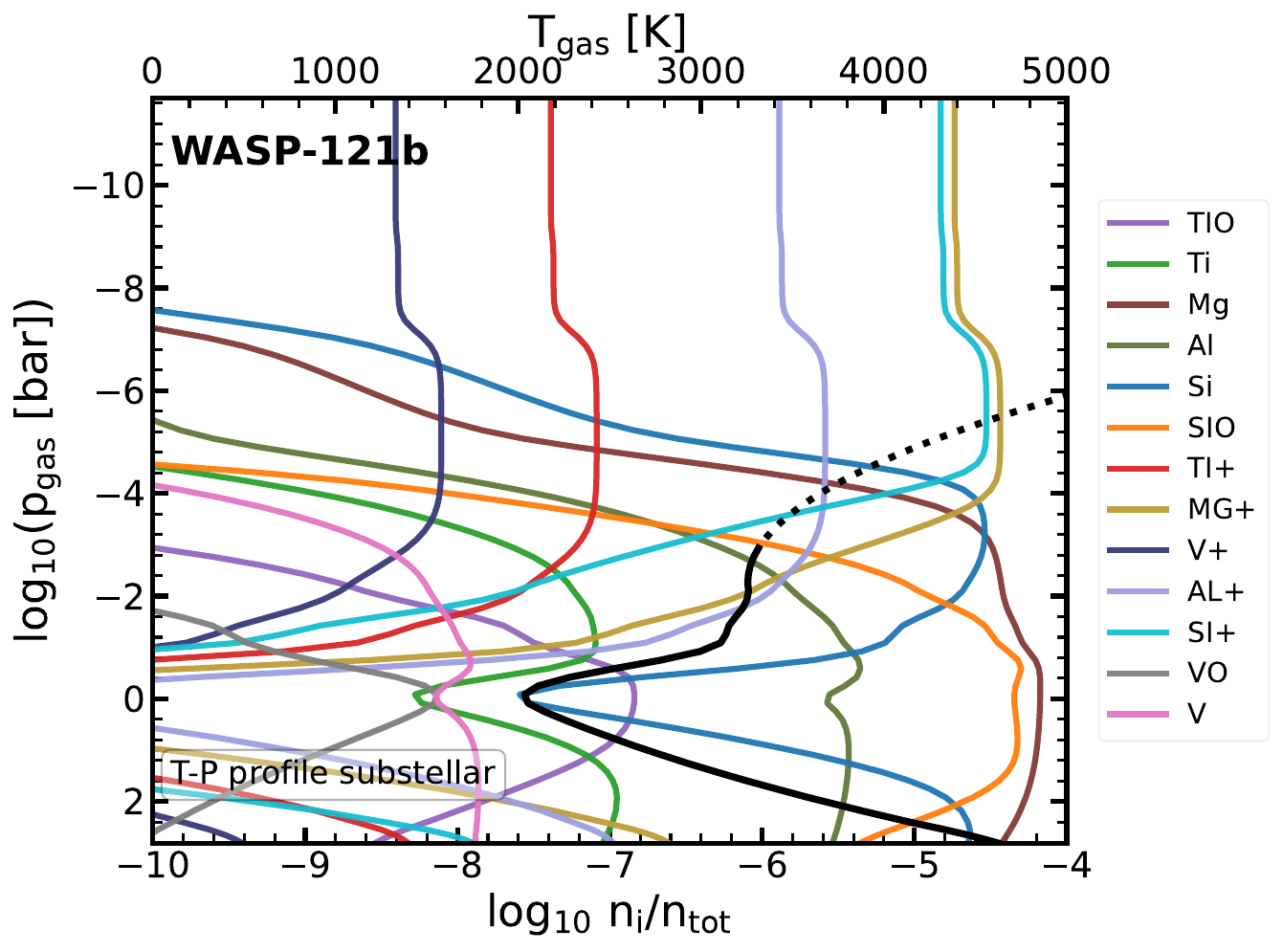}
\end{minipage}
\hspace{0.3cm}
\begin{minipage}{0.45\textwidth}
\centering
\includegraphics[width=1\linewidth]{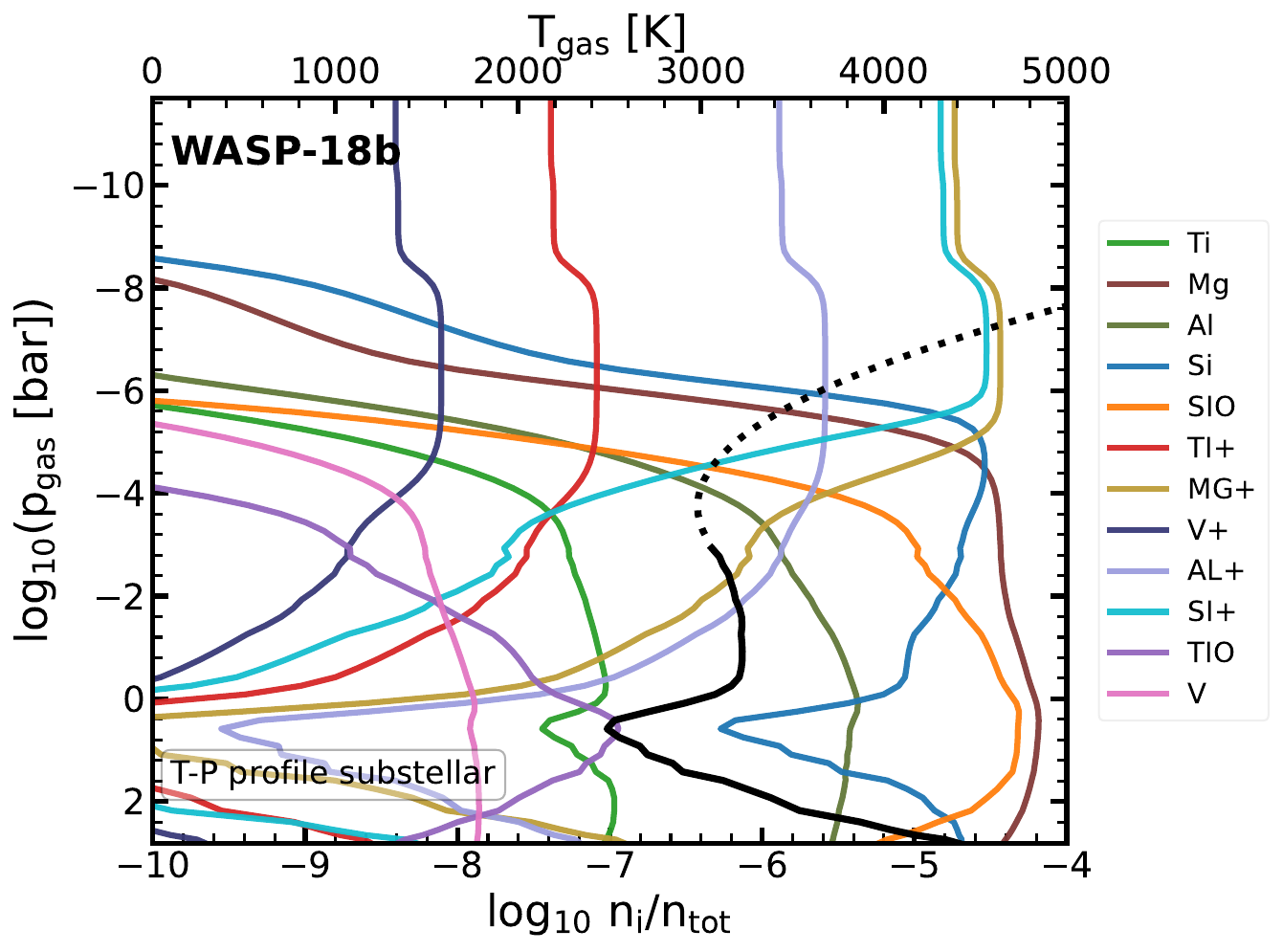}
\end{minipage}
\begin{minipage}{0.45\textwidth}
\centering
\includegraphics[width=1\linewidth]{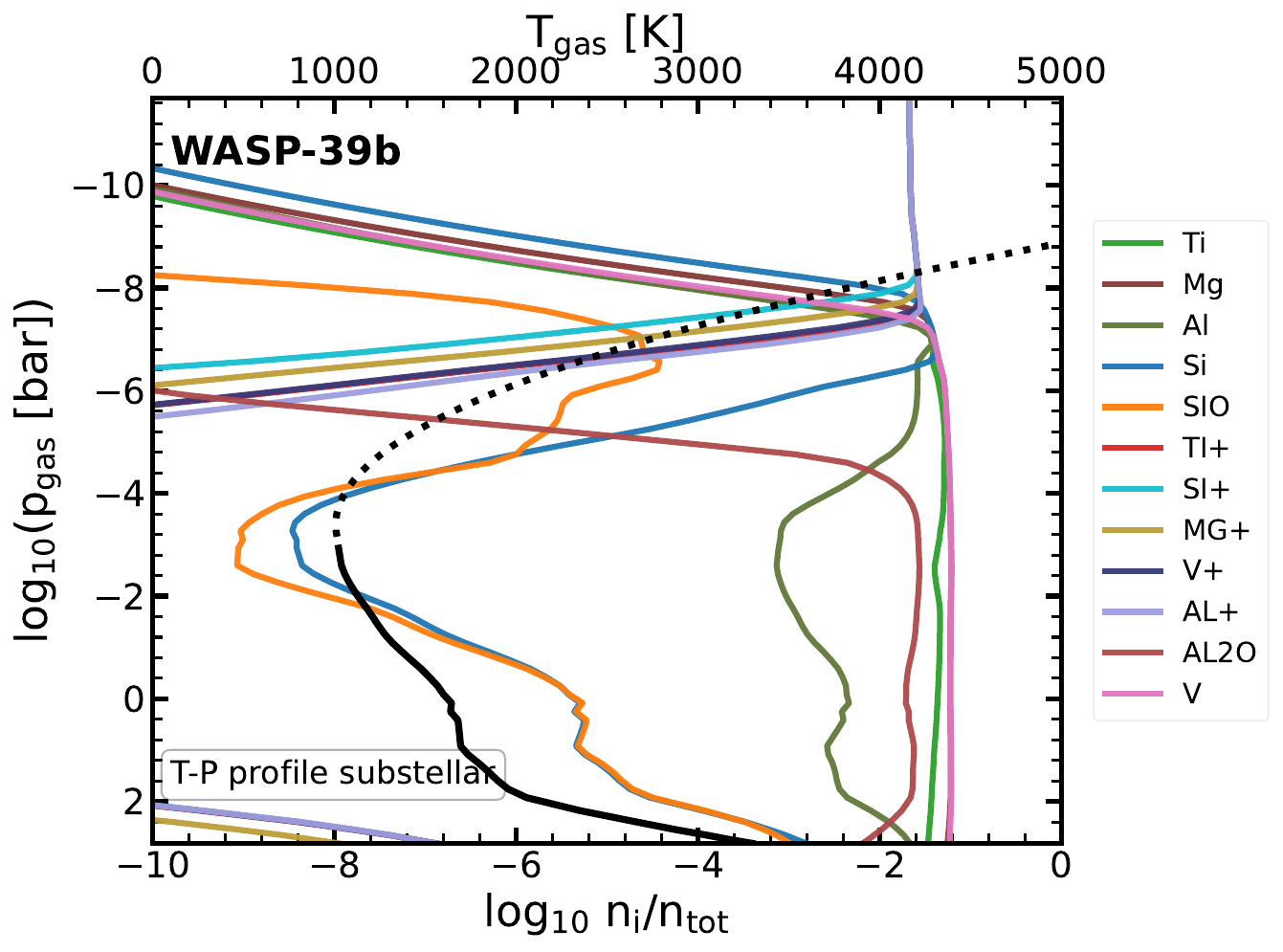}
\end{minipage}
\hspace{0.3cm}
\begin{minipage}{0.45\textwidth}
\centering
\includegraphics[width=1\linewidth]{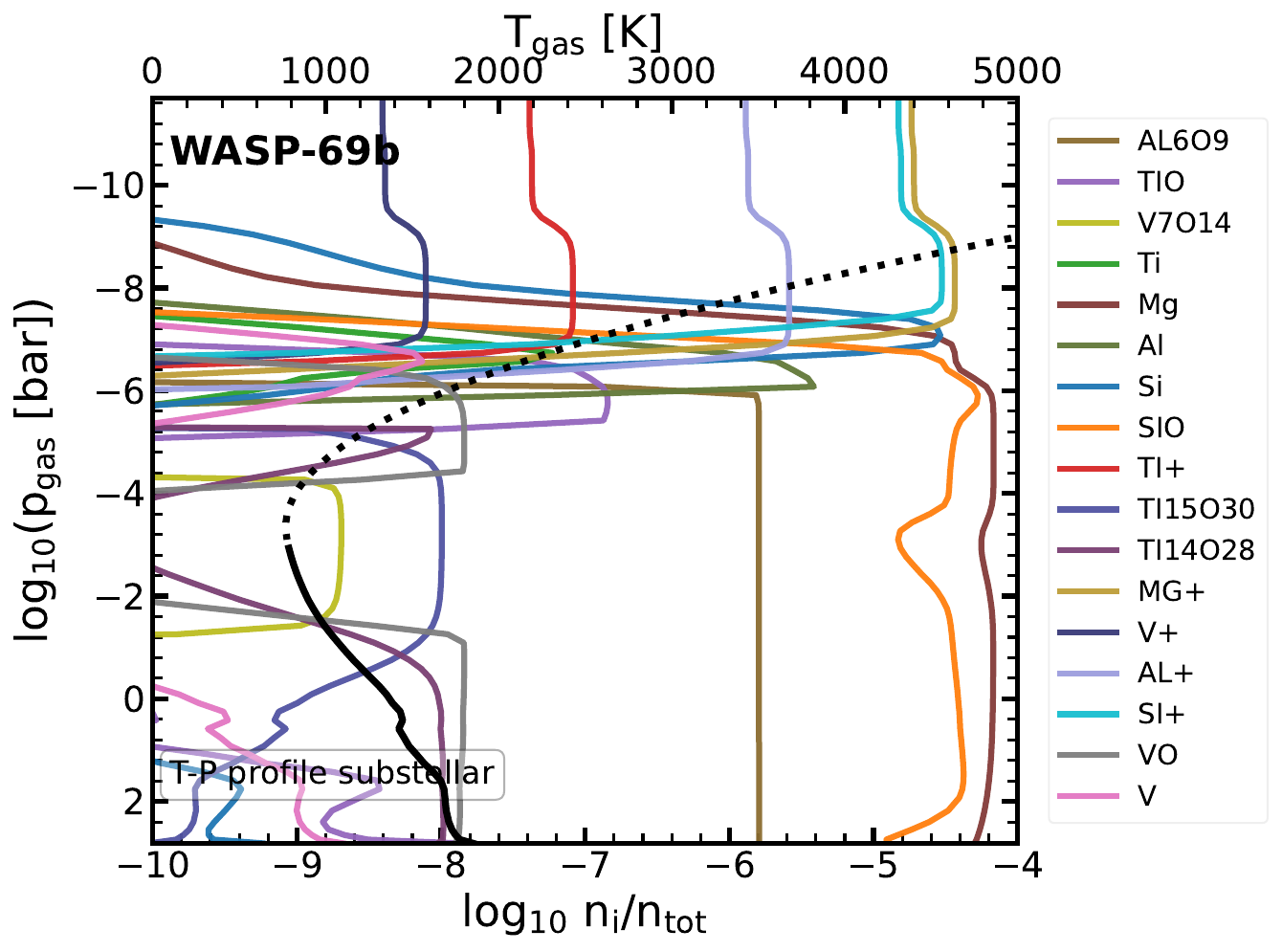}
\end{minipage}
\caption{Gas-phase concentrations at the substellar point, $n _{\rm i}$/$n_{\rm tot}$, of the most abundant Al-, Ti-, Mg-, Si-, and V-bearing metals, ions, and metal oxide clusters, shown as a function of the $T_{\rm gas}$–$p_{\rm gas}$ profile, where the bold black segment corresponds to GCM calculations and the dotted black segment indicates the extrapolated low-pressure regime.}
\label{Substellar}
\end{figure*}

As discussed in Sect.~\ref{section:hazesinexoplanets}, metal ions dominate the outer atmospheric regions at the substellar point (day side). In Fig.~\ref{Substellar}, we therefore plot the vertical concentration profiles of the most abundant neutral metal, metal ion, and metal oxide clusters at the substellar location. For each metal (Al, Ti, V, Si, and Mg), we identify the most abundant species by examining all corresponding metals, ions, and clusters individually. This allows us to assess which species dominate across the pressure ranges covered by both the original GCM domain and the extrapolated upper atmosphere. For both UHJs, WASP-121 b and WASP-18 b, no larger clusters are stable at any altitude. The upper atmosphere ($p_{\rm gas} \lesssim 10^{-5}$~bar) is dominated by metal ions for all considered species, while the deeper atmosphere is primarily composed of neutral metals and simple metal monoxides, in particular SiO and TiO. A similar behavior is observed for the HJ WASP-39 b, where metal ions dominate the upper atmosphere at pressures $p_{\rm gas} \lesssim 10^{-7}$~bar, and neutral metals together with simple metal oxide clusters prevail at lower atmosphere. As discussed in Sect.~\ref{section:hazesinexoplanets}, the enhanced metallicity of WASP-39 b further disfavors the stability of larger clusters. In contrast, the WJ WASP-69 b exhibits significant abundances of larger clusters ((Al$_2$O$_3$)$_3$, (VO$_2$)$_7$, (TiO$_2$)$_{14}$, and (TiO$_2$)$_{15}$) in the deeper atmosphere, where the temperature regime supports the thermochemical stability of high-temperature condensate precursors. However, in the upper atmosphere ($p_{\rm gas} \lesssim 10^{-8}$~bar), these clusters dissociate into neutral metal and oxygen atoms. With decreasing pressure, ionization becomes efficient, and metal ions once again dominate the chemical composition. Notably, an intermediate pressure layer ($10^{-8} \lesssim$ $p_{\rm gas} \lesssim 10^{-5}$~bar) is present in which metal ions coexist with neutral metals and simple clusters. This layer illustrates the gradual transition from cluster species to neutral metals and finally to metal ions with decreasing pressure in the atmosphere of WASP-69 b.

\begin{figure*}[!htbp]
\centering
\begin{minipage}{0.45\textwidth}
\centering
\includegraphics[width=1\linewidth]{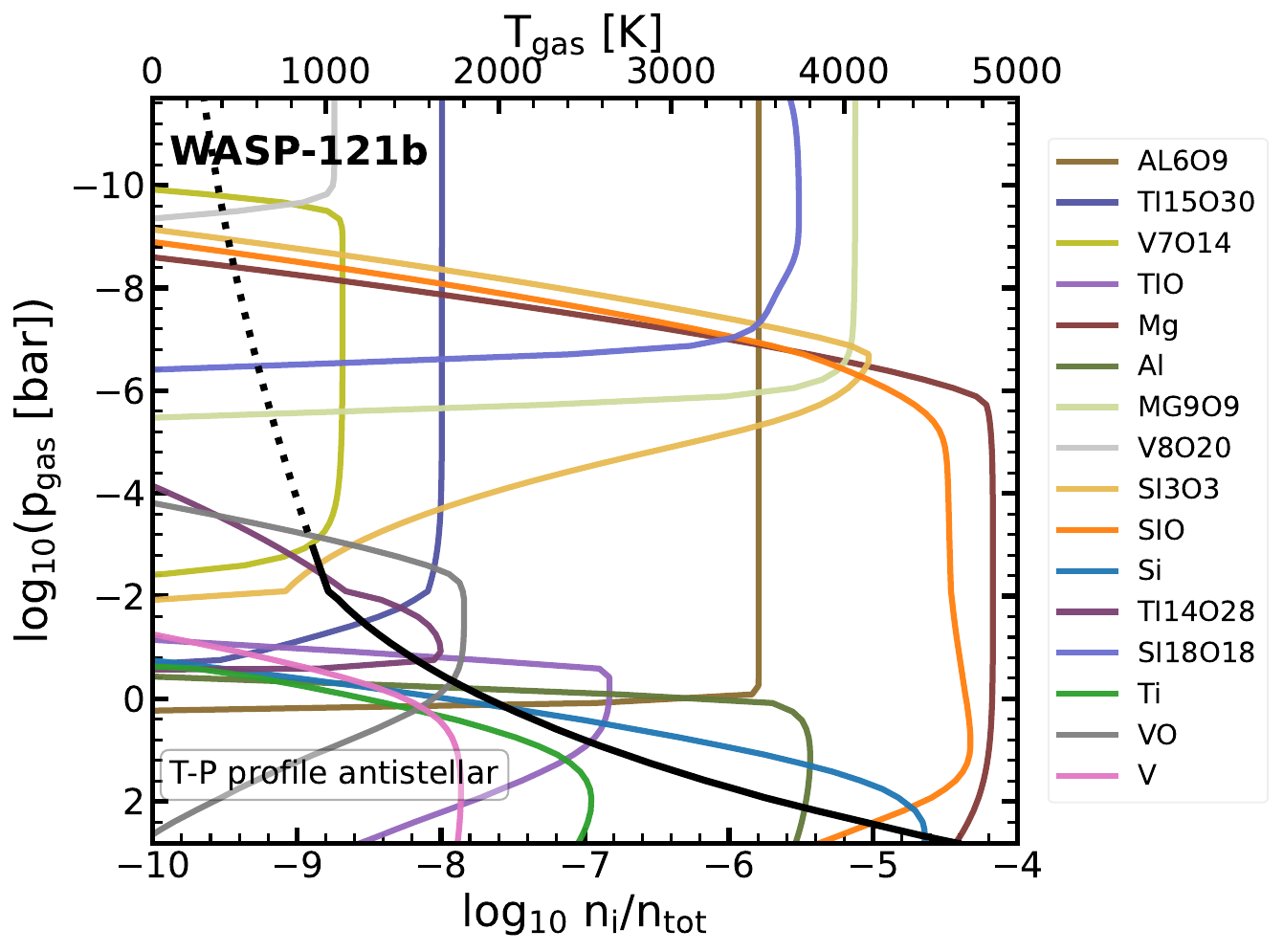}
\end{minipage}
\hspace{0.3cm}
\begin{minipage}{0.45\textwidth}
\centering
\includegraphics[width=1\linewidth]{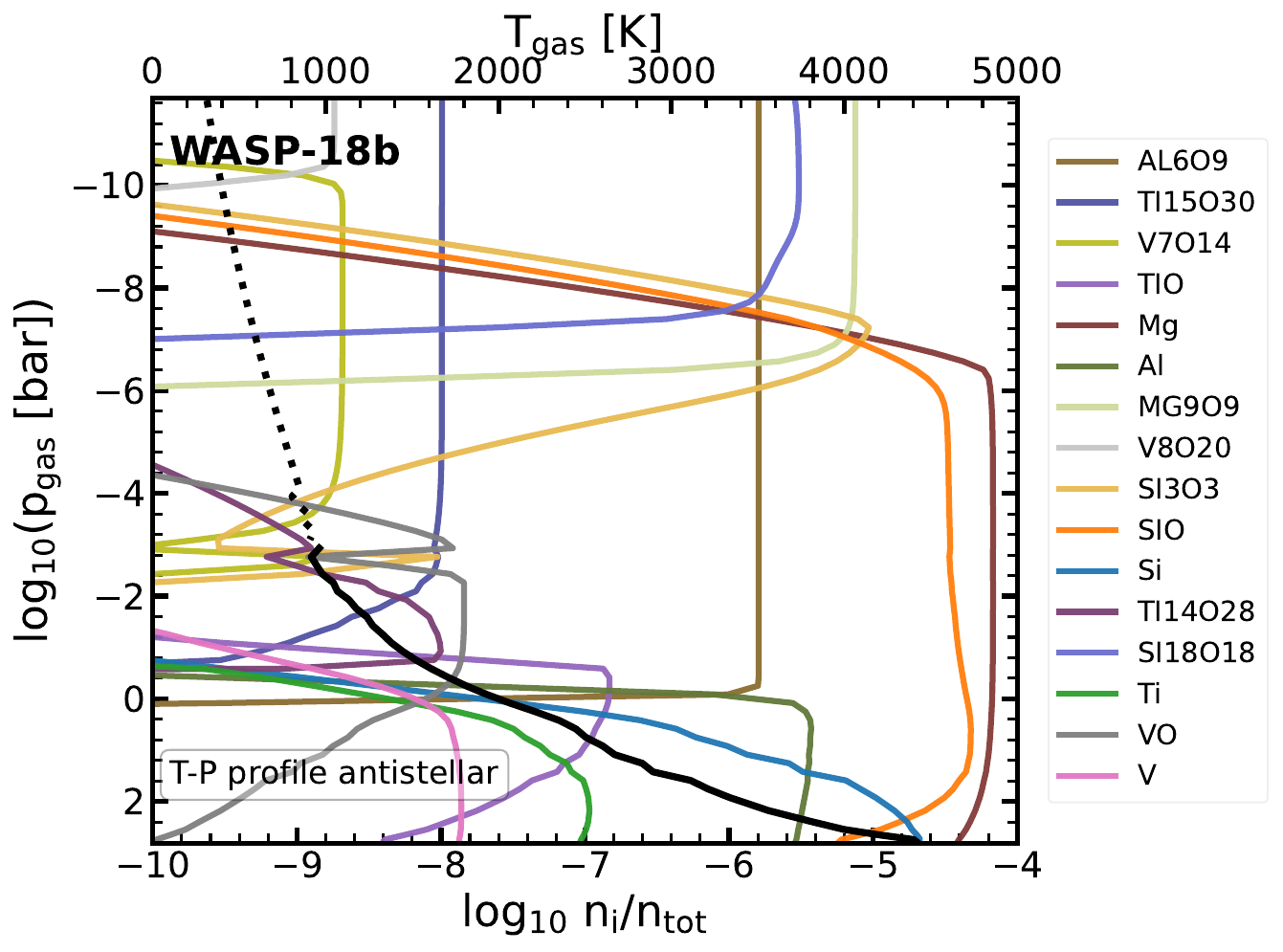}
\end{minipage}

\begin{minipage}{0.45\textwidth}
\centering
\includegraphics[width=1\linewidth]{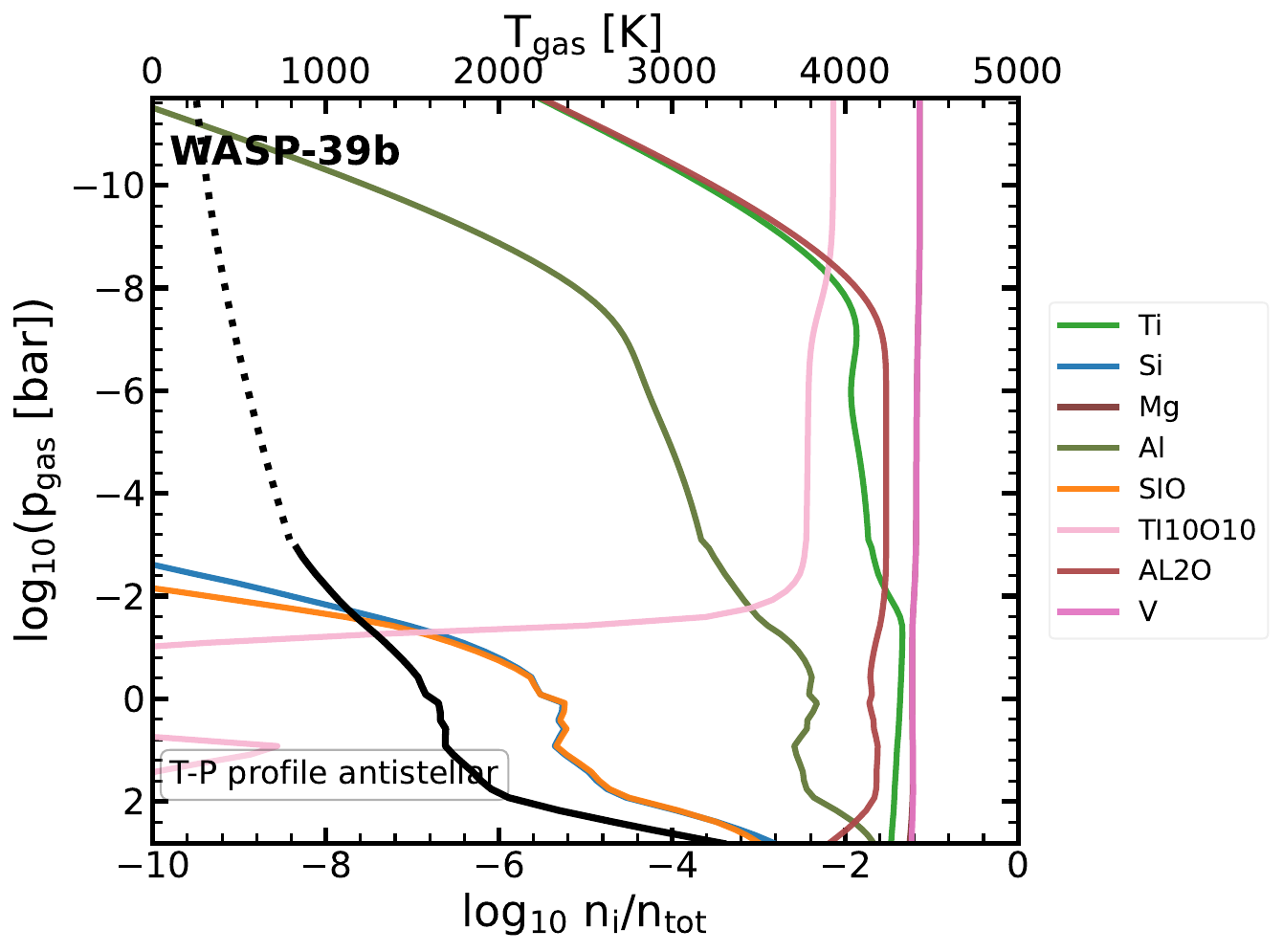}
\end{minipage}
\hspace{0.3cm}
\begin{minipage}{0.45\textwidth}
\centering
\includegraphics[width=1\linewidth]{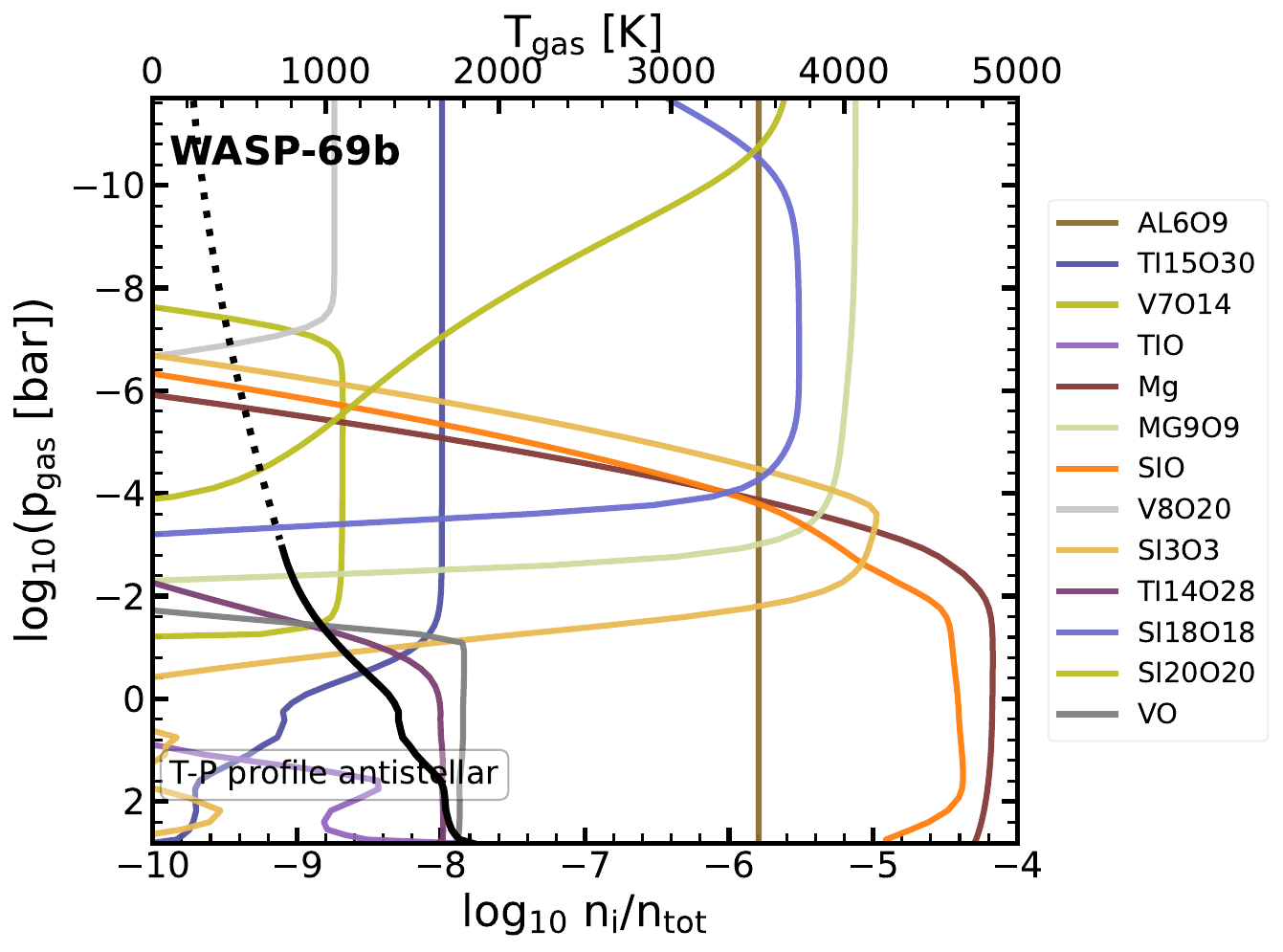}
\end{minipage}
\caption{Gas-phase concentrations at the antistellar point, $n _{\rm i}$/$n_{\rm tot}$, of the most abundant Al-, Ti-, Mg-, Si-, and V-bearing metals, ions, and metal oxide clusters, shown as a function of the $T_{\rm gas}$–$p_{\rm gas}$ profile, where the bold black segment corresponds to GCM calculations and the dotted black segment indicates the extrapolated low-pressure regime.}
\label{Antistellar}
\end{figure*}

In contrast to the substellar point, the antistellar (nightside) regions of all considered planets are dominated by metal oxide cluster species (Fig.~\ref{Antistellar}). For both UHJs (WASP-121 b and WASP-18 b), larger clusters ((Al$_2$O$_3$)$_3$, (TiO$_2$)$_{14}$, and (TiO$_2$)$_{15}$) are already thermochemically stable in the deep atmosphere at pressures near 1~bar, where high-temperature condensate precursors persist, while V-bearing clusters also become stable at $p_{\rm gas} \lesssim 10^{-3}$~bar. At $p_{\rm gas} \lesssim 10^{-6}$~bar, additional contributions from low-temperature condensate precursors, such as (MgO)$\rm\rm_N$ and (SiO)$\rm\rm_N$, become significant, leading to a chemically diverse cluster population in the upper nightside atmosphere of both the UHJs. WASP-39 b exhibits a markedly different behavior. Across nearly all pressure levels, the atmosphere is dominated by neutral metals or simple metal monoxides, such as Al$_2$O. This behavior is consistent with the suppressing effect of high metallicity on the stability of large clusters, as discussed in Sect.~\ref{section:hazesinexoplanets}. Notably, however, (TiO)$_{10}$ clusters are present throughout the atmosphere, which contrasts with the absence of large clusters observed in section \ref{equatorialslicepalnets}. This exception arises because TiO is a triplet in its ground electronic configuration, possesses open d-orbitals, and exhibits flexible bonding. Open-shell metal oxides polymerize more efficiently, as each added TiO unit provides a substantial enthalpy gain that can offset the universal entropic penalty associated with aggregation \citep{plane2012cosmic}. In contrast, the other metal oxide clusters considered in this study are thermodynamically saturated (closed-shell), such that the entropic penalty of aggregation is not compensated, suppressing the stability of larger clusters under chemical equilibrium conditions \citep{gail2014physics, woitkeggchem2018, plane2013nucleation}. Finally, WASP-69 b shows the emergence of stable clusters in the deeper atmosphere ($p_{\rm gas} \sim 1$~bar), with low-temperature condensate precursors becoming thermochemically favoured at pressures around $p_{\rm gas} \sim 10^{-2}$~bar. 
Except for Mg-bearing species, neutral metals do not dominate the lower atmosphere, where simple metal monoxides are instead prevalent.

\subsection{Metal oxide clusters at Morning Terminators and Metal/Metal- monoxides at Evening Terminators}
\begin{figure*}[!htbp]
\centering
\begin{minipage}{0.45\textwidth}
\centering
\includegraphics[width=1\linewidth]{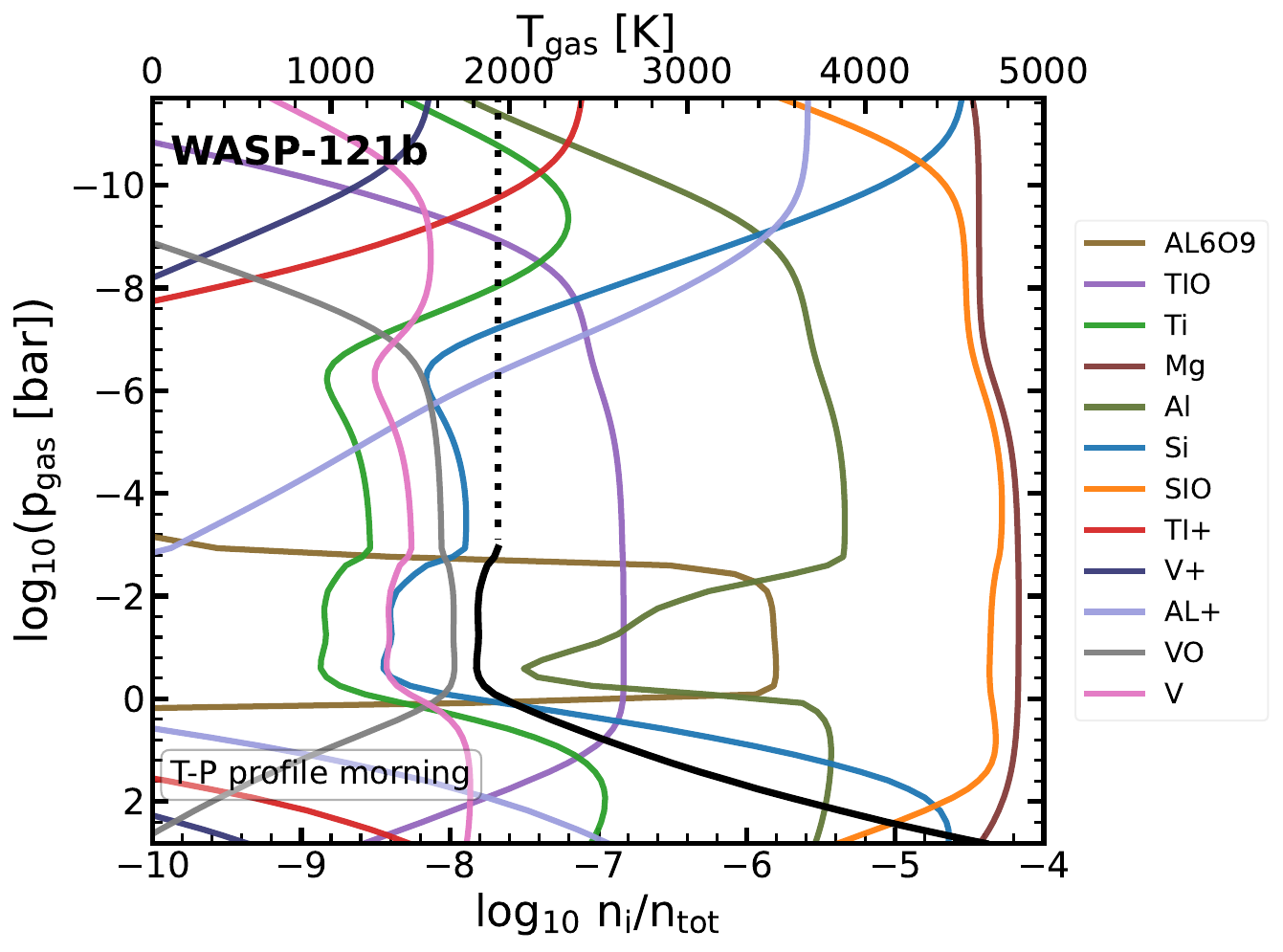}
\end{minipage}
\hspace{0.3cm}
\begin{minipage}{0.45\textwidth}
\centering
\includegraphics[width=1\linewidth]{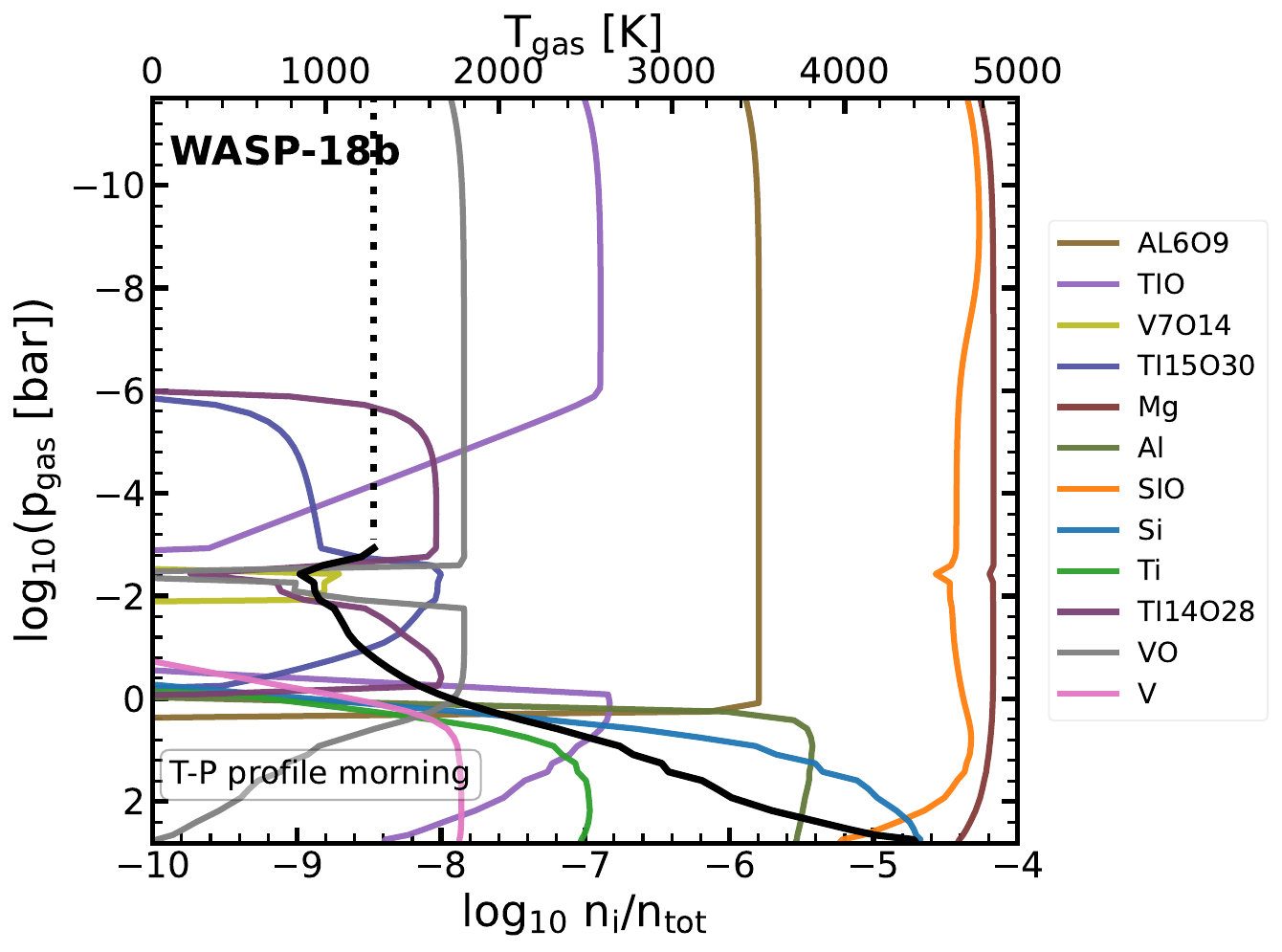}
\end{minipage}
\hspace{0.3cm}
\begin{minipage}{0.45\textwidth}
\centering
\includegraphics[width=1\linewidth]{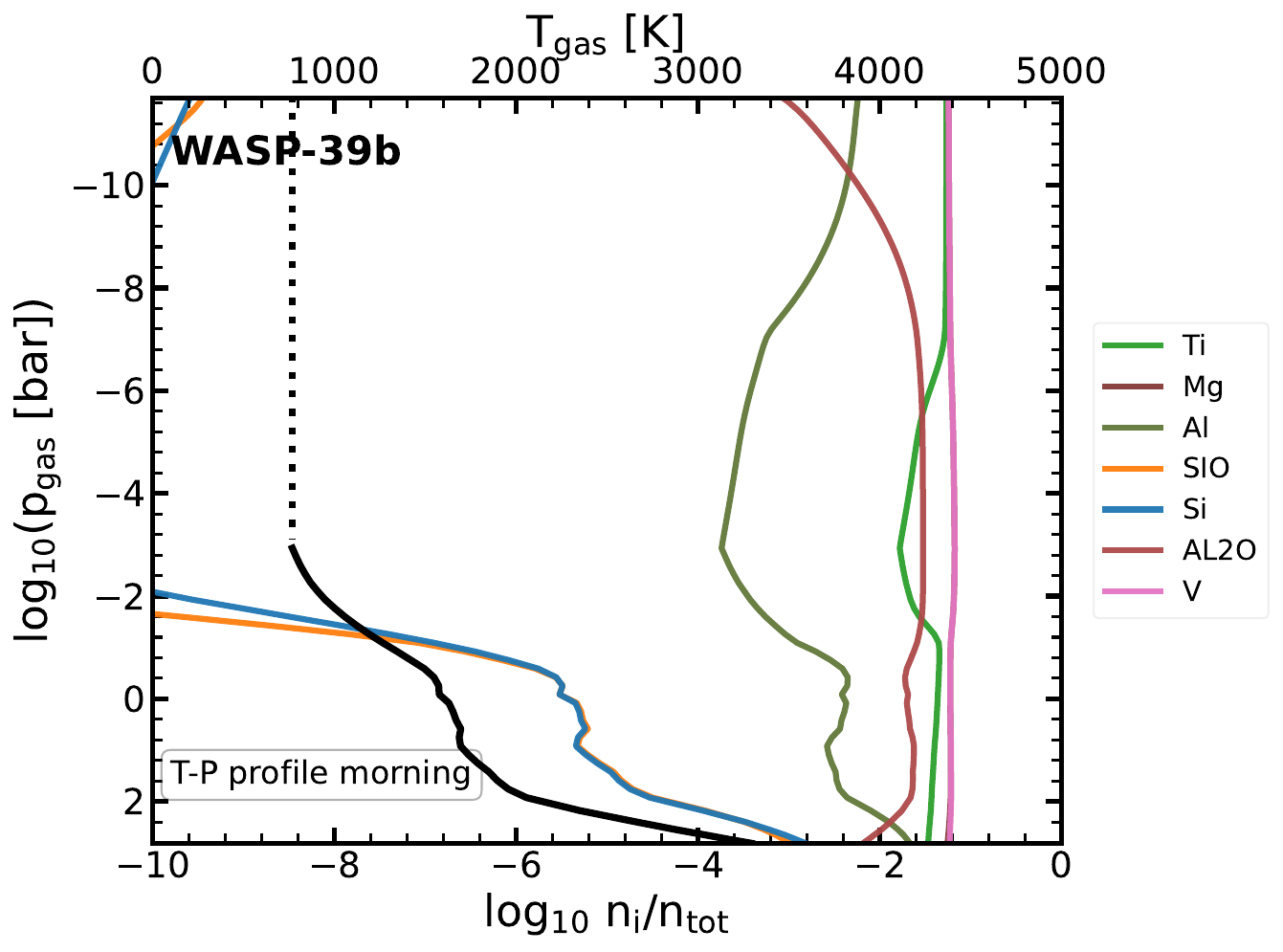}
\end{minipage}
\hspace{0.3cm}
\begin{minipage}{0.45\textwidth}
\centering
\includegraphics[width=1\linewidth]{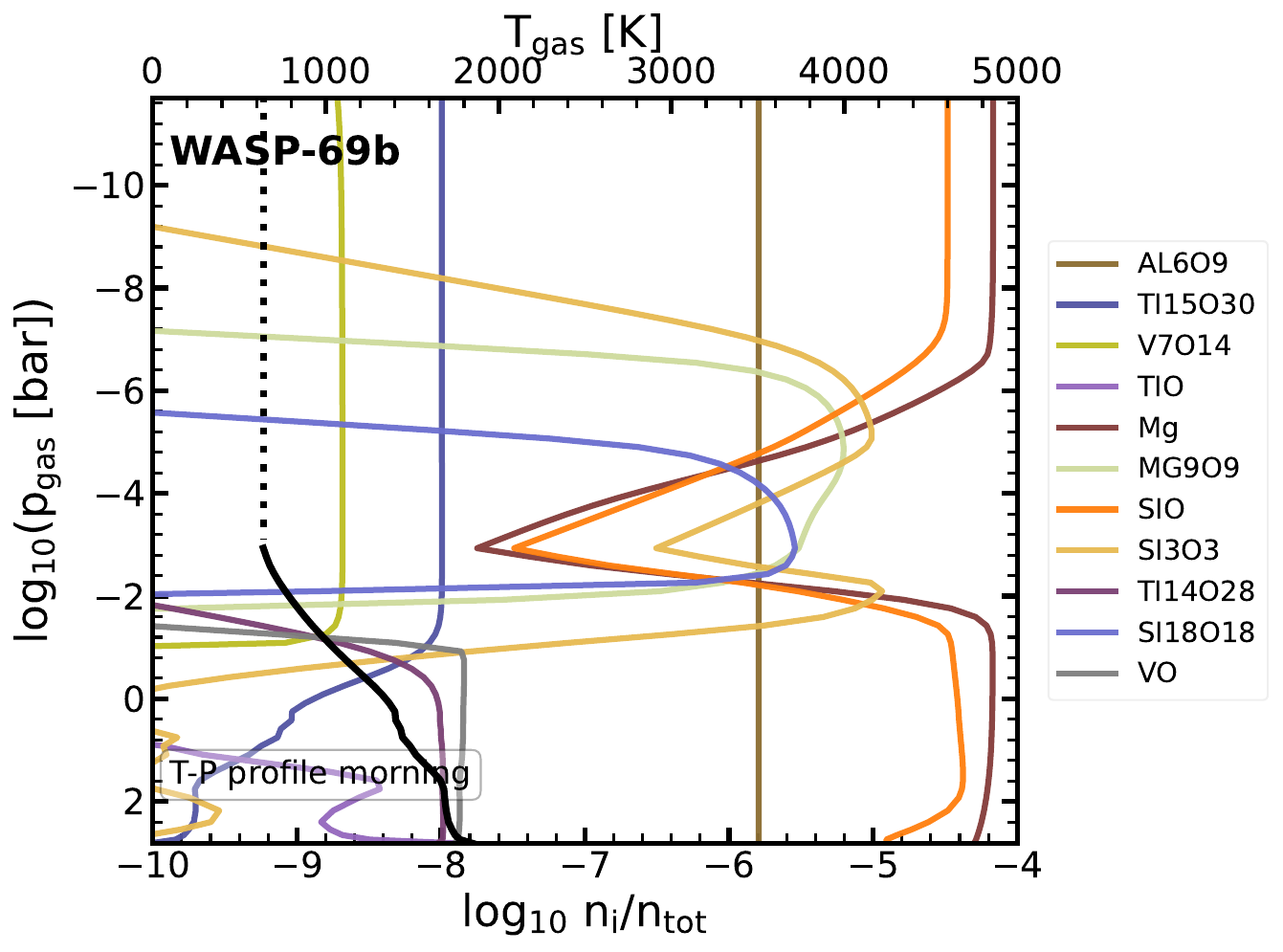}
\end{minipage}
\caption{Gas-phase concentrations at the morning terminator, $n _{\rm i}$/$n_{\rm tot}$, of the most abundant Al-, Ti-, Mg-, Si-, and V-bearing metals, ions, and metal oxide clusters, shown as a function of the $T_{\rm gas}$–$p_{\rm gas}$ profile, where the bold black segment corresponds to GCM calculations and the dotted black segment indicates the extrapolated low-pressure regime.}
\label{morningt}
\end{figure*}

As shown in Sect.~\ref{section:hazesinexoplanets}, metal oxide clusters are thermochemically favoured at the morning terminators, whereas the evening terminators are dominated by metal monoxides. In this section, we extend that analysis by combining the original GCM domain with the extrapolated upper atmosphere and examining the behavior of metal oxide clusters at both terminator regions for the planets considered in this study. For each metal considered in this study (Al, Ti, V, Si, and Mg), Fig.~\ref{morningt} presents the vertical concentration profiles of the most abundant neutral metal, metal ion, and cluster species at the morning terminator for all planets examined.

For WASP-121 b, the only large cluster that remains thermochemically stable is (Al$_2$O$_3$)$_3$, present over the pressure range $p_{\rm gas} \sim 1$ to $10^{-3}$~bar. Metal ions are stable both in the upper and deeper atmospheric layers, at pressures below $10^{-3}$~bar and above 10~bar, respectively, while neutral metals and metal oxide clusters persist throughout the atmosphere. For WASP-18 b, large clusters of high-temperature condensate precursors are thermochemically favoured at $p_{\rm gas} \sim 10^{-1}$~bar—specifically (TiO$_2$)$_{14}$ and (Al$_2$O$_3$)$_3$—and at $p_{\rm gas} \sim 1$~bar, where (TiO$_2$)$_{14}$, (TiO$_2$)$_{15}$, (VO$_2$)$_7$, and (Al$_2$O$_3$)$_3$ remain abundant. These clusters persist upward to the top of the atmosphere. Metal monoxides associated with low-temperature condensate precursors dominate intermediate pressure regions, while the deeper atmosphere is primarily composed of neutral metal species. Additionally, metal ions are absent on the morning side of WASP-18 b. For WASP-39 b, larger clusters are not thermochemically favoured at the morning terminator, consistent with the suppressing effect of high metallicity discussed in Sect.~\ref{section:hazesinexoplanets}. Instead, the atmosphere is dominated by neutral metal species and simple metal monomers throughout the entire pressure range. WASP-69 b exhibits particularly favorable conditions for stable clusters at the morning terminator. As shown in Fig.~\ref{morningt}, the considered species exhibit thermochemically stable large clusters starting in the deepest atmospheric layers, with (Al$_2$O$_3$)$_3$ being the most abundant. This indicates that high-temperature condensate precursors ((TiO$_2$)$_{14}$, (TiO$_2$)$_{15}$, and (Al$_2$O$_3$)$_3$) remain thermochemically favoured across much of the atmospheric column. In addition, low-temperature condensate precursors become abundant within the pressure range $10^{-7} \lesssim p_{\rm gas} \lesssim 10^{-2}$~bar. 

Except for Mg-bearing species, neutral metals are not thermochemically favoured in any atmospheric layer, highlighting the conditions in WASP-69 b that stabilise large clusters efficiently.

\begin{figure*}[!htbp]
\centering
\begin{minipage}{0.45\textwidth}
\centering
\includegraphics[width=1\linewidth]{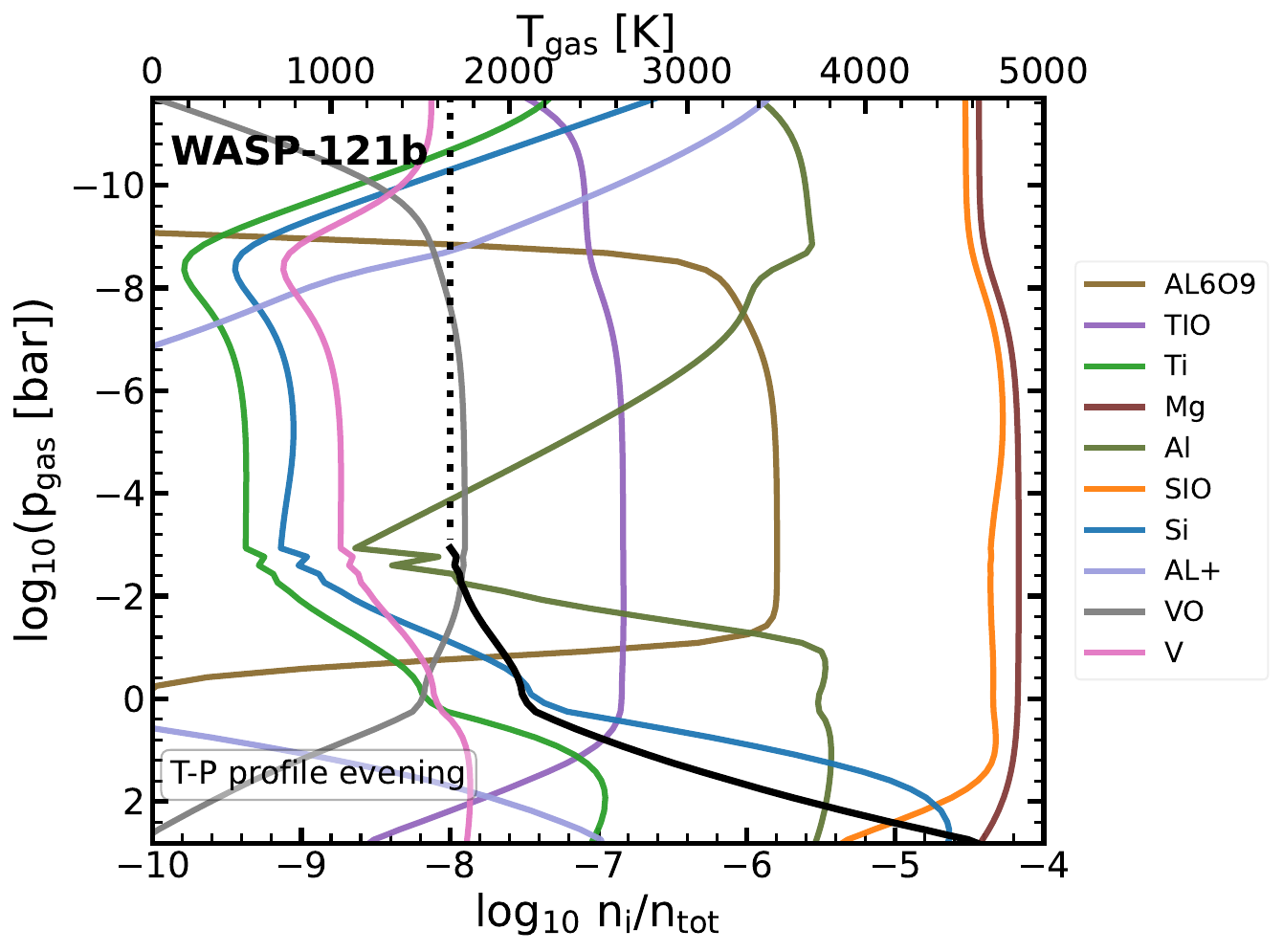}
\end{minipage}
\hspace{0.3cm}
\begin{minipage}{0.45\textwidth}
\centering
\includegraphics[width=1\linewidth]{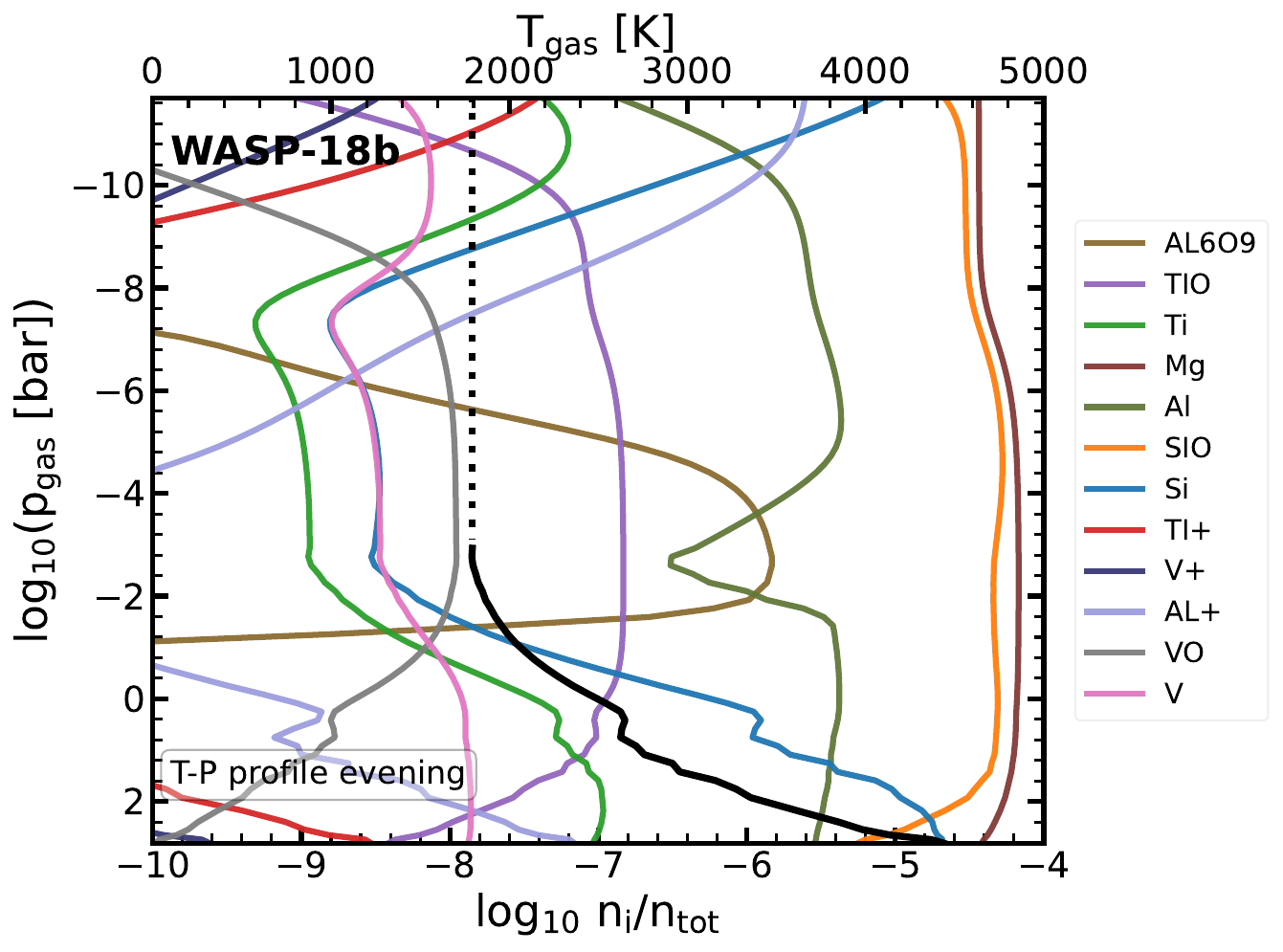}
\end{minipage}
\begin{minipage}{0.45\textwidth}
\centering
\includegraphics[width=1\linewidth]{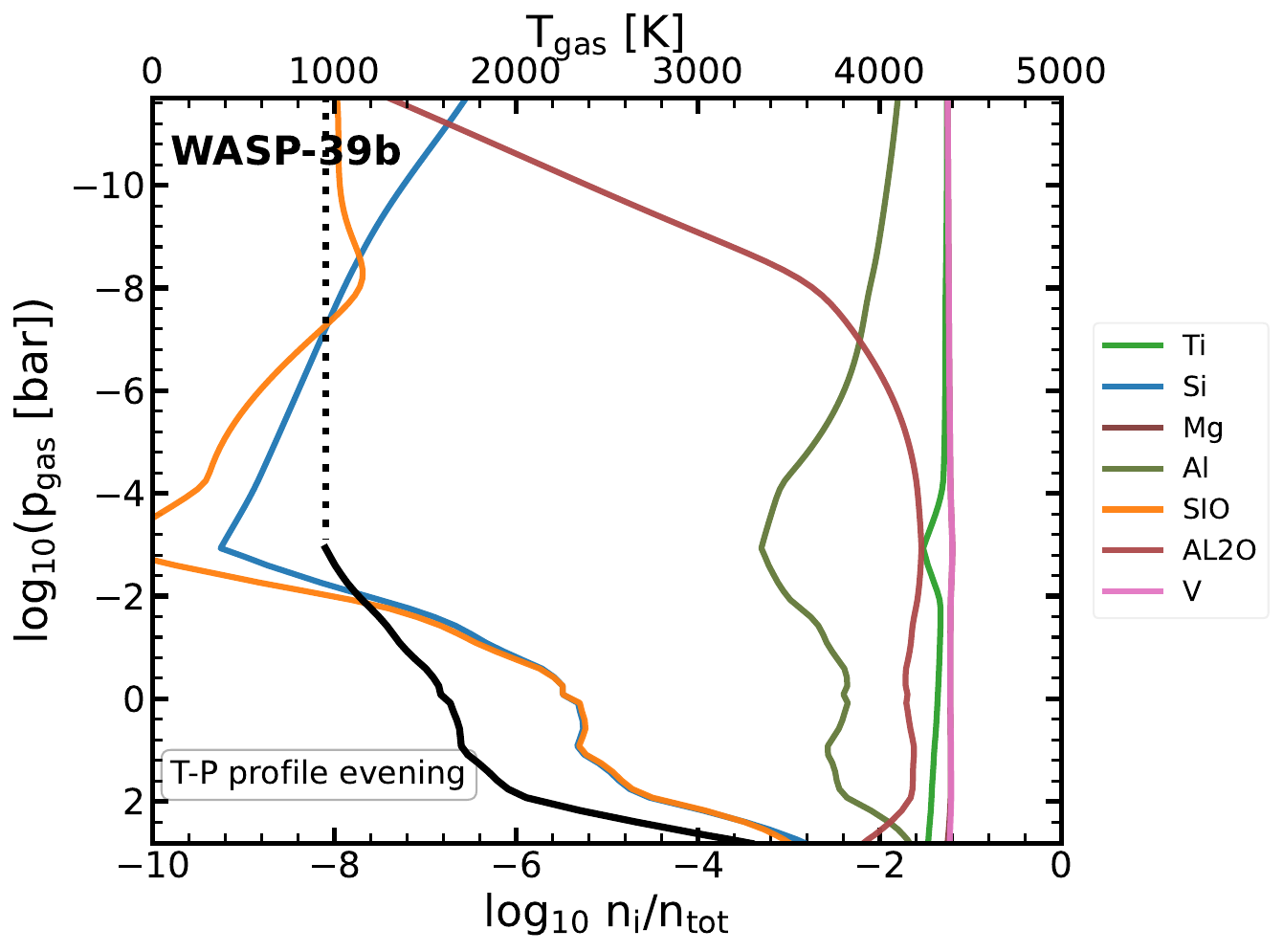}
\end{minipage}
\hspace{0.3cm}
\begin{minipage}{0.45\textwidth}
\centering
\includegraphics[width=1\linewidth]{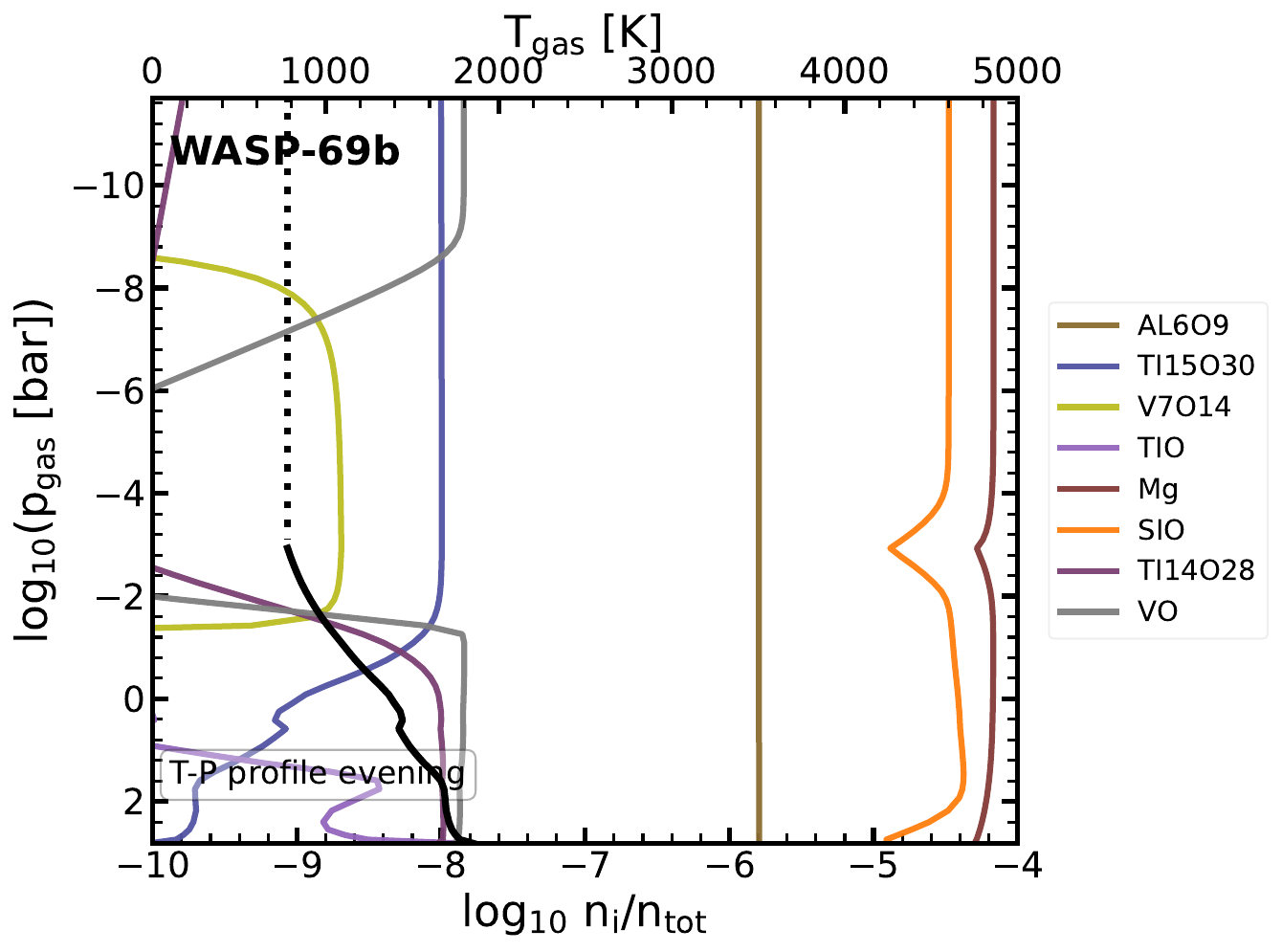}
\end{minipage}
\caption{Gas-phase concentrations at the evening terminator, $n _{\rm i}$/$n_{\rm tot}$, of the most abundant Al-, Ti-, Mg-, Si-, and V-bearing metals, ions, and metal oxide clusters, shown as a function of the $T_{\rm gas}$–$p_{\rm gas}$ profile, where the bold black segment corresponds to GCM calculations and the dotted black segment indicates the extrapolated low-pressure regime.}
\label{Eveningt}
\end{figure*}

Fig.~\ref{Eveningt} shows the vertical concentration profiles of the most abundant neutral metal, metal ion, and metal oxide clusters at the evening terminator for all planets considered in this study. For both UHJs, WASP-121 b and WASP-18 b, the upper atmosphere is dominated by metal ions, while the deeper atmosphere is primarily composed of neutral metal species. Interestingly, at the deepest atmospheric levels, metal ions reappear and become abundant again. This vertical behavior differs from the trends observed elsewhere in this study, where higher temperatures typically favor ionization and lower temperatures favor neutral species. The apparent discrepancy arises from the combined effects of temperature and pressure. As altitude increases, gas pressure decreases, lowering collisional rates and reducing the thermochemical stability of neutral metals and metal oxides. This results in ion-dominated upper atmospheres, even at lower temperatures. Conversely, in the deepest atmospheric layers, extremely high temperatures drive the thermal dissociation and ionization of neutral metals, resulting in the re-emergence of ions. We also find that (Al$_2$O$_3$)$_3$ remains stable in the middle atmospheric layers, spanning $p_{\rm gas} \sim 1$ to $10^{-9}$~bar for WASP-121 b and $p_{\rm gas} \sim 1$ to $10^{-7}$~bar for WASP-18 b, indicating that high-temperature condensate precursors can persist even at the evening terminator under favourable conditions.

For WASP-39 b, the evening terminator atmosphere is dominated by neutral metal species, with simple metal monoxides also contributing significantly, consistent with the suppression of large-cluster stability in high-metallicity environments. WASP-69 b stands out as the only planet in our sample where large metal oxide clusters—primarily high-temperature condensate precursors—remain thermochemically stable throughout the entire atmospheric column at the evening terminator. In addition, simple metal monoxides dominate parts of the lower atmosphere ($p_{\rm gas} \lesssim 10^{-2}$~bar), further highlighting the persistence of stable clusters under the thermodynamic conditions present in WASP-69 b.

\section{Discussion}
\citet{faherty2025silicate} reported the first detection of SiH$_4$ in the atmosphere of a brown dwarf, identifying it as an important gas-phase silicon reservoir and a potential precursor to silicate formation. Equilibrium chemistry calculations \citep{woitkeggchem2018,2020A&A...637A..59A} similarly predict SiH$_4$ to dominate at $T_{\rm gas} \lesssim 900\,\mathrm{K}$ when only small molecules are considered. Our results demonstrate that the inclusion of larger (SiO)$_N$ clusters fundamentally alters this picture: silicon is preferentially partitioned into clusters that become thermodynamically favored over SiH$_4$. This shift highlights the critical role of molecular clusters in regulating elemental reservoirs and suggests that models neglecting cluster formation may systematically overestimate SiH$_4$ abundances but also delay the predicted onset of silicate cloud formation in exoplanets, brown dwarfs, and oxygen-rich AGB stars. Cluster chemistry is, hence, not a secondary effect but a key component in determining the chemical structure of these environments.

Studies of meteoric metals in the Earth's atmosphere show that metal atoms and ions (e.g., Mg, Fe, Ca, Na) are rapidly converted into molecular reservoirs and subsequently into clusters and particles, which control their lifetimes and vertical distributions \citep{Plane2003, Plane2015, Plane2018}. This demonstrates that the partitioning of elements is governed by the formation and growth of molecular complexes rather than by isolated gas-phase species. Extending this framework to refractory oxides, bottom-up nucleation models \citep{2003A&A...407..191J,gobrecht2022bottom,sindel2022revisiting,lecoqhelena2024vanadium} indicate that condensation proceeds through the stepwise growth of stable clusters. Our results support this view by showing that the thermodynamic stability of larger metal oxide clusters strongly influences when and where gas-phase species are depleted. This provides a direct link between gas-phase chemistry and the emergence of cloud or dust particles, implying that the composition and spatial distribution of clouds are tightly coupled to cluster stability across different atmospheric regimes.

An important implication of this work is the connection between equilibrium cluster thermochemistry and the kinetics of nucleation. In classical nucleation theory (CNT), the formation of a critical cluster \citep{goeres1996chemistry} determines the nucleation barrier and thus the rate of particle formation. However, CNT relies on bulk approximations that may not be valid at the small cluster sizes relevant for astrophysical environments \citep{gail2014physics}. Our results suggest that the discrete stability of specific cluster sizes, as derived from quantum chemical data, can lead to deviations from classical predictions, potentially modifying both nucleation pathways and rates. This highlights the need to couple accurate cluster thermodynamics with kinetic models to achieve a more realistic description of cloud formation. Future applications of these results include time-dependent and multidimensional models of exoplanet atmospheres and AGB outflows \citep{2023A&A...669A.155F,2025A&A...699A.148A,2016A&A...594A..48L,2018MNRAS.481..194L,2023arXiv230108492C,helling2023exoplanet,PlaschzugW121b,2025arXiv251101548C}, where both chemical and dynamical processes shape cloud formation. In this context, uncertainties in thermochemical data and cluster stability may propagate into predictions of cloud properties and observable signatures. Therefore, combining accurate thermodynamic datasets with efficient nucleation and transport modelling will be essential for robust interpretation of current and future observations.

\section{Conclusion}
\label{sec:conclusion}
This study utilizes our thermochemical data for eight metal oxide cluster families, covering their respective size and accuracy limits. The largest clusters for each considered species are (TiO$_2$)$_{15}$, (SiO)$_{20}$, (TiO)$_{10}$, (Al$_2$O$_3$)$_{10}$, (MgO)$_{10}$, (VO)$_{10}$, (VO$_2$)$_{10}$, and (V$_2$O$_5$)$_4$. While sufficient for a first exploration of cluster stability, larger cluster data are needed for more rigorous derivation of nucleation rates relevant to exoplanet cloud formation. Analysis of (Al$_2$O$_3$)$\rm\rm_N$ and (MgO)$\rm\rm_N$ clusters at $p_{\rm gas} = 1$~bar shows that (Al$_2$O$_3$)$_3$ is the dominant species below $T_{\rm gas} \approx 2100$~K, reflecting stability of high-temperature aluminum oxide clusters. MgO clusters are thermochemically stable at lower temperatures, with larger clusters favoured only below $T_{\rm gas} \sim 600$~K, where (MgO)$_9$ (Magic cluster) is the dominant species. While (Al$_2$O$_3$)$\rm\rm_N$ remains stable at higher temperatures, neutral Mg atoms persist to higher temperatures, dissociating to Mg$^+$ near $T_{\rm gas} \sim 5000$~K, compared to Al dissociating to Al$^+$ at $\sim 4000$~K. Vibrational absorption spectra computed for (Al$_2$O$_3$)$\rm\rm_N$ at $T_{\rm gas} \sim 1000$~K and (MgO)$_9$ at $T_{\rm gas} \sim 250$~K show dense features in $\lambda \approx 10$–$50~\mu$m for aluminum oxides, and in $\lambda \approx 15$–$45~\mu$m for MgO, with only a few features at longer wavelengths associated with the largest clusters. Variations in altitude under an isothermal assumption affect only the magnitude of absorption, while spectral feature positions remain unchanged.

This study further highlights the structured stability of molecular clusters as a function of $T_{\rm gas}$ and $p_{\rm gas}$, considering all species in a combined atmosphere with solar metal abundances. Among the clusters analyzed, (TiO$_2$)$\rm_N$ and (Al$_2$O$_3$)$\rm_N$ are particularly thermochemically stable, maintaining larger cluster forms even under very high temperatures and pressures, indicating that they are likely key contributors to cloud-seed populations in hot exoplanetary atmospheres. In contrast, (MgO)$\rm\rm_N$ and (SiO)$\rm\rm_N$ clusters remain stable only under comparatively cooler conditions. Some species, such as TiO monoxides, cease contributing to potential cluster growth at high temperatures despite high monoxide densities, emphasizing the importance of thermal stability in sustaining cluster populations. Even clusters present at low concentrations, like (TiO)$\rm\rm_N$, can act as critical species in chemical growth pathways, potentially influencing nucleation rates under favorable thermodynamic conditions. The study also demonstrates that the vibrational absorption spectra of these clusters depend on both size and composition, providing a potential means to distinguish between different species in observational data. For example, (TiO$_2$)$\rm\rm_N$ and (Al$_2$O$_3$)$\rm\rm_N$ produce dense spectral features in the mid-infrared, while (MgO)$\rm\rm_N$ and (SiO)$\rm\rm_N$ are more prominent under cooler conditions. Absorption varies with $T_{\rm gas}$ and $p_{\rm gas}$, indicating that cloud signatures in exoplanet spectra are inherently altitude-dependent. Many key features fall within the 8–28~$\mu$m wavelength range accessible to \texttt{JWST}’s MIRI instrument, highlighting promising observational targets for probing metal oxide clouds in exoplanetary atmospheres. However, observations near 8~$\mu$m may be complicated by overlapping contributions from MgO and SiO monoxides.

Finally, by combining 3D GCM outputs with extrapolated thermodynamic profiles, this study systematically explores the stability and dissociation of metal oxide clusters, neutral metals, and metal ions across different atmospheric regions of UHJs (WASP-121 b and WASP-18 b), HJ (WASP-39 b), and WJ (WASP-69 b). The results demonstrate that longitudinal temperature asymmetries and vertical pressure gradients strongly control the chemical state of refractory species.
\begin{itemize}
\item In the UHJs WASP-121 b and WASP-18 b, extreme dayside temperatures and low collisional rates suppress the thermochemical stability of large clusters at the substellar point and evening terminator, where the upper atmospheres are dominated by metal ions and the deeper layers by neutral metals and simple metal monoxides. Stable clusters in these planets are instead favoured on the antistellar hemisphere and at the morning terminator, where lower temperatures allow high-temperature condensate precursors to persist as larger aggregates. Even in these favourable regions, cluster stability is restricted to relatively narrow pressure ranges, and the extrapolated upper atmospheres are consistently dominated by rapid dissociation into metal ions as pressure decreases.
\item WASP-39 b exhibits markedly different behavior due to its enhanced metallicity. Despite increased elemental abundances, large metal oxide clusters remain largely absent throughout both the GCM and extrapolated domains, with the atmosphere dominated by neutral metals and simple metal monoxides across all locations. This result demonstrates that higher metallicity alone does not guarantee the thermochemical stability of larger clusters, as chemical equilibrium favors smaller species that minimize Gibbs free energy. Only TiO retains stable larger-cluster forms in WASP-39 b, reflecting its open-shell bonding, which provides sufficient enthalpic stabilization to offset the entropic penalty of aggregation. This behavior underscores that molecular bonding characteristics, rather than elemental abundance alone, govern the stability and persistence of clusters in exoplanetary atmospheres.
\item The WJ WASP-69 b provides the most favorable conditions for metal oxide cluster stability among the planets studied. Moderate temperatures and higher collisional efficiencies allow both high- and low-temperature condensate precursors to remain stable as larger clusters over extended pressure ranges, particularly on the nightside and at the morning terminator. In WASP-69 b, metal oxide clusters persist from deep atmospheric layers into the extrapolated upper atmosphere before eventually dissociating into neutral metals and metal ions at very low pressures. The presence of well-defined transition layers, where clusters can coexist with neutral metals and ions, illustrates a gradual chemical evolution largely absent in the hotter planets. These results indicate that WJ atmospheres may provide optimal conditions for the persistence of larger clusters and the initial stages of cloud-seed formation, which ultimately drive condensation processes.
\end{itemize}

Looking forward, this study highlights both the progress made and the challenges that remain in understanding the stability of metal oxide clusters in exoplanetary atmospheres. Our combined analysis of GCM and extrapolated regions demonstrates that cluster persistence is highly sensitive to local thermodynamic conditions, planetary regime, and molecular chemistry. UHJs are dominated by ionized species with limited cluster stability, high-metallicity atmospheres can suppress the persistence of large clusters via thermodynamic constraints, and WJs provide the most favorable conditions for sustained metal oxide cluster populations. Additionally, the spectral interpretation of hot gaseous exoplanets may require updated gas-phase opacity treatments that include metal ions, which become abundant on the hot daysides of these planets. Despite these insights, the limited availability of accurate thermochemical data for larger clusters constrains detailed predictions of nucleation rates and cloud properties. Future work should focus on extending both computational and experimental datasets to include larger cluster sizes and additional species, particularly those present at low concentrations but potentially critical as intermediates in cluster growth pathways. Moreover, the temperature- and pressure-dependent spectral diversity of these clusters calls for detailed simulations to improve the interpretation of observational data from facilities such as \texttt{JWST}. Ultimately, bridging molecular-level cluster chemistry with global atmospheric models will enable more accurate predictions of cloud formation, metal oxide cluster distribution, and their observational signatures, advancing our understanding of exoplanetary atmospheres and the conditions under which clouds form.

\newpage
\vspace{12pt} 
\noindent{\bf Acknowledgments:}
D.B., Ch.H., and A.R. acknowledge funding from the Austrian Science Fund (FWF) under grant No.~10.55776/PAT3166823. D.G. thanks the Swedish National Infrastructure for Computing (SNIC) at C3SE, partially funded by the Swedish Research Council (grants No.~2018-05973 and 2022-06725) and NAISS resources. H.L.M. acknowledges support from the Swiss National Science Foundation (grant No.~200021-231596) and, together with L.C., thanks the Vienna Scientific Cluster for computational resources. We thank Thomas Ley for collaboration on (TiO)$\rm_N$ QCC data and the anonymous referee for valuable feedback.

\bibliographystyle{aasjournal}
\bibliography{references}

@article{helling2019exoplanet,
  title={Exoplanet clouds},
  author={Helling, Christiane},
  journal={Annual Review of Earth and Planetary Sciences},
  volume={47},
  number={1},
  pages={583--606},
  year={2019},
  publisher={Annual Reviews}
}

@article{sindel2022revisiting,
  title={Revisiting fundamental properties of TiO2 nanoclusters as condensation seeds in astrophysical environments},
  author={Sindel, JP and Gobrecht, D and Helling, Ch and Decin, L},
  journal={Astronomy \& Astrophysics},
  volume={668},
  pages={A35},
  year={2022},
  publisher={EDP Sciences}
}

@article{gobrecht2022bottom,
  title={Bottom-up dust nucleation theory in oxygen-rich evolved stars-I. Aluminium oxide clusters},
  author={Gobrecht, David and Plane, John MC and Bromley, Stefan T and Decin, Leen and Cristallo, Sergio and Sekaran, Sanjay},
  journal={Astronomy \& Astrophysics},
  volume={658},
  pages={A167},
  year={2022},
  publisher={EDP Sciences}
}

@article{asplund2009chemical,
  title={The chemical composition of the Sun},
  author={Asplund, Martin and Grevesse, Nicolas and Sauval, A Jacques and Scott, Pat},
  journal={Annual review of astronomy and astrophysics},
  volume={47},
  number={1},
  pages={481--522},
  year={2009},
  publisher={Annual Reviews}
}

@ARTICLE{lecoqhelena2024vanadium,
       author = {{Lecoq-Molinos}, H. and {Gobrecht}, D. and {Sindel}, J.~P. and {Helling}, Ch. and {Decin}, L.},
        title = "{Vanadium oxide clusters in substellar atmospheres: A quantum chemical study}",
      journal = {\aap},
         year = 2024,
        month = oct,
       volume = {690},
          eid = {A34},
        pages = {A34},
          doi = {10.1051/0004-6361/202347693},
archivePrefix = {arXiv},
       eprint = {2401.02784},
 primaryClass = {astro-ph.EP},
       adsurl = {https://ui.adsabs.harvard.edu/abs/2024A&A...690A..34L}
}

@ARTICLE{woitkeggchem2018,
       author = {{Woitke}, P. and {Helling}, Ch. and {Hunter}, G.~H. and {Millard}, J.~D. and {Turner}, G.~E. and {Worters}, M. and {Blecic}, J. and {Stock}, J.~W.},
        title = "{Equilibrium chemistry down to 100 K. Impact of silicates and phyllosilicates on the carbon to oxygen ratio}",
      journal = {\aap},
         year = 2018,
        month = jun,
       volume = {614},
          eid = {A1},
        pages = {A1},
          doi = {10.1051/0004-6361/201732193},
archivePrefix = {arXiv},
       eprint = {1712.01010},
 primaryClass = {astro-ph.EP},
       adsurl = {https://ui.adsabs.harvard.edu/abs/2018A&A...614A...1W}
}

@article{sindel2023infrared,
  title={Infrared spectra of TiO2 clusters for hot Jupiter atmospheres},
  author={Sindel, JP and Helling, Ch and Gobrecht, D and Chubb, KL and Decin, L},
  journal={Astronomy \& Astrophysics},
  volume={680},
  pages={A65},
  year={2023},
  publisher={EDP Sciences}
}

@article{spanget2015ir,
  title={IR intensity: Lorentz epsilon curve from'Gaussian'FREQ output},
  author={Spanget-Larsen, Jens},
  year={2015}
}

@book{gail2014physics,
  title={Physics and chemistry of circumstellar dust shells},
  author={Gail, Hans-Peter and Sedlmayr, Erwin},
  number={52},
  year={2014},
  publisher={Cambridge University Press}
}

@article{boulangier2019developing,
  title={Developing a self-consistent AGB wind model--II. Non-classical, non-equilibrium polymer nucleation in a chemical mixture},
  author={Boulangier, Jels and Gobrecht, David and Decin, Leen and de Koter, Alex and Yates, Jeremy},
  journal={Monthly Notices of the Royal Astronomical Society},
  volume={489},
  number={4},
  pages={4890--4911},
  year={2019},
  publisher={Oxford University Press}
}

@article{komatsu2018first,
  title={First evidence for silica condensation within the solar protoplanetary disk},
  author={Komatsu, Mutsumi and Fagan, Timothy J and Krot, Alexander N and Nagashima, Kazuhide and Petaev, Michail I and Kimura, Makoto and Yamaguchi, Akira},
  journal={Proceedings of the National Academy of Sciences},
  volume={115},
  number={29},
  pages={7497--7502},
  year={2018},
  publisher={National Academy of Sciences}
}

@article{gail2013seed,
  title={Seed particle formation for silicate dust condensation by SiO nucleation},
  author={Gail, H-P and Wetzel, Steffen and Pucci, Annemarie and Tamanai, Akemi},
  journal={Astronomy \& Astrophysics},
  volume={555},
  pages={A119},
  year={2013},
  publisher={EDP Sciences}
}

@article{andres1998time,
  title={A time-averaged inventory of subaerial volcanic sulfur emissions},
  author={Andres, RJ and Kasgnoc, AD},
  journal={Journal of Geophysical Research: Atmospheres},
  volume={103},
  number={D19},
  pages={25251--25261},
  year={1998},
  publisher={Wiley Online Library}
}

@article{gail1984dust,
  title={Dust formation in stellar winds. I-A rapid computational method and application to graphite condensation},
  author={Gail, H-P and Keller, R and Sedlmayr, E},
  journal={Astronomy and Astrophysics (ISSN 0004-6361), vol. 133, no. 2, April 1984, p. 320-332. Sponsorship: Deutsche Forschungsgemeinschaft.},
  volume={133},
  pages={320--332},
  year={1984}
}

@article{lee2018dust,
  title={Dust in brown dwarfs and extra-solar planets-VI. Assessing seed formation across the brown dwarf and exoplanet regimes},
  author={Lee, Elspeth KH and Blecic, Jasmina and Helling, Ch},
  journal={Astronomy \& Astrophysics},
  volume={614},
  pages={A126},
  year={2018},
  publisher={EDP Sciences}
}

@article{hudson1993cloud,
  title={Cloud condensation nuclei},
  author={Hudson, James G},
  journal={Journal of Applied Meteorology and Climatology},
  volume={32},
  number={4},
  pages={596--607},
  year={1993}
}

@article{bromley2016under,
  title={Under what conditions does (SiO) N nucleation occur? A bottom-up kinetic modelling evaluation},
  author={Bromley, Stefan T and Martin, Juan Carlos Gomez and Plane, John MC},
  journal={Physical Chemistry Chemical Physics},
  volume={18},
  number={38},
  pages={26913--26922},
  year={2016},
  publisher={Royal Society of Chemistry}
}

@article{patzer2014density,
  title={A density functional study of small TixCy (x, y= 1--4) molecules and their thermochemical properties},
  author={Patzer, ABC and Chang, Ch and S{\"u}lzle, D},
  journal={Chemical Physics Letters},
  volume={612},
  pages={39--44},
  year={2014},
  publisher={Elsevier}
}

@article{plane2013nucleation,
  title={On the nucleation of dust in oxygen-rich stellar outflows},
  author={Plane, John MC},
  journal={Philosophical Transactions of the Royal Society A: Mathematical, Physical and Engineering Sciences},
  volume={371},
  number={1994},
  pages={20120335},
  year={2013},
  publisher={The Royal Society Publishing}
}

@article{chang2013small,
  title={Small Fe bearing ring molecules of possible astrophysical interest: molecular properties and rotational spectra},
  author={Chang, C and Patzer, ABC and Kegel, WH and Chandra, S},
  journal={Astrophysics and Space Science},
  volume={347},
  pages={315--325},
  year={2013},
  publisher={Springer}
}

@article{chang2005inorganic,
  title={Inorganic cage molecules encapsulating Kr: A computational study},
  author={Chang, Ch and Patzer, ABC and Sedlmayr, E and S{\"u}lzle, D},
  journal={Physical Review B—Condensed Matter and Materials Physics},
  volume={72},
  number={23},
  pages={235402},
  year={2005},
  publisher={APS}
}

@article{lam2015atomistic,
  title={Atomistic mechanisms for the nucleation of aluminum oxide nanoparticles},
  author={Lam, Julien and Amans, David and Dujardin, Christophe and Ledoux, Gilles and Allouche, Abdul-Rahman},
  journal={The Journal of Physical Chemistry A},
  volume={119},
  number={33},
  pages={8944--8949},
  year={2015},
  publisher={ACS Publications}
}

@article{gail1986primary,
  title={The primary condensation process for dust around late M-type stars},
  author={Gail, H-P and Sedlmayr, E},
  journal={Astronomy and Astrophysics (ISSN 0004-6361), vol. 166, no. 1-2, Sept. 1986, p. 225-236.},
  volume={166},
  pages={225--236},
  year={1986}
}

@article{nuth1981vibrational,
  title={Vibrational disequilibrium in low pressure clouds},
  author={Nuth, JOSEPH A and Donn, Bertram},
  journal={Astrophysical Journal, Part 1, vol. 247, Aug. 1, 1981, p. 925-935.},
  volume={247},
  pages={925--935},
  year={1981}
}

@article{nuth1982experimental,
  title={Experimental studies of the vapor phase nucleation of refractory compounds. I. The condensation of SiO},
  author={Nuth, Joseph A and Donn, Bertram},
  journal={The Journal of Chemical Physics},
  volume={77},
  number={5},
  pages={2639--2646},
  year={1982},
  publisher={American Institute of Physics}
}

@article{nuth2006silicates,
  title={Silicates do nucleate in oxygen-rich circumstellar outflows: new vapor pressure data for SiO},
  author={Nuth III, Joseph A and Ferguson, Frank T},
  journal={The Astrophysical Journal},
  volume={649},
  number={2},
  pages={1178},
  year={2006},
  publisher={IOP Publishing}
}

@article{reber2006silicon,
  title={Silicon oxide nanoparticles reveal the origin of silicate grains in circumstellar environments},
  author={Reber, Arthur C and Clayborne, Penee A and Reveles, J Ulises and Khanna, Shiv N and Castleman, AW and Ali, Ashraf},
  journal={Nano letters},
  volume={6},
  number={6},
  pages={1190--1195},
  year={2006},
  publisher={ACS Publications}
}

@article{reber2008sio,
  title={From SiO molecules to silicates in circumstellar space: atomic structures, growth patterns, and optical signatures of Si n O m clusters},
  author={Reber, Arthur C and Paranthaman, Selvarengan and Clayborne, Pene{\'e} A and Khanna, Shiv N and Castleman Jr, A Welford},
  journal={ACS nano},
  volume={2},
  number={8},
  pages={1729--1737},
  year={2008},
  publisher={ACS Publications}
}

@phdthesis{lecoq2025microphysics,
  title        = {Microphysics of Cloud Formation: The Path to Heterogeneous Nucleation},
  author       = {Lecoq Molinos, Helena},
  school       = {KU Leuven},
  year         = {2025},
  type         = {PhD thesis},
  note         = {Advisors: Leen Decin, Christiane Helling, and David Gobrecht}
}

@article{chen2014structures,
  title={Structures and stabilities of (MgO) n nanoclusters},
  author={Chen, Mingyang and Felmy, Andrew R and Dixon, David A},
  journal={The Journal of Physical Chemistry A},
  volume={118},
  number={17},
  pages={3136--3146},
  year={2014},
  publisher={ACS Publications}
}

@ARTICLE{2001A&A...376..194H,
       author = {{Helling}, Ch. and {Oevermann}, M. and {L{\"u}ttke}, M.~J.~H. and {Klein}, R. and {Sedlmayr}, E.},
        title = "{Dust in brown dwarfs. I. Dust formation under turbulent conditions on microscopic scales}",
      journal = {\aap},
         year = 2001,
        month = sep,
       volume = {376},
        pages = {194-212},
          doi = {10.1051/0004-6361:20010937},
       adsurl = {https://ui.adsabs.harvard.edu/abs/2001A&A...376..194H}
}

@INPROCEEDINGS{1999IAUS..191..233J,
       author = {{Jeong}, K.~S. and {Winters}, J.~M. and {Sedlmayr}, E.},
        title = "{Dust formation in oxygen-rich circumstellar shells around long--period variables}",
    booktitle = {Asymptotic Giant Branch Stars},
         year = 1999,
       editor = {{Le Bertre}, T. and {Lebre}, A. and {Waelkens}, C.},
       series = {IAU Symposium},
       volume = {191},
        month = jan,
        pages = {233},
       adsurl = {https://ui.adsabs.harvard.edu/abs/1999IAUS..191..233J}
}

@ARTICLE{1986A&A...166..225G,
       author = {{Gail}, H. -P. and {Sedlmayr}, E.},
        title = "{The primary condensation process for dust around late M-type stars}",
      journal = {\aap},
         year = 1986,
        month = sep,
       volume = {166},
       number = {1-2},
        pages = {225-236},
       adsurl = {https://ui.adsabs.harvard.edu/abs/1986A&A...166..225G}
}

@ARTICLE{2003A&A...407..191J,
       author = {{Jeong}, K.~S. and {Winters}, J.~M. and {Le Bertre}, T. and {Sedlmayr}, E.},
        title = "{Self-consistent modeling of the outflow from the O-rich Mira IRC -20197}",
      journal = {\aap},
         year = 2003,
        month = aug,
       volume = {407},
        pages = {191-206},
          doi = {10.1051/0004-6361:20030693},
       adsurl = {https://ui.adsabs.harvard.edu/abs/2003A&A...407..191J}
}

@inproceedings{goeres1996chemistry,
  title={Chemistry and thermodynamics of the nucleation in R CrB star shells},
  author={Goeres, A},
  booktitle={Hydrogen Deficient Stars},
  volume={96},
  pages={69},
  year={1996}
}

@book{vehkamaki2006classical,
  title={Classical nucleation theory in multicomponent systems},
  author={Vehkam{\"a}ki, Hanna},
  year={2006},
  publisher={Springer}
}

@ARTICLE{1998FaDi..109..303G,
       author = {{Gail}, H. -P. and {Sedlmayr}, E.},
        title = "{Inorganic dust formation in astrophysical environments}",
      journal = {Faraday Discussions},
         year = 1998,
        month = jan,
       volume = {109},
        pages = {303},
          doi = {10.1039/a709290c},
       adsurl = {https://ui.adsabs.harvard.edu/abs/1998FaDi..109..303G}
}

@INPROCEEDINGS{2023ASSP...59...89G,
       author = {{Gobrecht}, David and {Plane}, John M.~C. and {Bromley}, Stefan T. and {Decin}, Leen and {Cristallo}, Sergio and {Sekeran}, Sanjay},
        title = "{The Corundum Conundrum}",
    booktitle = {European Conference on Laboratory Astrophysics ECLA2020. The Interplay of Dust},
         year = 2023,
        month = jan,
        pages = {89-93},
          doi = {10.1007/978-3-031-29003-9_10},
       adsurl = {https://ui.adsabs.harvard.edu/abs/2023ASSP...59...89G}
}

@MISC{2024jwst.prop.6045B,
       author = {{Baeyens}, Robin and {Barat}, Saugata and {Decin}, Leen and {Desert}, Jean-Michel and {Gobrecht}, David and {Helling}, Christiane and {Lecoq-Molinos}, Helena and {Savel}, Arjun Baliga and {Shivkumar}, Hinna and {Sikora}, James and {Sindel}, Jan Philip},
        title = "{Detecting ongoing gas-to-solid nucleation on the ultra-hot planet WASP-76 b}",
 howpublished = {JWST Proposal. Cycle 3, ID. \#6045},
         year = 2024,
        month = feb,
        pages = {6045},
       adsurl = {https://ui.adsabs.harvard.edu/abs/2024jwst.prop.6045B}
}

@ARTICLE{2024RASTI...3..636C,
       author = {{Chubb}, Katy L. and {Robert}, S{\'e}verine and {Sousa-Silva}, Clara and {Yurchenko}, Sergei N. and {Allard}, Nicole F. and {Boudon}, Vincent and {Buldyreva}, Jeanna and {Bultel}, Benjamin and {Coustenis}, Athena and {Foltynowicz}, Aleksandra and {Gordon}, Iouli E. and {Hargreaves}, Robert J. and {Helling}, Christiane and {Hill}, Christian and {Hrodmarsson}, Helgi Rafn and {Karman}, Tijs and {Lecoq-Molinos}, Helena and {Migliorini}, Alessandra and {Rey}, Micha{\"e}l and {Richard}, Cyril and {Sadiek}, Ibrahim and {Schmidt}, Fr{\'e}d{\'e}ric and {Sokolov}, Andrei and {Stefani}, Stefania and {Tennyson}, Jonathan and {Venot}, Olivia and {Wright}, Sam O.~M. and {Arenales-Lope}, Rosa and {Barstow}, Joanna K. and {Bocchieri}, Andrea and {Carrasco}, Nathalie and {Dubey}, Dwaipayan and {Egorov}, Oleg and {Mu{\~n}oz}, Antonio Garc{\'\i}a and {Gharib-Nezhad}, Ehsan (Sam) and {Gkouvelis}, Leonardos and {Gr{\"u}bel}, Fabian and {Irwin}, Patrick Gerard Joseph and {Kn{\'\i}{\v{z}}ek}, Anton{\'\i}n and {Lewis}, David A. and {Lodge}, Matt G. and {Ma}, Sushuang and {Martins}, Zita and {Molaverdikhani}, Karan and {Morello}, Giuseppe and {Nikitin}, Andrei and {Panek}, Emilie and {Rengel}, Miriam and {Rinaldi}, Giovanna and {Skinner}, Jack W. and {Tinetti}, Giovanna and {van Kempen}, Tim A. and {Yang}, Jingxuan and {Zingales}, Tiziano},
        title = "{Data availability and requirements relevant for the Ariel space mission and other exoplanet atmosphere applications}",
      journal = {RAS Techniques and Instruments},
         year = 2024,
        month = jan,
       volume = {3},
       number = {1},
        pages = {636-690},
          doi = {10.1093/rasti/rzae039},
archivePrefix = {arXiv},
       eprint = {2404.02188},
 primaryClass = {astro-ph.IM},
       adsurl = {https://ui.adsabs.harvard.edu/abs/2024RASTI...3..636C}
}

@ARTICLE{2020A&A...639A...3C,
       author = {{Chubb}, Katy L. and {Min}, Michiel and {Kawashima}, Yui and {Helling}, Christiane and {Waldmann}, Ingo},
        title = "{Aluminium oxide in the atmosphere of hot Jupiter WASP-43b}",
      journal = {\aap},
         year = 2020,
        month = jul,
       volume = {639},
          eid = {A3},
        pages = {A3},
          doi = {10.1051/0004-6361/201937267},
archivePrefix = {arXiv},
       eprint = {2004.13679},
 primaryClass = {astro-ph.EP},
       adsurl = {https://ui.adsabs.harvard.edu/abs/2020A&A...639A...3C}
}

@article{baeyens2024detecting,
  title={Detecting ongoing gas-to-solid nucleation on the ultra-hot planet WASP-76 b},
  author={Baeyens, Robin and Barat, Saugata and Decin, Leen and Desert, Jean-Michel and Gobrecht, David and Helling, Christiane and Lecoq-Molinos, Helena and Savel, Arjun Baliga and Shivkumar, Hinna and Sikora, James and others},
  journal={JWST Proposal. Cycle 3},
  pages={6045},
  year={2024}
}

@INPROCEEDINGS{1996ASPC...96...69G,
       author = {{Goeres}, A.},
        title = "{Chemistry and thermodynamics of the nucleation in R CrB star shells}",
    booktitle = {Hydrogen Deficient Stars},
         year = 1996,
       editor = {{Jeffery}, C.~S. and {Heber}, U.},
       series = {Astronomical Society of the Pacific Conference Series},
       volume = {96},
        month = jan,
        pages = {69},
       adsurl = {https://ui.adsabs.harvard.edu/abs/1996ASPC...96...69G}
}

@ARTICLE{2019AREPS..47..583H,
       author = {{Helling}, Christiane},
        title = "{Exoplanet Clouds}",
      journal = {Annual Review of Earth and Planetary Sciences},
         year = 2019,
        month = may,
       volume = {47},
        pages = {583-606},
          doi = {10.1146/annurev-earth-053018-060401},
archivePrefix = {arXiv},
       eprint = {1812.03793},
 primaryClass = {astro-ph.EP},
       adsurl = {https://ui.adsabs.harvard.edu/abs/2019AREPS..47..583H}
}

@ARTICLE{2020A&A...636A..71H,
       author = {{Herbort}, O. and {Woitke}, P. and {Helling}, Ch. and {Zerkle}, A.},
        title = "{The atmospheres of rocky exoplanets. I. Outgassing of common rock and the stability of liquid water}",
      journal = {\aap},
         year = 2020,
        month = apr,
       volume = {636},
          eid = {A71},
        pages = {A71},
          doi = {10.1051/0004-6361/201936614},
archivePrefix = {arXiv},
       eprint = {2003.03628},
 primaryClass = {astro-ph.EP},
       adsurl = {https://ui.adsabs.harvard.edu/abs/2020A&A...636A..71H}
}

@ARTICLE{2013A&A...555A.119G,
       author = {{Gail}, H. -P. and {Wetzel}, S. and {Pucci}, A. and {Tamanai}, A.},
        title = "{Seed particle formation for silicate dust condensation by SiO nucleation}",
      journal = {\aap},
         year = 2013,
        month = jul,
       volume = {555},
          eid = {A119},
        pages = {A119},
          doi = {10.1051/0004-6361/201321807},
archivePrefix = {arXiv},
       eprint = {1305.2879},
 primaryClass = {astro-ph.SR},
       adsurl = {https://ui.adsabs.harvard.edu/abs/2013A&A...555A.119G}
}

@ARTICLE{1997A&A...320..553K,
       author = {{Koehler}, T.~M. and {Gail}, H. -P. and {Sedlmayr}, E.},
        title = "{MgO dust nucleation in M-Stars: calculation of cluster properties and nucleation rates.}",
      journal = {\aap},
         year = 1997,
        month = apr,
       volume = {320},
        pages = {553-567},
       adsurl = {https://ui.adsabs.harvard.edu/abs/1997A&A...320..553K}
}

@article{wang2018magic,
  title={Magic number colloidal clusters as minimum free energy structures},
  author={Wang, Junwei and Mbah, Chrameh Fru and Przybilla, Thomas and Apeleo Zubiri, Benjamin and Spiecker, Erdmann and Engel, Michael and Vogel, Nicolas},
  journal={Nature communications},
  volume={9},
  number={1},
  pages={5259},
  year={2018},
  publisher={Nature Publishing Group UK London}
}

@article{harbola1992magic,
  title={Magic numbers for metallic clusters and the principle of maximum hardness.},
  author={Harbola, Manoj K},
  journal={Proceedings of the National Academy of Sciences},
  volume={89},
  number={3},
  pages={1036--1039},
  year={1992}
}

@article{helling2013modelling,
  title={Modelling the formation of atmospheric dust in brown dwarfs and planetary atmospheres},
  author={Helling, Christiane and Fomins, Aleksejs},
  journal={Philosophical Transactions of the Royal Society A: Mathematical, Physical and Engineering Sciences},
  volume={371},
  number={1994},
  pages={20110581},
  year={2013},
  publisher={The Royal Society Publishing}
}

@article{mahapatra2017cloud,
  title={Cloud formation in metal-rich atmospheres of hot super-Earths like 55 Cnc e and CoRoT7b},
  author={Mahapatra, G and Helling, Ch and Miguel, Y},
  journal={Monthly Notices of the Royal Astronomical Society},
  volume={472},
  number={1},
  pages={447--464},
  year={2017},
  publisher={Oxford University Press}
}

@article{evans2016detection,
  title={Detection of H2O and evidence for TiO/VO in an ultra-hot exoplanet atmosphere},
  author={Evans, Thomas M and Sing, David K and Wakeford, Hannah R and Nikolov, Nikolay and Ballester, Gilda E and Drummond, Benjamin and Kataria, Tiffany and Gibson, Neale P and Amundsen, David S and Spake, Jessica},
  journal={The Astrophysical Journal Letters},
  volume={822},
  number={1},
  pages={L4},
  year={2016},
  publisher={IOP Publishing}
}

@article{parmentier2018thermal,
  title={From thermal dissociation to condensation in the atmospheres of ultra hot Jupiters: WASP-121b in context},
  author={Parmentier, Vivien and Line, Mike R and Bean, Jacob L and Mansfield, Megan and Kreidberg, Laura and Lupu, Roxana and Visscher, Channon and D{\'e}sert, Jean-Michel and Fortney, Jonathan J and Deleuil, Magalie and others},
  journal={Astronomy \& Astrophysics},
  volume={617},
  pages={A110},
  year={2018},
  publisher={EDP Sciences}
}

@article{cortes2020tramos,
  title={TraMoS-V. Updated ephemeris and multi-epoch monitoring of the hot Jupiters WASP-18Ab, WASP-19b, and WASP-77Ab},
  author={Cort{\'e}s-Zuleta, P{\'\i}a and Rojo, Patricio and Wang, Songhu and Hinse, Tobias C and Hoyer, Sergio and Sanhueza, Bastian and Correa-Amaro, Patricio and Albornoz, Julio},
  journal={Astronomy \& Astrophysics},
  volume={636},
  pages={A98},
  year={2020},
  publisher={EDP Sciences}
}

@article{coulombe2023broadband,
  title={A broadband thermal emission spectrum of the ultra-hot Jupiter WASP-18b},
  author={Coulombe, Louis-Philippe and Benneke, Bj{\"o}rn and Challener, Ryan and Piette, Anjali AA and Wiser, Lindsey S and Mansfield, Megan and MacDonald, Ryan J and Beltz, Hayley and Feinstein, Adina D and Radica, Michael and others},
  journal={Nature},
  volume={620},
  number={7973},
  pages={292--298},
  year={2023},
  publisher={Nature Publishing Group UK London}
}

@article{espinoza2024inhomogeneous,
  title={Inhomogeneous terminators on the exoplanet WASP-39 b},
  author={Espinoza, N{\'e}stor and Steinrueck, Maria E and Kirk, James and MacDonald, Ryan J and Savel, Arjun B and Arnold, Kenneth and Kempton, Eliza M-R and Murphy, Matthew M and Carone, Ludmila and Zamyatina, Maria and others},
  journal={Nature},
  volume={632},
  number={8027},
  pages={1017--1020},
  year={2024},
  publisher={Nature Publishing Group UK London}
}

@article{bangera2025kinetic,
  title={Kinetic and Photochemical Disequilibrium in the Potentially Carbon-rich Atmosphere of the Warm Jupiter WASP-69 b},
  author={Bangera, Nidhi and Helling, Christiane and Guilluy, Gloria and Cubillos, Patricio and Fossati, Luca and Giacobbe, Paolo and Rimmer, Paul and Kitzmann, Daniel},
  journal={The Astrophysical Journal},
  volume={980},
  number={1},
  pages={147},
  year={2025},
  publisher={IOP Publishing}
}

@article{helling2023exoplanet,
  title={Exoplanet weather and climate regimes with clouds and thermal ionospheres-A model grid study in support of large-scale observational campaigns},
  author={Helling, Christiane and Samra, Dominic and Lewis, David and Calder, Robb and Hirst, Georgina and Woitke, Peter and Baeyens, Robin and Carone, Ludmila and Herbort, Oliver and Chubb, Katy L},
  journal={Astronomy \& Astrophysics},
  volume={671},
  pages={A122},
  year={2023},
  publisher={EDP Sciences}
}

@article{kataria2016atmospheric,
  title={The atmospheric circulation of a nine-hot-Jupiter sample: probing circulation and chemistry over a wide phase space},
  author={Kataria, Tiffany and Sing, David K and Lewis, Nikole K and Visscher, Channon and Showman, Adam P and Fortney, Jonathan J and Marley, Mark S},
  journal={The Astrophysical Journal},
  volume={821},
  number={1},
  pages={9},
  year={2016},
  publisher={IOP Publishing}
}

@article{agundez2020chemical,
  title={Chemical equilibrium in AGB atmospheres: successes, failures, and prospects for small molecules, clusters, and condensates},
  author={Ag{\'u}ndez, Marcelino and Mart{\'\i}nez, Jos{\'e} I and de Andres, PL and Cernicharo, Jos{\'e} and Mart{\'\i}n-Gago, Jos{\'e} A},
  journal={Astronomy \& Astrophysics},
  volume={637},
  pages={A59},
  year={2020},
  publisher={EDP Sciences}
}

@article{doye2002entropic,
  title={Entropic effects on the structure of Lennard-Jones clusters},
  author={Doye, Jonathan PK and Calvo, Florent},
  journal={The Journal of chemical physics},
  volume={116},
  number={19},
  pages={8307--8317},
  year={2002},
  publisher={American Institute of Physics}
}

@article{helling2021cloud,
  title={Cloud property trends in hot and ultra-hot giant gas planets (WASP-43b, WASP-103b, WASP-121b, HAT-P-7b, and WASP-18b)},
  author={Helling, Ch and Lewis, D and Samra, D and Carone, L and Graham, V and Herbort, O and Chubb, KL and Min, M and Waters, R and Parmentier, V and others},
  journal={Astronomy \& Astrophysics},
  volume={649},
  pages={A44},
  year={2021},
  publisher={EDP Sciences}
}

@article{madhusudhan2009temperature,
  title={A temperature and abundance retrieval method for exoplanet atmospheres},
  author={Madhusudhan, N and Seager, Sara},
  journal={The Astrophysical Journal},
  volume={707},
  number={1},
  pages={24},
  year={2009},
  publisher={IOP Publishing}
}

@article{munoz2007physical,
  title={Physical and chemical aeronomy of HD 209458b},
  author={Mu{\~n}oz, A Garc{\'\i}a},
  journal={Planetary and Space Science},
  volume={55},
  number={10},
  pages={1426--1455},
  year={2007},
  publisher={Elsevier}
}

@article{yelle2004aeronomy,
  title={Aeronomy of extra-solar giant planets at small orbital distances},
  author={Yelle, Roger V},
  journal={Icarus},
  volume={170},
  number={1},
  pages={167--179},
  year={2004},
  publisher={Elsevier}
}

@article{plane2012cosmic,
  title={Cosmic dust in the earth's atmosphere},
  author={Plane, John MC},
  journal={Chemical Society Reviews},
  volume={41},
  number={19},
  pages={6507--6518},
  year={2012},
  publisher={Royal Society of Chemistry}
}

@article{carone2023wasp,
  title={WASP-39b: exo-Saturn with patchy cloud composition, moderate metallicity, and underdepleted S/O},
  author={Carone, Ludmila and Lewis, David A and Samra, Dominic and Schneider, Aaron D and Helling, Christiane},
  journal={arXiv preprint arXiv:2301.08492},
  year={2023}
}

@article{carone2018stratosphere,
  title={Stratosphere circulation on tidally locked ExoEarths},
  author={Carone, Ludmila and Keppens, Rony and Decin, Leen and Henning, Th},
  journal={Monthly Notices of the Royal Astronomical Society},
  volume={473},
  number={4},
  pages={4672--4685},
  year={2018},
  publisher={Oxford University Press}
}

@article{lee2023modelling,
  title={Modelling dynamically driven global cloud formation microphysics in the HAT-P-1b atmosphere},
  author={Lee, Elspeth KH},
  journal={Monthly Notices of the Royal Astronomical Society},
  volume={524},
  number={2},
  pages={2918--2933},
  year={2023},
  publisher={Oxford University Press}
}

@ARTICLE{DelineW18b,
       author = {{Deline}, A. and {Cubillos}, P.~E. and {Carone}, L. and {Demory}, B.-O. and {Lendl}, M. and {Benz}, W. and {Brandeker}, A. and {G{\"u}nther}, M.~N. and {Heitzmann}, A. and {Barros}, S.~C.~C. and {Kreidberg}, L. and {Bruno}, G. and {Kitzmann}, D. and {Bonfanti}, A. and {Farnir}, M. and {Persson}, C.~M. and {Sousa}, S.~G. and {Wilson}, T.~G. and {Ehrenreich}, D. and {Singh}, V. and {Iro}, N. and {Alibert}, Y. and {Alonso}, R. and {B{\'a}rczy}, T. and {Barrado Navascues}, D. and {Baumjohann}, W. and {Bergomi}, M. and {Billot}, N. and {Borsato}, L. and {Broeg}, C. and {Busch}, M.-D. and {Collier Cameron}, A. and {Correia}, A.~C.~M. and {Csizmadia}, Sz. and {Davies}, M.~B. and {Deleuil}, M. and {Delrez}, L. and {Demangeon}, O.~D.~S. and {Derekas}, A. and {Edwards}, B. and {Erikson}, A. and {Fortier}, A. and {Fossati}, L. and {Fridlund}, M. and {Gandolfi}, D. and {Gazeas}, K. and {Gillon}, M. and {G{\"u}del}, M. and {Hasiba}, J. and {Helling}, Ch. and {Isaak}, K.~G. and {Kiss}, L.~L. and {Korth}, J. and {Lam}, K.~W.~F. and {Laskar}, J. and {Lecavelier des {\'E}tangs}, A. and {Magrin}, D. and {Maxted}, P.~F.~L. and {Mer{\'\i}n}, B. and {Mordasini}, C. and {Nascimbeni}, V. and {Olofsson}, G. and {Ottensamer}, R. and {Pagano}, I. and {Pall{\'e}}, E. and {Peter}, G. and {Piazza}, D. and {Piotto}, G. and {Pollacco}, D. and {Queloz}, D. and {Ragazzoni}, R. and {Rando}, N. and {Ratti}, F. and {Rauer}, H. and {Ribas}, I. and {Santos}, N.~C. and {Scandariato}, G. and {S{\'e}gransan}, D. and {Simon}, A.~E. and {Smith}, A.~M.~S. and {Stalport}, M. and {Sulis}, S. and {Szab{\'o}}, Gy. M. and {Udry}, S. and {Van Grootel}, V. and {Venturini}, J. and {Villaver}, E. and {Walton}, N.~A. and {Westerdorff}, K.},
        title = "{Dark skies of the slightly eccentric WASP-18 b from its optical-to-infrared dayside emission}",
      journal = {\aap},
         year = 2025,
        month = jul,
       volume = {699},
          eid = {A150},
        pages = {A150},
          doi = {10.1051/0004-6361/202450939},
archivePrefix = {arXiv},
       eprint = {2505.01544},
 primaryClass = {astro-ph.EP},
       adsurl = {https://ui.adsabs.harvard.edu/abs/2025A&A...699A.150D}
}

@ARTICLE{SteinrueckW39b,
       author = {{Steinrueck}, Maria E. and {Savel}, Arjun B. and {Christie}, Duncan A. and {Carone}, Ludmila and {Tsai}, Shang-Min and {Ak{\i}n}, Can and {Kennedy}, Thomas D. and {Kiefer}, Sven and {Lewis}, David A. and {Rauscher}, Emily and {Samra}, Dominic and {Zamyatina}, Maria and {Arnold}, Kenneth and {Baeyens}, Robin and {Gkouvelis}, Leonardos and {Haegele}, David and {Helling}, Christiane and {Mayne}, Nathan J. and {Powell}, Diana and {Roman}, Michael T. and {Beltz}, Hayley and {Espinoza}, N{\'e}stor and {Heng}, Kevin and {Iro}, Nicolas and {Kempton}, Eliza M. -R. and {Kreidberg}, Laura and {Kirk}, James and {Murphy}, Matthew M. and {Rackham}, Benjamin V. and {Tan}, Xianyu},
        title = "{Limb Asymmetries on WASP-39b: A Multi-GCM Comparison of Chemistry, Clouds, and Hazes}",
      journal = {arXiv e-prints},
         year = 2025,
        month = sep,
          eid = {arXiv:2509.21588},
        pages = {arXiv:2509.21588},
          doi = {10.48550/arXiv.2509.21588},
archivePrefix = {arXiv},
       eprint = {2509.21588},
 primaryClass = {astro-ph.EP},
       adsurl = {https://ui.adsabs.harvard.edu/abs/2025arXiv250921588S}
}

@ARTICLE{PlaschzugW121b,
       author = {{Plaschzug}, Alexander and {Reza}, Amit and {Carone}, Ludmila and {Gernjak}, Sebastian and {Helling}, Christiane},
        title = "{Accelerating exoplanet climate modelling: A machine learning approach to complement 3D GCM grid simulations}",
      journal = {arXiv e-prints},
         year = 2025,
        month = aug,
          eid = {arXiv:2508.10827},
        pages = {arXiv:2508.10827},
          doi = {10.48550/arXiv.2508.10827},
archivePrefix = {arXiv},
       eprint = {2508.10827},
 primaryClass = {astro-ph.EP},
       adsurl = {https://ui.adsabs.harvard.edu/abs/2025arXiv250810827P}
}

@article{Carone2020,
	adsurl = {https://ui.adsabs.harvard.edu/abs/2020MNRAS.496.3582C},
	archiveprefix = {arXiv},
	author = {{Carone}, Ludmila and {Baeyens}, Robin and {Molli{\`e}re}, Paul and {Barth}, Patrick and {Vazan}, Allona and {Decin}, Leen and {Sarkis}, Paula and {Venot}, Olivia and {Henning}, Thomas},
	doi = {10.1093/mnras/staa1733},
	eprint = {1904.13334},
	journal = {\mnras},
	month = aug,
	number = {3},
	pages = {3582-3614},
	primaryclass = {astro-ph.EP},
	title = {{Equatorial retrograde flow in WASP-43b elicited by deep wind jets?}},
	volume = {496},
	year = 2020}

@ARTICLE{Schneider2022a,
       author = {{Schneider}, Aaron David and {Carone}, Ludmila and {Decin}, Leen and {J{\o}rgensen}, Uffe Gr{\r{a}}e and {Molli{\`e}re}, Paul and {Baeyens}, Robin and {Kiefer}, Sven and {Helling}, Christiane},
        title = "{Exploring the deep atmospheres of HD 209458b and WASP-43b using a non-gray general circulation model}",
      journal = {\aap},
         year = 2022,
        month = aug,
       volume = {664},
          eid = {A56},
        pages = {A56},
          doi = {10.1051/0004-6361/202142728},
archivePrefix = {arXiv},
       eprint = {2202.09183},
 primaryClass = {astro-ph.EP},
       adsurl = {https://ui.adsabs.harvard.edu/abs/2022A&A...664A..56S}
}

@article{brogi2023roasting,
  title={The roasting marshmallows program with IGRINS on Gemini South I: composition and climate of the ultrahot Jupiter WASP-18 b},
  author={Brogi, Matteo and Emeka-Okafor, Vanessa and Line, Michael R and Gandhi, Siddharth and Pino, Lorenzo and Kempton, Eliza M-R and Rauscher, Emily and Parmentier, Vivien and Bean, Jacob L and Mace, Gregory N and others},
  journal={The Astronomical Journal},
  volume={165},
  number={3},
  pages={91},
  year={2023},
  publisher={IOP Publishing}
}

@article{faherty2025silicate,
  title={Silicate precursor silane detected in cold low-metallicity brown dwarf},
  author={Faherty, Jacqueline K and Meisner, Aaron M and Burningham, Ben and Visscher, Channon and Line, Michael and Su{\'a}rez, Genaro and Gagn{\'e}, Jonathan and Alejandro Merchan, Sherelyn and Rothermich, Austin James and Burgasser, Adam J and others},
  journal={Nature},
  volume={645},
  number={8079},
  pages={62--66},
  year={2025},
  publisher={Nature Publishing Group UK London}
}

@article{Plane2015,
  author    = {Plane, John M. C. and Feng, Wuhu and Dawkins, Erin C. M.},
  title     = {The Mesosphere and Metals: Chemistry and Changes},
  journal   = {Chemical Reviews},
  year      = {2015},
  volume    = {115},
  number    = {10},
  pages     = {4497--4541},
  doi       = {10.1021/cr500501m},
  url       = {https://doi.org/10.1021/cr500501m},
  publisher = {American Chemical Society}
}

@article{Plane2003,
  author    = {Plane, John M. C.},
  title     = {Atmospheric Chemistry of Meteoric Metals},
  journal   = {Chemical Reviews},
  year      = {2003},
  volume    = {103},
  number    = {12},
  pages     = {4963--4984},
  doi       = {10.1021/cr0205309},
  url       = {https://doi.org/10.1021/cr0205309},
  publisher = {American Chemical Society}
}

@article{Plane2018,
  author    = {Plane, J. M. C. and Carrillo-Sánchez, J. D. and Mangan, T. P. and Crismani, M. and Schneider, N. M. and Määttänen, A.},
  title     = {Meteoric Metal Chemistry in the Martian Atmosphere},
  journal   = {Journal of Geophysical Research: Planets},
  year      = {2018},
  volume    = {123},
  number    = {3},
  pages     = {695--707},
  doi       = {10.1002/2017JE005510},
  url       = {https://doi.org/10.1002/2017JE005510}
}

@ARTICLE{2021A&A...649A..44H,
       author = {{Helling}, Ch. and {Lewis}, D. and {Samra}, D. and {Carone}, L. and {Graham}, V. and {Herbort}, O. and {Chubb}, K.~L. and {Min}, M. and {Waters}, R. and {Parmentier}, V. and {Mayne}, N.},
        title = "{Cloud property trends in hot and ultra-hot giant gas planets (WASP-43b, WASP-103b, WASP-121b, HAT-P-7b, and WASP-18b)}",
      journal = {\aap},
         year = 2021,
        month = may,
       volume = {649},
          eid = {A44},
        pages = {A44},
          doi = {10.1051/0004-6361/202039911},
archivePrefix = {arXiv},
       eprint = {2102.11688},
 primaryClass = {astro-ph.EP},
       adsurl = {https://ui.adsabs.harvard.edu/abs/2021A&A...649A..44H}
}

@ARTICLE{2022PSJ.....3...82B,
       author = {{Blecic}, Jasmina and {Harrington}, Joseph and {Cubillos}, Patricio E. and {Oliver Bowman}, M. and {Rojo}, Patricio M. and {Stemm}, Madison and {Challener}, Ryan C. and {Himes}, Michael D. and {Foster}, Austin J. and {Dobbs-Dixon}, Ian and {Foster}, Andrew S.~D. and {Lust}, Nathaniel B. and {Blumenthal}, Sarah D. and {Bruce}, Dylan and {Loredo}, Thomas J.},
        title = "{An Open-source Bayesian Atmospheric Radiative Transfer (BART) Code. III. Initialization, Atmospheric Profile Generator, Post-processing Routines}",
      journal = {\psj},
         year = 2022,
        month = apr,
       volume = {3},
       number = {4},
          eid = {82},
        pages = {82},
          doi = {10.3847/PSJ/ac3515},
archivePrefix = {arXiv},
       eprint = {2104.12525},
 primaryClass = {astro-ph.EP},
       adsurl = {https://ui.adsabs.harvard.edu/abs/2022PSJ.....3...82B}
}

@ARTICLE{2020A&A...642A..28M,
       author = {{Min}, Michiel and {Ormel}, Chris W. and {Chubb}, Katy and {Helling}, Christiane and {Kawashima}, Yui},
        title = "{The ARCiS framework for exoplanet atmospheres. Modelling philosophy and retrieval}",
      journal = {\aap},
         year = 2020,
        month = oct,
       volume = {642},
          eid = {A28},
        pages = {A28},
          doi = {10.1051/0004-6361/201937377},
archivePrefix = {arXiv},
       eprint = {2006.12821},
 primaryClass = {astro-ph.EP},
       adsurl = {https://ui.adsabs.harvard.edu/abs/2020A&A...642A..28M}
}

@article{lee2015dust,
  title={Dust in brown dwarfs and extra-solar planets-IV. Assessing TiO2 and SiO nucleation for cloud formation modelling},
  author={Lee, E and Helling, Ch and Giles, H and Bromley, ST},
  journal={Astronomy \& Astrophysics},
  volume={575},
  pages={A11},
  year={2015},
  publisher={EDP Sciences}
}

@ARTICLE{2010ApJ...710.1395D,
       author = {{Dobbs-Dixon}, Ian and {Cumming}, Andrew and {Lin}, D.~N.~C.},
        title = "{Radiative Hydrodynamic Simulations of HD209458b: Temporal Variability}",
      journal = {\apj},
         year = 2010,
        month = feb,
       volume = {710},
       number = {2},
        pages = {1395-1407},
          doi = {10.1088/0004-637X/710/2/1395},
archivePrefix = {arXiv},
       eprint = {1001.0982},
 primaryClass = {astro-ph.EP},
       adsurl = {https://ui.adsabs.harvard.edu/abs/2010ApJ...710.1395D}
}

@ARTICLE{2020A&A...637A..59A,
       author = {{Ag{\'u}ndez}, M. and {Mart{\'\i}nez}, J.~I. and {de Andres}, P.~L. and {Cernicharo}, J. and {Mart{\'\i}n-Gago}, J.~A.},
        title = "{Chemical equilibrium in AGB atmospheres: successes, failures, and prospects for small molecules, clusters, and condensates}",
      journal = {\aap},
         year = 2020,
        month = may,
       volume = {637},
          eid = {A59},
        pages = {A59},
          doi = {10.1051/0004-6361/202037496},
archivePrefix = {arXiv},
       eprint = {2004.00519},
 primaryClass = {astro-ph.SR},
       adsurl = {https://ui.adsabs.harvard.edu/abs/2020A&A...637A..59A}
}

@ARTICLE{2025A&A...699A.148A,
       author = {{Ahmad}, A. and {Freytag}, B. and {H{\"o}fner}, S.},
        title = "{Multi-mode pulsations in AGB stars: Insights from 3D RHD CO5BOLD simulations}",
      journal = {\aap},
         year = 2025,
        month = jul,
       volume = {699},
          eid = {A148},
        pages = {A148},
          doi = {10.1051/0004-6361/202554160},
archivePrefix = {arXiv},
       eprint = {2502.11978},
 primaryClass = {astro-ph.SR},
       adsurl = {https://ui.adsabs.harvard.edu/abs/2025A&A...699A.148A}
}

@ARTICLE{2023arXiv230108492C,
       author = {{Carone}, Ludmila and {Lewis}, David A. and {Samra}, Dominic and {Schneider}, Aaron D. and {Helling}, Christiane},
        title = "{WASP-39b: exo-Saturn with patchy cloud composition, moderate metallicity, and underdepleted S/O}",
      journal = {arXiv e-prints},
         year = 2023,
        month = jan,
          eid = {arXiv:2301.08492},
        pages = {arXiv:2301.08492},
          doi = {10.48550/arXiv.2301.08492},
archivePrefix = {arXiv},
       eprint = {2301.08492},
 primaryClass = {astro-ph.EP},
       adsurl = {https://ui.adsabs.harvard.edu/abs/2023arXiv230108492C}
}

@ARTICLE{2016A&A...594A..48L,
       author = {{Lee}, E. and {Dobbs-Dixon}, I. and {Helling}, Ch. and {Bognar}, K. and {Woitke}, P.},
        title = "{Dynamic mineral clouds on HD 189733b. I. 3D RHD with kinetic, non-equilibrium cloud formation}",
      journal = {\aap},
         year = 2016,
        month = oct,
       volume = {594},
          eid = {A48},
        pages = {A48},
          doi = {10.1051/0004-6361/201628606},
archivePrefix = {arXiv},
       eprint = {1603.09098},
 primaryClass = {astro-ph.EP},
       adsurl = {https://ui.adsabs.harvard.edu/abs/2016A&A...594A..48L}
}

@ARTICLE{2018MNRAS.481..194L,
       author = {{Lines}, S. and {Manners}, J. and {Mayne}, N.~J. and {Goyal}, J. and {Carter}, A.~L. and {Boutle}, I.~A. and {Lee}, Elspeth and {Helling}, Ch and {Drummond}, B. and {Acreman}, D.~M. and {Sing}, D.~K.},
        title = "{Exonephology: transmission spectra from a 3D simulated cloudy atmosphere of HD 209458b}",
      journal = {\mnras},
         year = 2018,
        month = nov,
       volume = {481},
       number = {1},
        pages = {194-205},
          doi = {10.1093/mnras/sty2275},
archivePrefix = {arXiv},
       eprint = {1808.05887},
 primaryClass = {astro-ph.EP},
       adsurl = {https://ui.adsabs.harvard.edu/abs/2018MNRAS.481..194L}
}

@ARTICLE{2023A&A...669A.155F,
       author = {{Freytag}, Bernd and {H{\"o}fner}, Susanne},
        title = "{Global 3D radiation-hydrodynamical models of AGB stars with dust-driven winds}",
      journal = {\aap},
         year = 2023,
        month = jan,
       volume = {669},
          eid = {A155},
        pages = {A155},
          doi = {10.1051/0004-6361/202244992},
archivePrefix = {arXiv},
       eprint = {2301.11836},
 primaryClass = {astro-ph.SR},
       adsurl = {https://ui.adsabs.harvard.edu/abs/2023A&A...669A.155F}
}

@ARTICLE{2025arXiv251101548C,
       author = {{Carone}, Ludmila and {Helling}, Christiane and {Gernjak}, Sebastian and {Leitner}, Hanna and {Janz}, Tamara},
        title = "{Exoplanet climate characterization with transit asymmetries -- A comprehensive population study from the optical to the infrared}",
      journal = {arXiv e-prints},
         year = 2025,
        month = nov,
          eid = {arXiv:2511.01548},
        pages = {arXiv:2511.01548},
          doi = {10.48550/arXiv.2511.01548},
archivePrefix = {arXiv},
       eprint = {2511.01548},
 primaryClass = {astro-ph.EP},
       adsurl = {https://ui.adsabs.harvard.edu/abs/2025arXiv251101548C}
}

@article{decin2017study,
  title={Study of the aluminium content in AGB winds using ALMA-Indications for the presence of gas-phase (Al2O3) n clusters},
  author={Decin, Leen and Richards, AMS and Waters, LBFM and Danilovich, Taissa and Gobrecht, David and Khouri, Theo and Homan, Ward and Bakker, JM and Van de Sande, Marie and Nuth, JA and others},
  journal={Astronomy \& Astrophysics},
  volume={608},
  pages={A55},
  year={2017},
  publisher={EDP Sciences}
}

@article{garai2025kelt,
  title={The KELT-7b atmospheric thermal-inversion conundrum revisited with CHEOPS, TESS, and additional data},
  author={Garai, Zolt{\'a}n and Krenn, Andreas and Cubillos, Patricio E and Bruno, Giovanni and Smith, AMS and Wilson, Thomas G and Brandeker, Alexis and G{\"u}nther, Maximilian N and Heitzmann, Alexis and Carone, Ludmila and others},
  journal={Astronomy \& Astrophysics},
  volume={700},
  pages={A5},
  year={2025},
  publisher={EDP Sciences}
}

@ARTICLE{2019ApJ...883....4S,
       author = {{Showman}, Adam P. and {Tan}, Xianyu and {Zhang}, Xi},
        title = "{Atmospheric Circulation of Brown Dwarfs and Jupiter- and Saturn-like Planets: Zonal Jets, Long-term Variability, and QBO-type Oscillations}",
      journal = {\apj},
         year = 2019,
        month = sep,
       volume = {883},
       number = {1},
          eid = {4},
        pages = {4},
          doi = {10.3847/1538-4357/ab384a},
archivePrefix = {arXiv},
       eprint = {1807.08433},
 primaryClass = {astro-ph.EP},
       adsurl = {https://ui.adsabs.harvard.edu/abs/2019ApJ...883....4S}
}

@ARTICLE{2021MNRAS.502.2198T,
       author = {{Tan}, Xianyu and {Showman}, Adam P.},
        title = "{Atmospheric circulation of brown dwarfs and directly imaged exoplanets driven by cloud radiative feedback: global and equatorial dynamics}",
      journal = {\mnras},
         year = 2021,
        month = apr,
       volume = {502},
       number = {2},
        pages = {2198-2219},
          doi = {10.1093/mnras/stab097},
archivePrefix = {arXiv},
       eprint = {2101.04417},
 primaryClass = {astro-ph.EP},
       adsurl = {https://ui.adsabs.harvard.edu/abs/2021MNRAS.502.2198T}
}

@ARTICLE{2018ApJ...869..107M,
       author = {{Mendon{\c{c}}a}, Jo{\~a}o M. and {Tsai}, Shang-min and {Malik}, Matej and {Grimm}, Simon L. and {Heng}, Kevin},
        title = "{Three-dimensional Circulation Driving Chemical Disequilibrium in WASP-43b}",
      journal = {\apj},
         year = 2018,
        month = dec,
       volume = {869},
       number = {2},
          eid = {107},
        pages = {107},
          doi = {10.3847/1538-4357/aaed23},
archivePrefix = {arXiv},
       eprint = {1808.00501},
 primaryClass = {astro-ph.EP},
       adsurl = {https://ui.adsabs.harvard.edu/abs/2018ApJ...869..107M}
}

@ARTICLE{2014A&A...561A...1M,
       author = {{Mayne}, Nathan J. and {Baraffe}, Isabelle and {Acreman}, David M. and {Smith}, Chris and {Browning}, Matthew K. and {Sk{\r{a}}lid Amundsen}, David and {Wood}, Nigel and {Thuburn}, John and {Jackson}, David R.},
        title = "{The unified model, a fully-compressible, non-hydrostatic, deep atmosphere global circulation model, applied to hot Jupiters. ENDGame for a HD 209458b test case}",
      journal = {\aap},
         year = 2014,
        month = jan,
       volume = {561},
          eid = {A1},
        pages = {A1},
          doi = {10.1051/0004-6361/201322174},
       adsurl = {https://ui.adsabs.harvard.edu/abs/2014A&A...561A...1M}
}

@article{guilluy2022gaps,
  title={The GAPS Programme at TNG-XXXVIII. Five molecules in the atmosphere of the warm giant planet WASP-69b detected at high spectral resolution},
  author={Guilluy, Gloria and Giacobbe, Paolo and Carleo, Ilaria and Cubillos, PE and Sozzetti, Alessandro and Bonomo, ALDO STEFANO and Brogi, M and Gandhi, S and Fossati, L and Nascimbeni, VALERIO and others},
  journal={Astronomy \& Astrophysics},
  volume={665},
  pages={A104},
  year={2022},
  publisher={EDP Sciences}
}

@ARTICLE{2013RSPTA.37110581H,
       author = {{Helling}, C. and {Fomins}, A.},
        title = "{Modelling the formation of atmospheric dust in brown dwarfs and planetary atmospheres}",
      journal = {Philosophical Transactions of the Royal Society of London Series A},
         year = 2013,
        month = jun,
       volume = {371},
       number = {1994},
        pages = {20110581-20110581},
          doi = {10.1098/rsta.2011.0581},
       adsurl = {https://ui.adsabs.harvard.edu/abs/2013RSPTA.37110581H}
}

@INPROCEEDINGS{2007ASPC..378..181P,
       author = {{Patzer}, A.~B.~C.},
        title = "{Molecular Clusters in Dust Nucleation Processes in Circumstellar Outflows of Oxygen--Rich AGB Stars}",
    booktitle = {Why Galaxies Care About AGB Stars: Their Importance as Actors and Probes},
         year = 2007,
       editor = {{Kerschbaum}, F. and {Charbonnel}, C. and {Wing}, R.~F.},
       series = {Astronomical Society of the Pacific Conference Series},
       volume = {378},
        month = nov,
        pages = {181},
       adsurl = {https://ui.adsabs.harvard.edu/abs/2007ASPC..378..181P}
}
\appendix
\section{Rotational Constants}
\begin{table}[h!]
\centering
\begin{tabular}{ccccc}
&&N&Rotational Constants (GHz)&\\
\end{tabular}\\
\begin{tabular}{cccc}
1&9.676&2.561&2.025\\
2&1.078&1.078&1.078\\
3&0.563&0.491&0.377\\
4&0.349&0.219&0.209\\
5&0.195&0.174&0.132\\
6&0.150&0.131&0.095\\
7&0.129&0.089&0.065\\
8&0.113&0.061&0.056
\end{tabular}
\caption{Rotational constants of (Al$_2$O$_3$)$_{\rm N}$ GM candidates as calculated in the QCC of \citet{gobrecht2022bottom}.}
\label{al2o3rotationalconstants}
\end{table}
\begin{table}[h!]
\centering
\begin{tabular}{ccccc}
&&N&Rotational Constants (GHz)&\\
\end{tabular}\\
\begin{tabular}{cccc}
1&16.865&&\\
2&7.527&7.111&3.656\\
3&2.560&2.550&1.278\\
4&1.659&1.659&1.658\\
5&1.306&0.844&0.726\\
6&0.749&0.746&0.578\\
7&0.720&0.409&0.409\\
8&0.766&0.275&0.275\\
9&0.367&0.311&0.310\\
10&0.605&0.147&0.147
\end{tabular}
\caption{Rotational constants of (MgO)$_{\rm N}$ GM candidates as calculated in the QCC of \citet{boulangier2019developing}.}
\label{mgorotationalconstants}
\end{table}
\newpage
\section{(VO$_2$)$\rm\rm_N$ Magic cluster}
\begin{figure}[ht!]
\centering
\begin{minipage}[b]{0.44\textwidth}
\centering
\includegraphics[width=\linewidth]{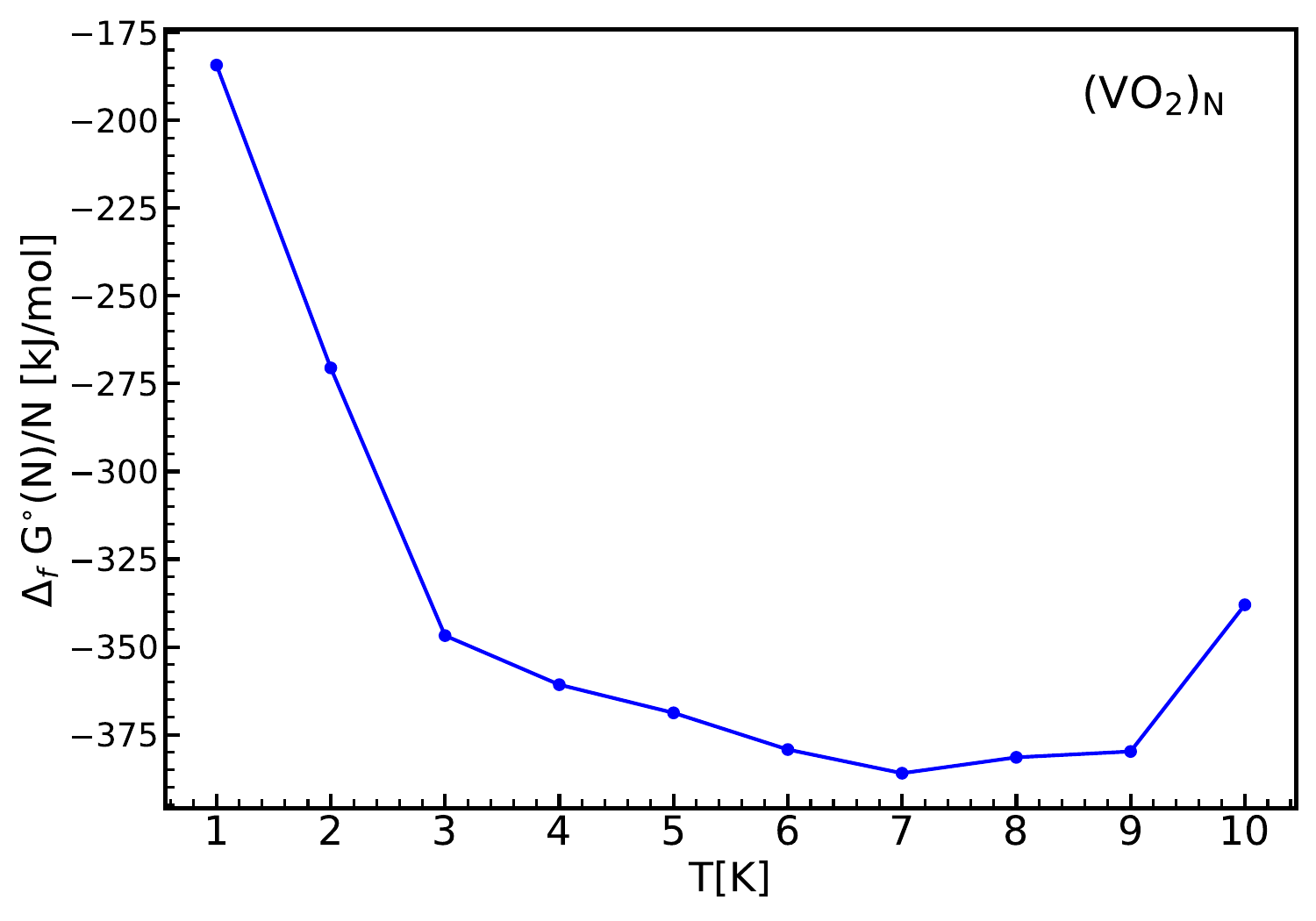}
\end{minipage}
\caption{$\Delta G^\circ_f(N)/\rm N$ of the GM candidate of (VO$_2$)$\rm\rm_N$ clusters at T${\rm gas}$ \(\sim 1000\) K.}
\label{VO2nmagiccluster}
\end{figure}

\section{ExoRad input parameters}
Table~\ref{Table:Input} lists the planetary and stellar parameters are detailed that are used for the 3D GCM \texttt{ExoRad} simulations of WASP-39 b, WASP-121 b, WASP-18 b and WASP-69 b. For the planets with high global temperatures, WASP-121 b and WASP-18 b, VO and TiO is in gasphase at the dayside and results in an upper atmosphere temperature inversion (Fig.~\ref{2Dtemp}). The high dayside temperatures lead to partial ionization of the dayside \citep{helling2021cloud,helling2019exoplanet} and necessitates the inclusion of magnetic field coupling that can be parameterized with a friction time scale and was applied for WASP-18 b to yield agreement with dayside eclipse measurements\citep{DelineW18b}. The same friction timescale was used not only for WASP-18 b but also for WASP-121 b simulations. See e.g. \citet{DelineW18b} for the impact of TiO and VO in gas phase and magnetic drag for the 3D climate of UHJs. The model set-up is described in more details in \citet{Carone2020}, the full radiative transfer method and opacity sources are described in \citet{Schneider2022a}.

\begin{table*}[h]
\caption{Planetary and stellar parameters for GCM  simulations used in this work}
\label{Exoradparameters}
\centering 
\def\arraystretch{1.3}
\label{Table:Input}
\begin{tabular}{c|c|c|c}
\textbf{WASP-39~b} & &  \textbf{WASP-121~b} & \\
\hline
Radius [R$_{\rm{Jup}}$] & 1.27 &  Radius [R$_{\rm{Jup}}$] & 1.865 \\
Mass [M$_{\rm{Jup}}$] & 0.28 &Mass [M$_{\rm{Jup}}$] & 1.184\\
Surface gravity g [m s$^{-2}$] & 4.30 &  Surface gravity g [m s$^{-2}$] & 8.44\\
Metallicity & 10 $\times$ Solar &  Metallicity & 1 $\times$ Solar\\
C/O & 0.55 &      C/O & 0.55\\
Stellar T$_{\rm eff}$ [K]& 5400 &  Stellar T$_{\rm eff}$ [K] & 6460\\
a [AU] &  0.0486 &   a [AU] & 0.02544\\
$T_{\rm global}$ [K] & 1117&  $T_{\rm global}$ [K] & 2360\\
$\tau_{\rm fric}$ [s] & N/A & $\tau_{\rm fric}$ [s] & $10^4$~s\\
$p_{\rm gas}$-range [bar] & $700 \ldots 10^{-4}$ &  $p_{\rm gas}$-range [bar] & $700 \ldots 10^{-4}$\\
\hline
\textbf{WASP-18~b} & \textbf{Value} &   \textbf{WASP-69~b} & \textbf{Value} \\
\hline
Radius [R$_{\rm{Jup}}$] & 1.191 &  Radius [R$_{\rm{Jup}}$] & 1.057\\
Mass [M$_{\rm{Jup}}$] & 10.43 & Mass [M$_{\rm{Jup}}$] & 0.26\\
Surface gravity g (m s$^{-2}$) & 190.5 &   Surface gravity g (m s$^{-2}$) & 5.77\\
Metallicity & 1 $\times$ Solar &         Metallicity & 1 $\times$ Solar\\
C/O & 0.55 &  C/O & 0.55\\
Stellar T$_{\rm eff}$ [K]& 6400 & Stellar T$_{\rm eff}$ [K]& 4715 \\
a [AU]& 0.02047 &   a [AU]&  0.04525\\
$T_{\rm global}$ [K] & 2392 &  $T_{\rm global}$ [K] & 964\\
$\tau_{\rm fric}$ [s] & $10^4$~s & $\tau_{\rm fric}$ [s] & N/A\\
$p_{\rm gas}$-range [bar] & $700 \ldots 10^{-4}$ & $p_{\rm gas}$-range [bar] & $700 \ldots 10^{-4}$\\
\end{tabular}
\tablecomments{Pressure layers  $p_{\rm gas}=10^{-4} \ldots 10^{-5}$ bar comprise ghost cells as described in \citet{Carone2020} and are not part of the computational volume. Solar element abundances from \citet{asplund2009chemical} are assumed when calculating the atmospheric metallicity.}  
\end{table*}

\section{3D GCM}
\begin{figure*}[!htbp]
\centering
\begin{minipage}{0.4\textwidth}
\centering
\includegraphics[width=0.9\linewidth]{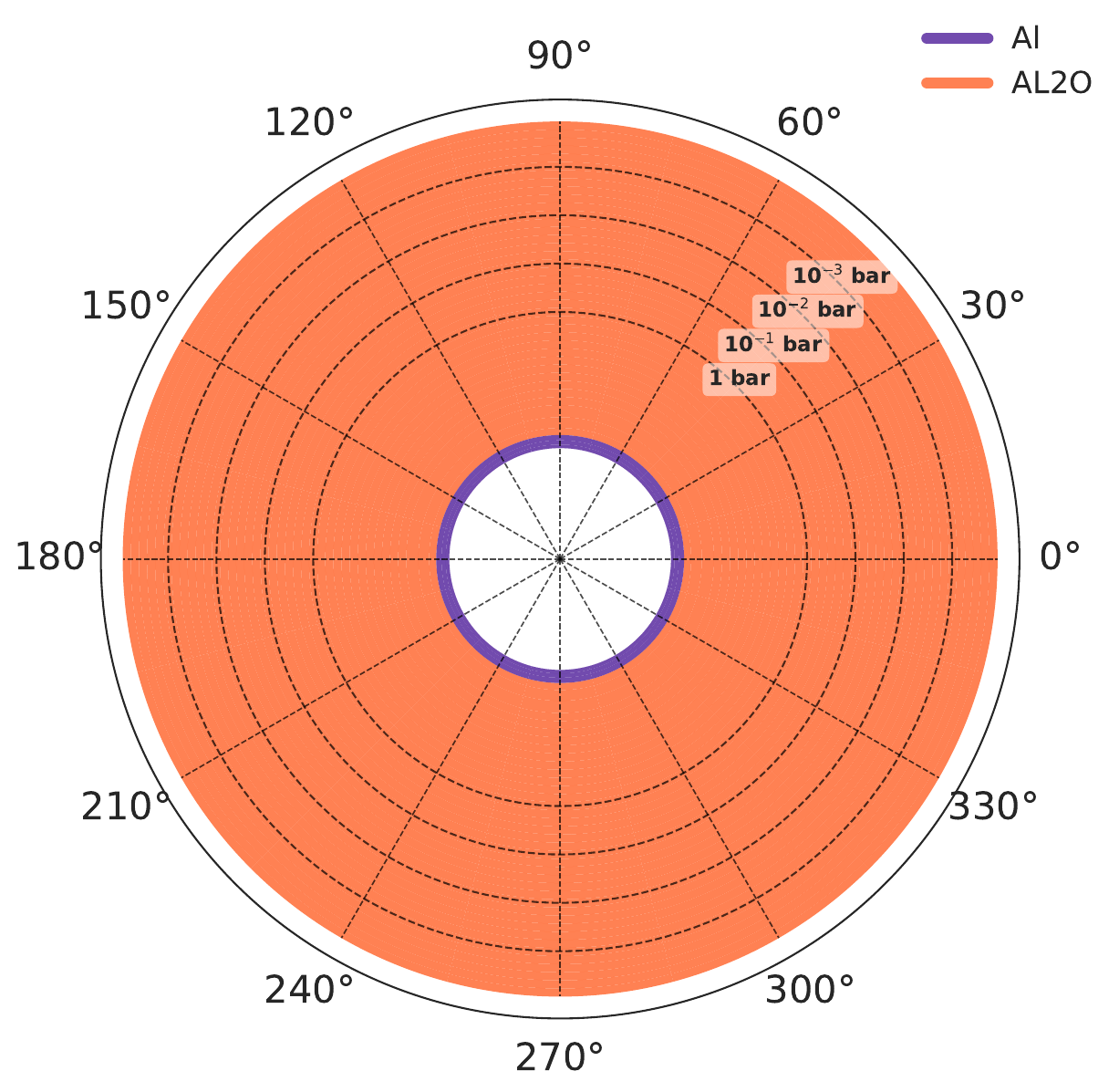}
\end{minipage}
\hspace{0.2cm}
\begin{minipage}{0.4\textwidth}
\centering
\includegraphics[width=0.9\linewidth]{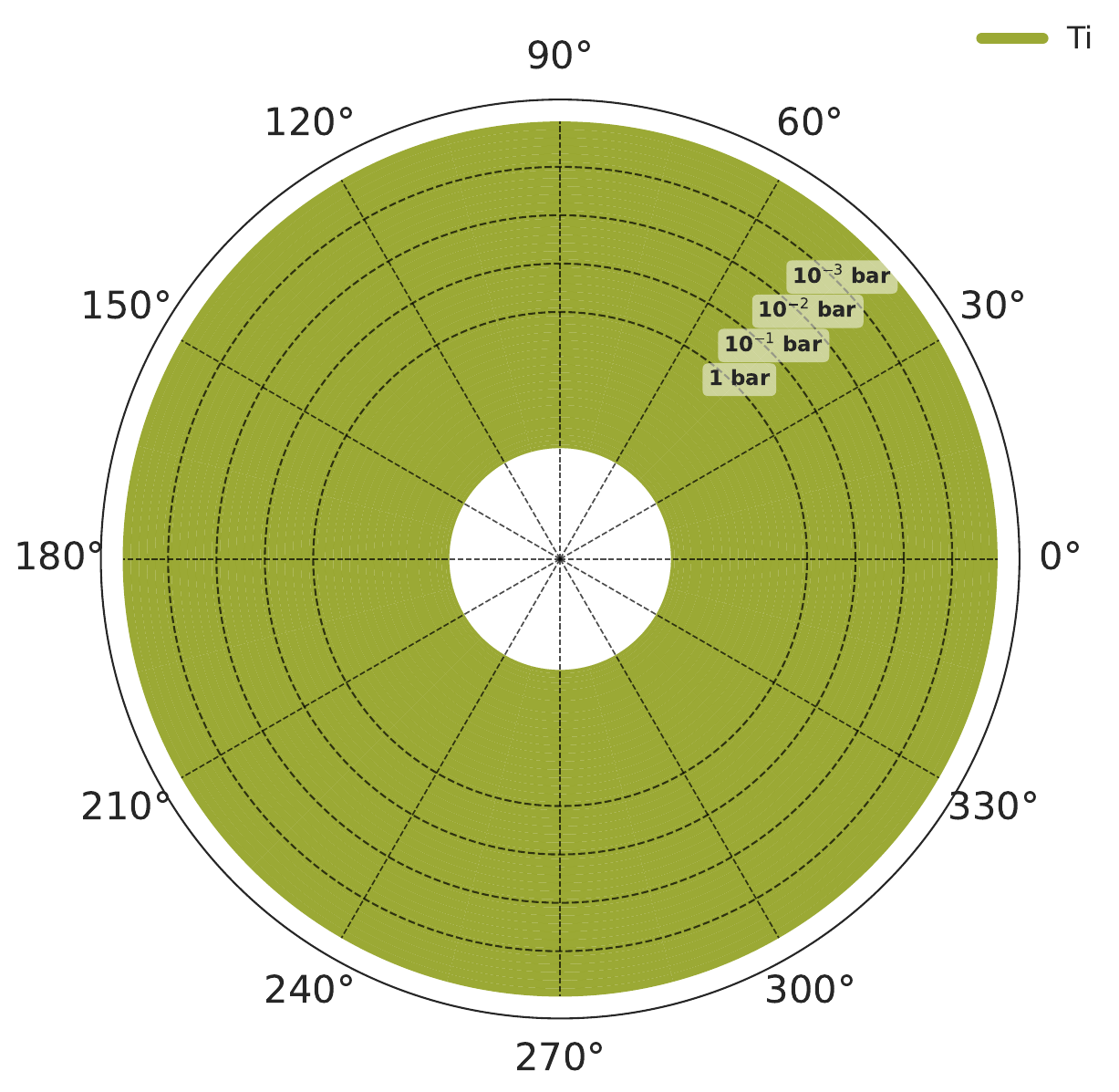}
\end{minipage}
\vspace{0.2cm}

\begin{minipage}{0.4\textwidth}
\centering
\includegraphics[width=0.9\linewidth]{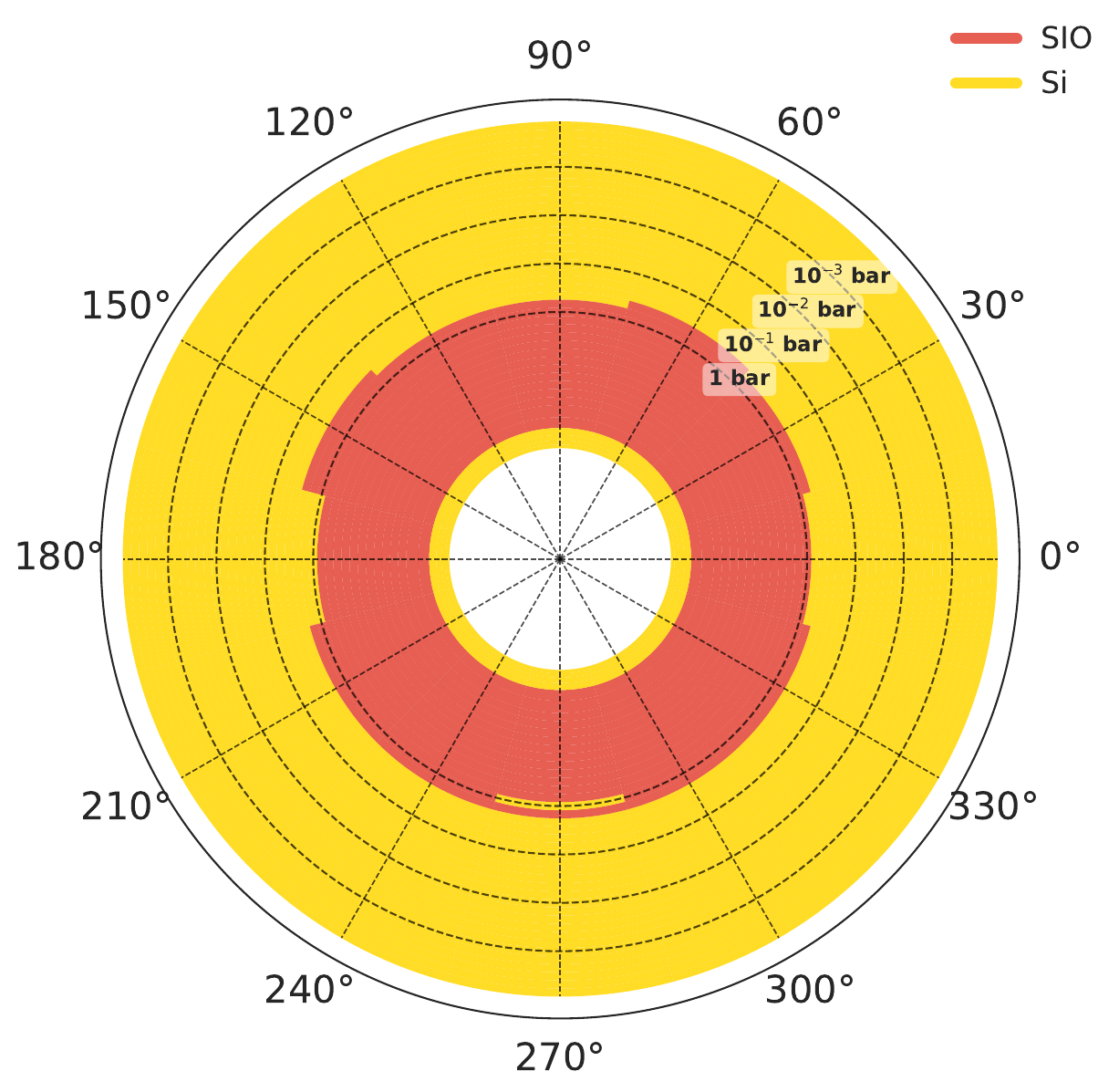}
\end{minipage}
\hspace{0.2cm}
\begin{minipage}{0.4\textwidth}
\centering
\includegraphics[width=0.9\linewidth]{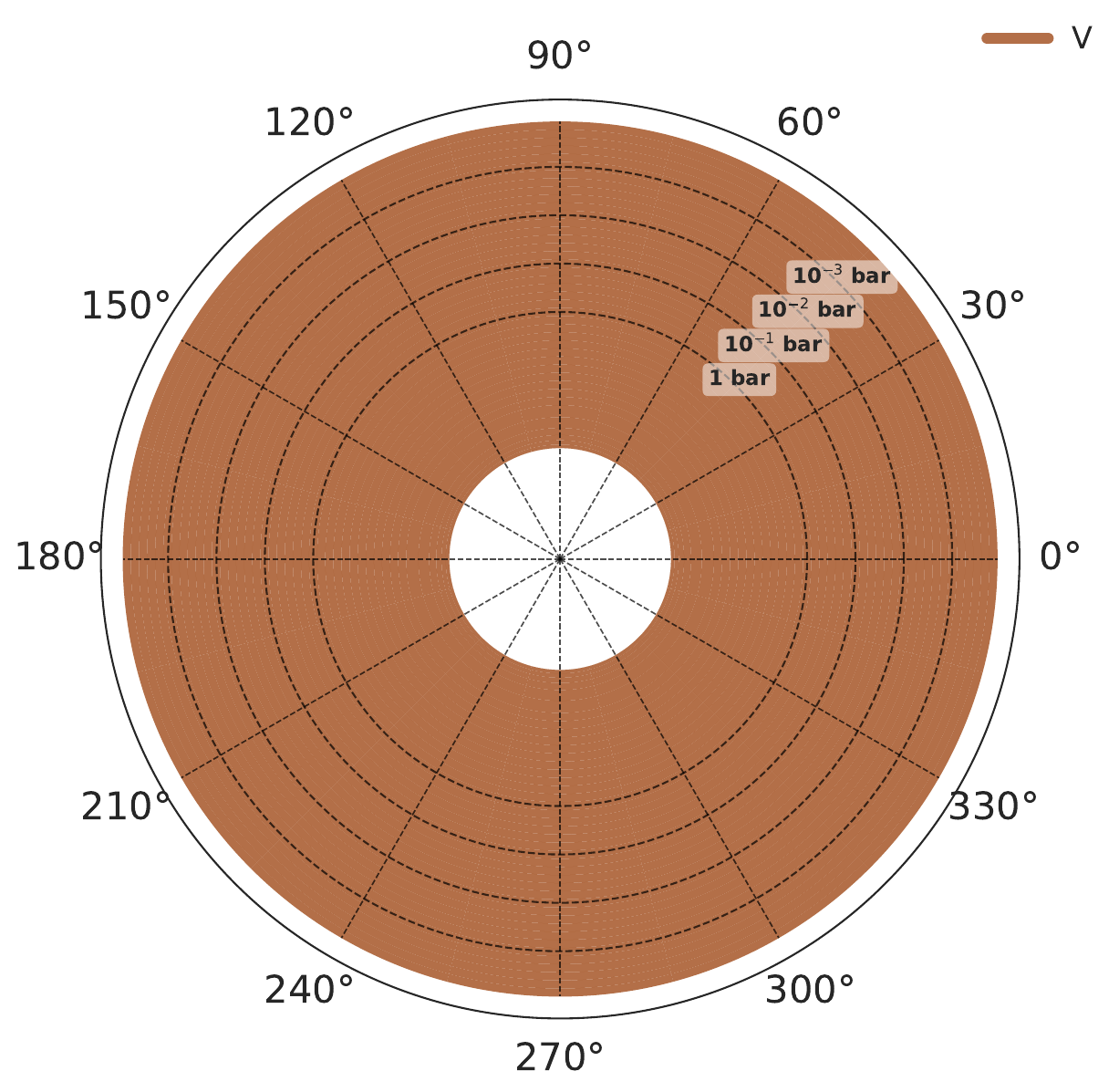}
\end{minipage}
\caption{2D equatorial plane slices ($\theta$ = 0$^\circ$) corresponding to 1D profiles extracted from the 3D GCM models. The panels show the most abundant species among metals, ions, and metal-oxide clusters for Al, Ti, Si, and V across WASP-39 b.}
\label{fig:GCMwap39b}
\end{figure*}
\end{document}